\documentclass[preprintnumbers, twocolumn,showkeys,superscriptaddress,showpacs,
nofootinbib,byrevtex,fleqn,prd,notitlepage]{revtex4-1}
\usepackage{bm,pifont}
\usepackage{amssymb}
\usepackage{amsmath}
\usepackage{amsfonts}
\usepackage{epsfig}
\usepackage{graphicx}
\usepackage{tabularx}
\usepackage{color}
\usepackage{slashed}
\newcommand{\seq}{\begin{subequations}}
\newcommand{\sen}{\end{subequations}}
\newcommand{\eq}{\begin{eqnarray}}
\newcommand{\en}{\end{eqnarray}}

\def\shiftdown#1{#1\llap{\lower.04ex\hbox{#1}}}

\def\nn{\nonumber}

\begin{document}
\title{Implication of the dark axion portal for the EDM of fermions and dark matter 
probing with NA64$e$, NA64$\mu$, LDMX, $\mbox{M}^3$,  and BaBar.}	
\date{\today}
	
\author{Alexey S. Zhevlakov  \footnote{{\bf e-mail}: zhevlakov@theor.jinr.ru}}
\affiliation{
Bogoliubov Laboratory of Theoretical Physics, JINR, 141980 Dubna, Russia} 
\affiliation{Matrosov Institute for System Dynamics and 
Control Theory SB RAS Lermontov str., 134, 664033, Irkutsk, Russia } \author{Dmitry V.~Kirpichnikov\footnote{{\bf e-mail}: kirpich@ms2.inr.ac.ru}}
\affiliation{Institute for Nuclear Research of the Russian Academy 
	of Sciences, 117312 Moscow, Russia} 
\author{Valery~E.~Lyubovitskij\footnote{{\bf e-mail}: 
		valeri.lyubovitskij@uni-tuebingen.de}} 
\affiliation{Institut f\"ur Theoretische Physik,
	Universit\"at T\"ubingen,
	Kepler Center for Astro and Particle Physics,
	Auf der Morgenstelle 14, D-72076 T\"ubingen, Germany}
\affiliation{Departamento de F\'\i sica y Centro Cient\'\i fico
	Tecnol\'ogico de Valpara\'\i so-CCTVal, Universidad T\'ecnica 
	Federico Santa Mar\'\i a, Casilla 110-V, Valpara\'\i so, Chile}
\affiliation{Millennium Institute for Subatomic Physics at
the High-Energy Frontier (SAPHIR) of ANID, \\
Fern\'andez Concha 700, Santiago, Chile}
\affiliation{Department of Physics, Tomsk State University,
634050 Tomsk, Russia} 
\affiliation{Tomsk Polytechnic University, 634050 Tomsk, Russia}

\begin{abstract}
The link between ordinary Standard model (SM) photon and both dark photon and axion like particle (ALP) can be 
introduced through the  dark axion portal coupling. 
Given the dark axion portal setup, in the present paper we refer the dark photon as the mediator between SM and dark
matter (DM) particles, implying that it decays predominantly into  pair of DM fermions. Furthermore,
we discuss in detail the  implication of the dark axion portal scenario for the lepton (electron and muon) fixed  target 
experiments. In particular, for the specific fixed target facility we study the  missing  energy signatures of  the dark 
photon production followed by its invisible decay into stable DM  particles.  We investigated the potential to probe dark axion 
portal vertices with regarding signatures and derive the expected sensitivities of NA64$e$,  NA64$\mu$, LDMX and 
$\mbox{M}^3$.  Moreover, we  estimated the expected reach of NA64$e$ from the projected statistics of the $J/\psi$ 
vector meson invisible  decays.  We also recasted BaBar monophoton  bounds for the specific dark axion portal scenario.
In addition,  we  modified the dark axion portal setup by including in the model both the hadron and lepton specific
ALP  couplings. As the result, we obtain the  bounds on the combination of fermion-specific couplings of ALP from 
the  fixed target  experiments.  We  discuss the implication of the modified  dark axion portal scenario for the
electric dipole moments (EDM) of SM fermions. In addition,  we derived  the novel constraints on the combination of the  
$CP$-violating neutron-specific ALP  couplings from the existing bounds on neutron EDM by taking into account the neutron anomalous 
magnetic  moment.
\end{abstract}

\maketitle

\section{introduction}

Axion and axion like particles (ALPs) arise naturally in the Standard Model (SM) extensions which
are  connected with $CP$--violation  problem in strong interaction physics \cite{Peccei:1977ur, DiLuzio:2020wdo}, 
can explain muon $(g-2)$ anomaly  \cite{Aoyama:2020ynm, Dorokhov:2014iva} and Dark Matter  (DM)
abundance~\cite{Boehm:2003hm,Dolan:2014ska,Hochberg:2018rjs}.
A more exotic cases of ALPs phenomenology that include a lepton flavor violation effects are 
studied  in  the  literature thoroughly~\cite{Han:2020dwo,Davoudiasl:2021mjy,Gninenko:2022ttd,Bauer:2020jbp}.
The ALPs and other dark sector particles have been extensively discussed recently in context of the experimental searches (see,  
e.~g., Refs.~\cite{Choi:2020rgn,Dusaev:2020gxi,NA64:2020qwq,Ishida:2020oxl,Sakaki:2020mqb,Brdar:2020dpr,Salnikov:2020urr,Bogorad:2019pbu,Kahn:2022uko,Darme:2020sjf,Dev:2021ofc,Abramowicz:2021zja,Fortin:2021cog,Asai:2021ehn,Balkin:2021jdr,Blinov:2021say,PrimEx:2010fvg,Gninenko:2016kpg,NA64:2016oww,Gninenko:2017yus,Gninenko:2019qiv,Banerjee:2019pds,Andreev:2021fzd,NA64:2021xzo,Blinov:2020epi,Beattie:2018xsk} 
and references therein). 

The link between SM and dark sector particles can be established through the idea of 
{\it portal}~\cite{Essig:2013lka}. 
For instance, that concept includes such scenarios as the Higgs portal~\cite{Arcadi:2019lka},  
the dark photon portal~\cite{Fortuna:2020wwx,Buras:2021btx,Kachanovich:2021eqa}, sterile neutrino portal~\cite{Escudero:2016tzx} and 
axion portal~\cite{Nomura:2008ru}. These portals offer systematic examination of the DM and also give  rise 
to novel experimental signatures.   

Recently, a new {\it dark axion portal} was suggested~\cite{Kaneta:2016wvf,Kaneta:2017wfh} 
to  provide the  dark photon  production mechanism in the  early Universe, 
implying the explanation of DM abundance due to the sufficiently light dark photon. 
Such portal  connects the axion or ALPs and dark photon via both axion--SM photon--dark photon  
and axion--dark photon--dark photon couplings. 

Now let us briefly summarize recent progress in the dark axion portal. In particular, an authors of 
Ref.~\cite{Gutierrez:2021gol} have developed the idea of dark axion portal in order to explain the DM cosmic 
density through the mixture 
of both axions and dark photons.  In addition, in Ref.~\cite{deNiverville:2018hrc} authors discussed in detail 
the implication of the  relevant portal for the leptonic $(g-2)$ anomalies, B--factories, fixed target neutrino 
experiments and beam dumps.  Also there has been previous study~\cite{deNiverville:2019xsx} of the 
monophoton signal for the  future experiments SHIP~\cite{Alekhin:2015byh} and FASER~\cite{Feng:2022inv} 
in the framework of dark axion portal.  
Moreover, the detailed  analysis of the regarding monophoton signatures and the expected sensitivities of 
reactor neutrino  experiments was presented in Ref.~\cite{Deniverville:2020rbv}. An authors of 
Ref.~\cite{Domcke:2021yuz} suggested the scenario that provides a novel link between the phenomenological 
dark axion portal, dark photons, and the hierarchy problem of the Higgs mass. In addition, the 
Ref.~\cite{Ge:2021cjz} proposed a new scenario of using the dark axion portal at one-loop level  
in order to explain recent result of the Fermilab Muon $(g-2)$ experiment on the muon anomalous magnetic moment. 

In the first part of this paper we develop the ideas presented in 
Refs.~\cite{Kirpichnikov:2020tcf,Kirpichnikov:2020lws} where
the implication of light sub--GeV bosons for  electric dipole moments (EDM) of 
fermions and $CP$--odd dark axion portal coupling was discussed in detail. 
In  particular,  we study the contribution of $CP$--violation vertices of SM
fermions and ALPs to EDM and derive novel constraint on combination of 
couplings of neutron with ALPs.  In addition, we argue that the EDM of 
fermions  can be induced by: (i) $CP$--odd  Yukawa--like  couplings of ALP,  (ii) 
$CP$--even  interaction of SM fermions and dark photon and (iii) $CP$--even dark 
axion  portal coupling. As a result, one can obtain the bounds on the  
combination  of regarding couplings. 

The second part of the paper develops the idea suggested in Ref.~\cite{deNiverville:2018hrc,deNiverville:2019xsx} that 
implies the probing the dark ALP portal scenario through the dark photon decaying predominantly to the DM particles. In 
this setup it is assumed that dark photon is the gauge field of the  hidden $U_D(1)$ 
group and thus it can serve the mediator between DM and SM particles via dark axion portal interaction.
We show that regarding scenario has a very broad phenomenological implication and 
can be probed via missing energy signatures in the existed, 
NA64$e$~\cite{Gninenko:2017yus,Gninenko:2019qiv,Banerjee:2019pds,NA64:2021xzo,Andreev:2021fzd} and
NA64$\mu$~\cite{Gninenko:2014pea,Gninenko:2018tlp,Kirpichnikov:2021jev,Sieber:2021fue}, and the projected, 
LDMX~\cite{Mans:2017vej,Berlin:2018bsc,LDMX:2018cma,Ankowski:2019mfd,Schuster:2021mlr,Akesson:2022vza}
and  $\mbox{M}^3$~\cite{Kahn:2018cqs,Capdevilla:2021kcf}, lepton fixed target facilities. We also 
recast the BaBar constraints on the dark ALP portal scenario with dark photon decaying mainly to  DM 
fermion pair.  In addition we address to $CP$--violating interaction of ALP with SM fermions in order to modify dark ALP 
portal scenario, as a result we obtain the expected bounds on the combination of hadron--specific and 
lepton--specific  couplings from NA64$e$, NA64$\mu$, LDMX and $\mbox{M}^3$.

The paper is organized as follows.
In Sec.~\ref{DescriptionSection} we provide the description of the considered modified dark ALP portal 
scenarios. In Sec.~\ref{EDMsectionLabel} we consider dark ALP portal applied to generation of EDM. In 
Sec.~\ref{ExperimentalBenchmark} we give a description of the missing energy signatures for the  
analysis of dark matter production at the fixed target experiments. In Sec.~\ref{MinimalSetupSection} 
we show that dark ALP portal couplings can be constrained  by using the data from $e^+e^-$ colliders and the 
experiments that exploit the  electron and muon beam impinging on the fixed target. 
In  Sec.~\ref{NonMinimalHadrophilicSect} and~Sec.\ref{NonMinimalLeptophilicSect} 
we obtain the constraints on combination of dark axion portal coupling with both hadron--specific and 
lepton--specific interactions, respectively. In Sec.~\ref{ResultsSection} 
we summarize  our main results.

\section{The description of the benchmark scenarios
\label{DescriptionSection}}
One of the possible connection between SM and dark sector particles is vector portal via the coupling of 
the SM and dark photons. One can describe this mixing by using the effective phenomenological Lagrangian 
\begin{equation}
\label{mixSD}
\mathcal{L}_{\mbox{\scriptsize vector portal}} = \frac{\epsilon}{2} F_{\mu\nu} F^{\prime\mu\nu}\,,  
\end{equation}
where $\epsilon$ is the kinetic mixing parameter 
between the two $U(1)$ gauge symmetries~\cite{Holdom:1985ag}, $F_{\mu\nu}$ and 
$F^{\prime}_{\mu\nu}$ are the strength tensors of SM electromagnetic and DM dark gauge (dark photon) 
fields, respectively. Nevertheless, in the present analysis we assume throughout the paper that 
$\epsilon \ll 1$, implying complete neglecting the vector portal mixing in Eq.~(\ref{mixSD}). 

In what follows, one can assume~\cite{Kaneta:2016wvf,Kaneta:2017wfh}  
that mixing between dark and electromagnetic photons can
be associated with ALPs interaction. 
The effective Lagrangian for such nonrenormalizable dark axion portal has form
\begin{equation}
\label{LDAP}
\mathcal{L}_{\mbox{\scriptsize dark\! axion\! portal}} = 
\frac{g_{a \gamma_D \gamma_D}}{4} a F^\prime_{\mu\nu} \widetilde{F}^{\prime\mu\nu}   
+  \frac{g_{a \gamma \gamma_D}}{2} a F_{\mu\nu} \widetilde{F}^{\prime\mu\nu}\,,  
\end{equation}
where the first term is the coupling between ALP and two dark photons, the second terms is the 
interaction  between ALP and both SM and dark photon.

First, we consider the  benchmark extension of the dark axion portal setup by exploiting  the 
dark matter (DM) Lagrangian, such that the dark photon serves the mediator between visible 
SM photon and hidden DM sector through the Lagrangian Eq.~(\ref{LDAP}). In particular, 
we specify throughout the  paper the Lagrangian  
\begin{equation}
     \mathcal{L} \supset \mathcal{L}_{\mbox{\scriptsize dark axion portal}}  + \bar{\chi}( \gamma^\mu  i \partial_\mu -  g_D \gamma^\mu  A_\mu'  + m_{\chi} )\chi,
 \label{MinimalSetupCoupling1}
\end{equation}
that is referred to  
\begin{itemize}
    \item  {\it the  minimal dark ALP portal scenario}. 
\end{itemize}
The field $\chi$ in Eq.~(\ref{MinimalSetupCoupling1}) is  a Dirac  DM  fermion of mass $m_\chi$ from the 
dark sector,  $g_D$ is the coupling constant between DM and dark photon $A^\prime_\mu$ field, 
that is associated with hidden charge of the $U_{D}(1)$ gauge group. 
Moreover, we assume that dark photon decays predominantly to dark fermion $\chi \bar{\chi}$ pair. 

Second, we also employ a simplified model framework, with $CP$-violation Yukawa coupling between ALP and
SM fermions than can be written as:
\begin{align}
      \mathcal{L} \supset & \mathcal{L}_{\mbox{\scriptsize dark axion portal}} 
   + \bar{\chi}( \gamma^\mu  i \partial_\mu -  g_D \gamma^\mu  A_\mu'  + m_{\chi} )\chi 
 \nn   \\ 
 & + \sum_{\psi} a \bar{\psi} (g_\psi^s+ig_\psi^p \gamma_5) \psi \,,
    \label{LeptophilicAndHadroPhilicLagr1} 
\end{align}
where the  couplings, $g_\psi^s$ and $g_\psi^a$, are 
restricted to be from two benchmark scenarios
\begin{itemize}
\item {\it the  non--minimal lepton--specific ALP setup}: 
$g_\psi^s \ne 0, \, g_\psi^p \ne 0$ for $\psi = e, \mu$, 
where $e$ and $\mu$ are electron and muon respectively, and  $g_\psi^s =g_\psi^p = 0$ for $\psi = n, p$, where 
$p$ and $n$ are proton and neutron respectively;
 \item {\it the non--minimal hadron--specific ALP setup}:  $g_\psi^s \ne 0, \, g_\psi^p \ne 0$ for $\psi = n, p$ and  
 $g_\psi^s =g_\psi^p = 0$ for $\psi = e, \mu$.
\end{itemize}

 The Yukawa--like couplings of the  
Lagrangian~(\ref{LeptophilicAndHadroPhilicLagr1}) may originate from an effective interation~\cite{Gunion:1989we,Djouadi:2005gj,Branco:2011iw,Chun:2021rtk} 
in framework of the the two--Higgs--doublet--model. 
 Moreover since such terms 
 violate the  $CP$ symmetry,
thus they can induce the  electric dipole moment (EDM) \cite{Dzuba:2018anu, Flambaum:2009mz} of 
fermions.  It is worth to estimate the bounds on the regarding couplings from the EDM of fermions in the 
framework of both the lepton--specific and  hadron--specific scenarios. That study is of particular interest 
of the  present paper. In Sec~\ref{EDMsectionLabel} we discuss  its  phenomenological implication. 
  
In addition, for the specific benchmark scenario one can obtain also the upper limit/expected sensitivity on either the coupling 
$g_{a\gamma\gamma_D}$ or  the  combination of couplings $|g_{a\gamma\gamma_D} g^{s}_\psi|$ and $|g_{a\gamma\gamma_D} g^{p}_\psi|$ 
%(see e.~g.~Eqs.~(\ref{LDAP}),~(\ref{MinimalSetupCoupling1}) and~(\ref{LeptophilicAndHadroPhilicLagr1})) 
from the null result of DM detection in the BaBar  $e^+e^-$ collider 
and in the fixed target experiments such as NA64$e$($\mu$) at  
CERN SPS and LDMX,  M$^3$ at FermiLab.  We discuss that in detail in 
Secs.~\ref{ExperimentalBenchmark},~\ref{MinimalSetupSection},~\ref{NonMinimalHadrophilicSect},  
and~\ref{NonMinimalLeptophilicSect}.

\section{EDM
\label{EDMsectionLabel}}

Let us consider the  bounds on ALPs couplings from the experimental constraints on EDM fermions 
(electron, muon, and neutron) generated by $CP$--violating Lagrangian~\cite{Yan:2019dar,OHare:2020wah}. We 
can estimate a contribution to EDM by calculation a simple Feynman diagrams (see Fig.~\ref{fig:002}) with
intermediate ALP exchange~\cite{Kirpichnikov:2020tcf}.
The contribution to fermion EDM with mass $m_\psi$ has a form: 
\begin{figure}[b]		
		\includegraphics[width=0.5\textwidth, trim={2cm 21.5cm 1cm 2cm},clip]{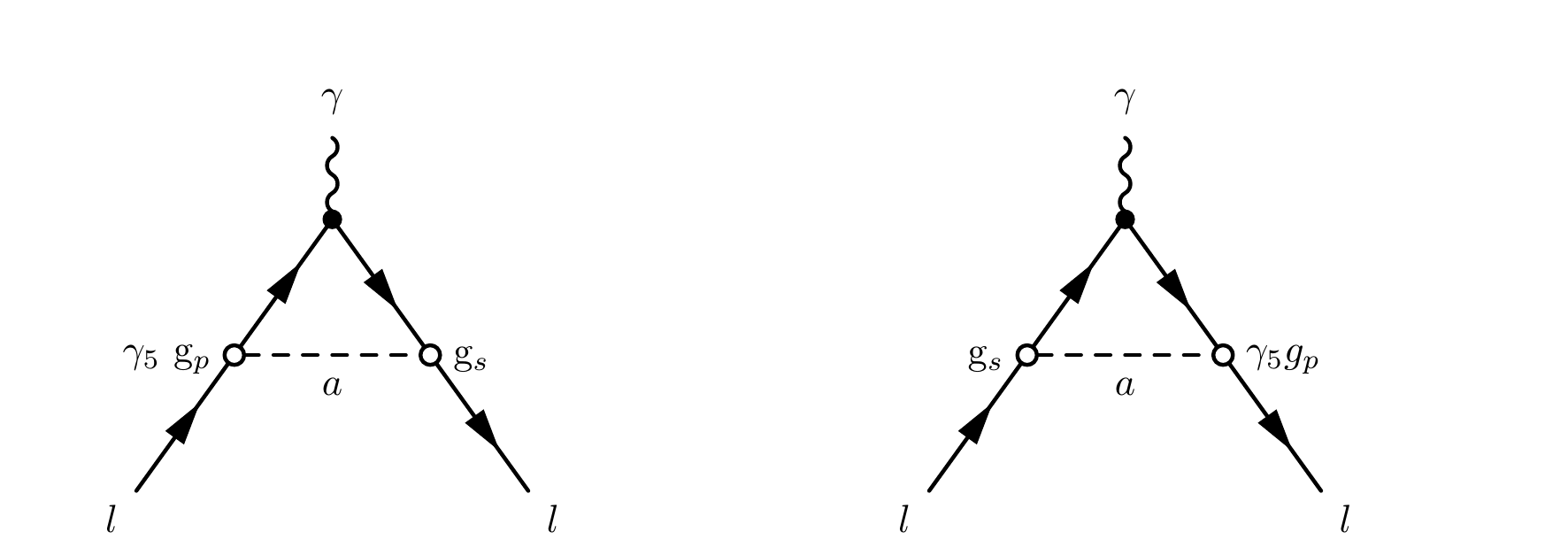}
	\caption{Feynman diagrams with taking into account ALPs which generate EDM of charge fermions, leptons.}
	\label{fig:002}
\end{figure}

\begin{figure}[b]		
		\includegraphics[width=0.5\textwidth, trim={2cm 21.5cm 1cm 2cm},clip]{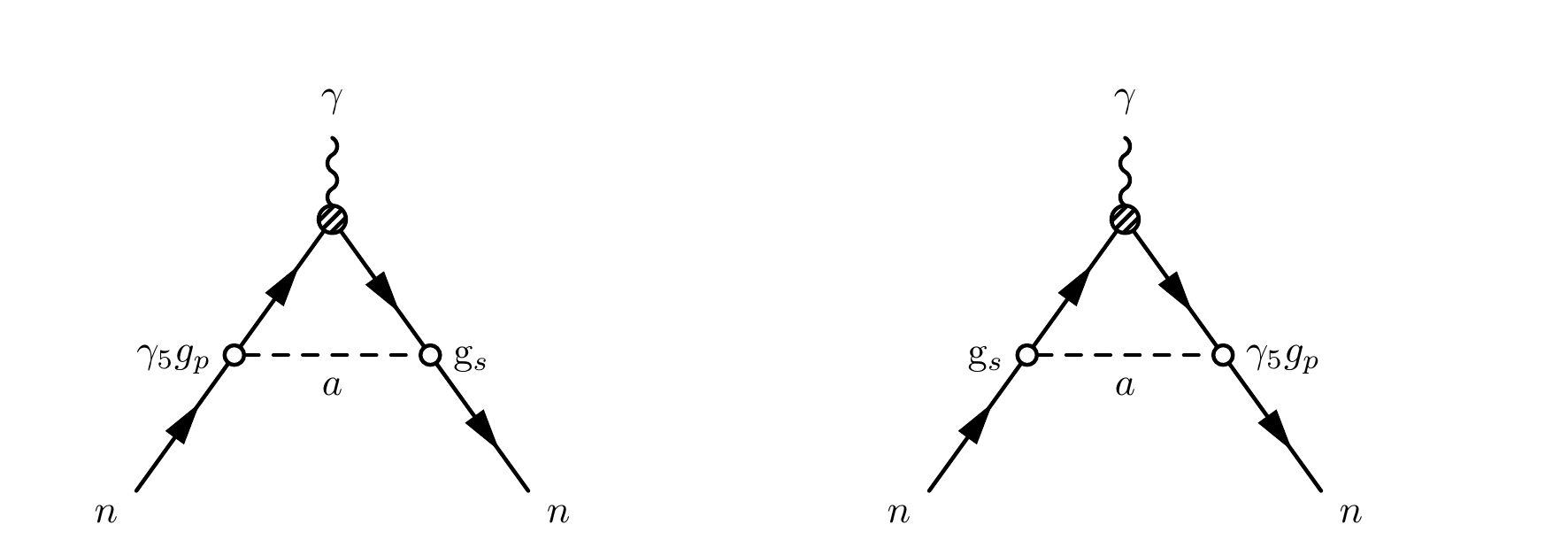}
	\caption{Feynman diagrams with taking into account ALPs which generate neutron EDM induced by 
	nonminimal electromagnetic couplings (anomalous magnetic moments) 
	with external electromagnetic field (shaded blob).}
	\label{fig:002N}
\end{figure}

\begin{align}
d_\psi = \frac{e g_\psi^s g_\psi^p}{8\pi^2 m_\psi} g_i(y)
\label{ALP_EDM_contr}
\end{align}
where index $i=1$ labels the lepton couplings $(\psi =e, \mu)$
 with the function $g_1(y)$  being
\eq
g_1(y)=\int^1_0 dx \frac{x^2}{x^2+y^2(1-x)},
\label{gy1}
\en
here $y=m_a/m_\psi$. For neutron case
($\psi = n$) we specify the index $i=2$ in Eq.~(\ref{ALP_EDM_contr}). 
Then one should account the neutron interaction with external  electromagnetic field $A_\mu$ through the 
anomalous magnetic moment (see diagrams on Fig.\ref{fig:002N}). The regarding  Lagrangian can be written 
in the following form 
\eq
\mathcal{L}_{ANN}=ieA_{\mu} \bar{N}\left(\gamma^\mu Q_N+\frac{i\sigma^{\mu\nu}q_{\nu}}{2m_N}k_N\right) N \,,
\en
where $m_N$ is the nucleon mass, $Q_N=\mathrm{diag}(1, 0)$ and 
and $k_N = \mathrm{diag}(k_p, k_n)$ are the matrices of the electic charges and anomalous magnetic moments of 
nucleons (proton $p$ and neutron $n$), respectively, with 
$k_p = 1.793$ and $k_n = -1.913$. Here,  
$\sigma^{\mu\nu}=\frac{i}{2}[\gamma^\mu,\gamma^\nu]$, where $\gamma^\alpha$ are Dirac gamma matrices. 
Using magnetic moment of neutron we can calculate its contribution to the EDM generated by ALP exchange 
at 1-loop level (see, e.~g., Eq.~(\ref{ALP_EDM_contr}) and Fig.~\ref{BarZeeALPPortal1} for detail), 
the regarding $g_2(y)$  structure of the Feynman integral can be written as
\begin{align}
  g_2(y)=     k_n\int^1_0  dx 
  \frac{(1-x)(1-x^2)}{(1-x)^2+y^2x},  
  \label{gy2}
\end{align} 
where $y=m_a/m_N$. Analytic expression for the loop integrals $g_i(y)$ is presented in Appendix~\ref{AppLoop}.

\begin{figure}[t!]		
	\includegraphics[width=0.49\textwidth, trim={0cm 0cm 0cm 0cm},clip]{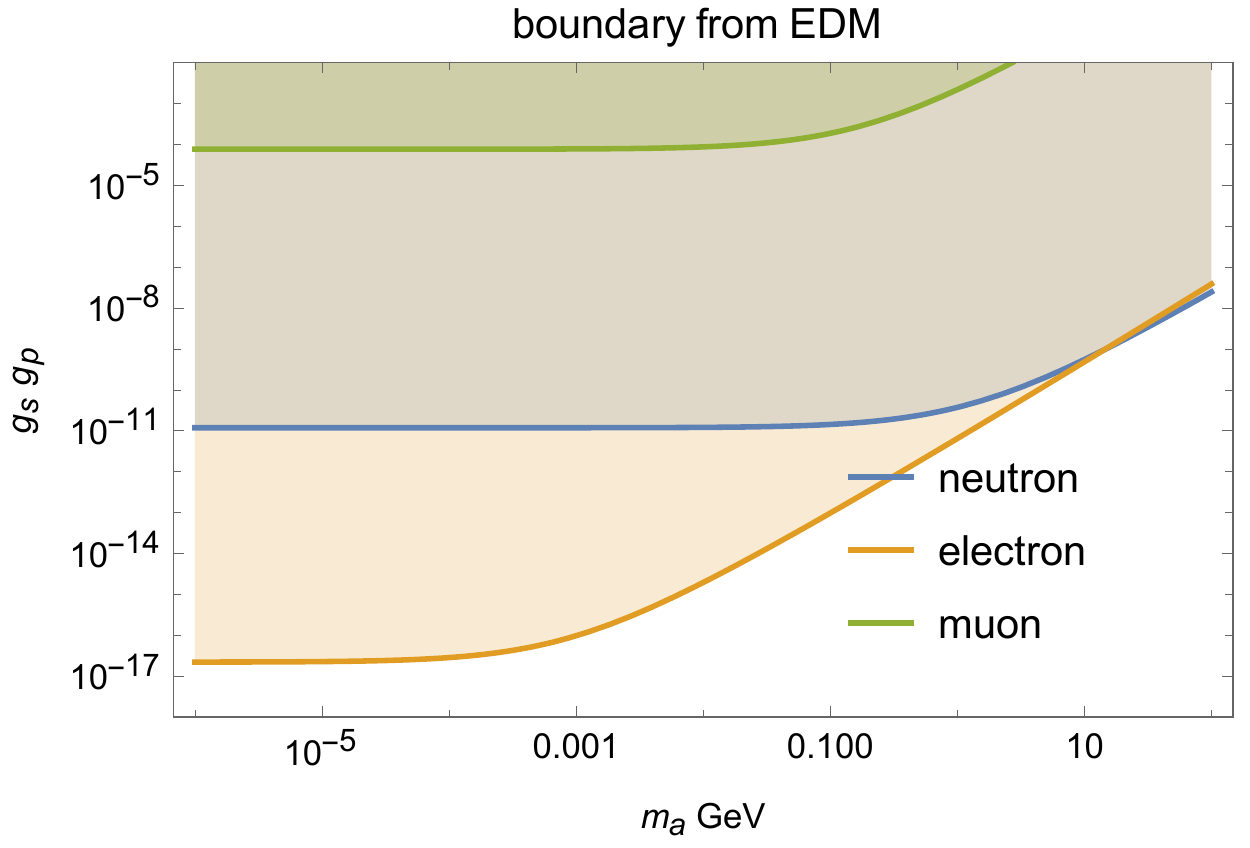}
	\caption{Boundaries for the product of scalar and pseudoscalar couplings from EDM neutron 
	and electron/muon as the function of the ALP mass. These boundaries correspond to leptophilic and hadrophilic scenarios.} 
	\label{from_edm}
\end{figure}

% \begin{figure}[!hb]
\begin{figure}[!t]
\centering
\includegraphics[width=0.5\textwidth, trim={4cm 21cm 1cm 2cm},clip]{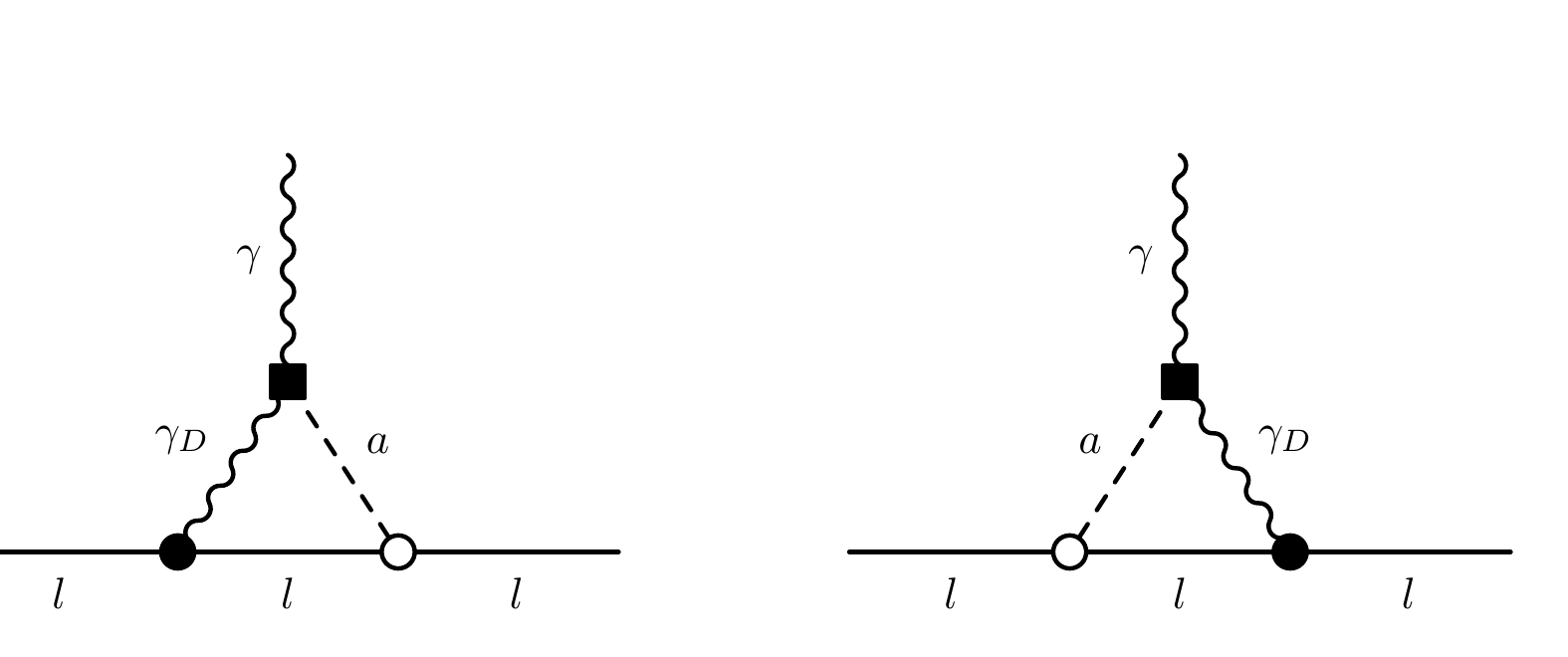}
\caption{Feynman diagrams which generate EDM terms due to axion with both  dark and SM photons 
(see, e.~g., Eq.~(\ref{LagrangianVectorAxionEDM}) for detail).    \label{BarZeeALPPortal1}}
\end{figure}

The boundaries for combination of couplings $|g^\psi_s g^\psi_p|$~\cite{Kirpichnikov:2020lws,Dzuba:2018anu} 
as a function of the  ALP mass $m_a$ are shown in 
Fig.~\ref{from_edm}. Difference of behavior different boundaries is
connected with mass of fermions and existed limits to EDM of ones ($|d_e|<1.1 \cdot 10^{-29} \,\,
\text{e} \cdot \text{cm}$, $|d_\mu|<1.8 \cdot 10^{-19} \,\,
\text{e} \cdot \text{cm}$  and $|d_n|<1.8 \cdot 10^{-26} \,\,\text{e} \cdot \text{cm}$ see PDG
Ref.~\cite{PDG20}).  Bounds  on $|g^n_s g^n_p|$ from neutron EDM for light mass 
of ALPs are proportional to $\bar{\theta}$, parameter of $CP$-violation of vacuum in the quantum 
chromodynamics (QCD)  which typical value is $\sim 10^{-10}$ \cite{Zhevlakov:2020bvr,Crewther:1979pi}. It is general for neutron EDM feature due the light  scalar/pseudoscalar boson exchange in one loop  \cite{Crewther:1979pi} or two 
loops  with $CP$--violation  vertex~\cite{Zhevlakov:2018rwo,Zhevlakov:2019ymi,Zhevlakov:2020bvr,Gutsche:2016jap}. 

However from EDM of fermions we can  estimate only the combination of ALPs couplings $g^\psi_s$ 
and $g^\psi_p$. Wherein we want to note that the Lagrangian with universal 
$CP$--violating couplings to the SM fermions
\begin{align}
\label{ALP}
\mathcal{L}_{\,  \not \! CP } \supset  \sum_\psi \bar{\psi} a (g_\psi^s+ig_\psi^p \gamma_5) \psi \,, 
\end{align} 
 plays important role in spin interaction. In particular, by exploiting the data on Schiff moments 
 of atoms or molecules its is straightforward to constrain the combination of couplings $|g^s_N g^p_e|$ 
 (see, e.~g., Refs.~\cite{Dzuba:2018anu,Maison:2022zaz,OHare:2020wah} for detail).

We would like to stress that the ALPs in the considered scenario could contribute 
to the fermion EDMs. Corresponding matrix element is induced by the $CP$--violating diagrams 
analogous to the Barr-Zee diagrams with ALPs exchange in loop (see Fig.\ref{BarZeeALPPortal1}).
This contribution is generated by an additional interaction Lagrangian containing three terms: 
(i) $P$-parity violating coupling of ALP with SM fermions, 
(ii) $P$-parity conserving coupling of ALP with SM photon and dark photon, 
and (iii) $P$-parity conserving coupling of dark photon with SM fermions, 
\begin{eqnarray}
\mathcal{L} \supset a \Big[ g_s \overline{\psi} \psi + \frac{g_{a \gamma \gamma_D}}{2} F_{\mu\nu} \widetilde{F}^{\prime\mu\nu} 
\Big] + e \epsilon \overline{\psi} \gamma^\mu A_\mu' \psi \,.  
\label{LagrangianVectorAxionEDM}
\end{eqnarray}

The detailed analysis of present bounds 
would require the consideration of both the vector portal and dark ALP portal 
scenarios. Such analysis is beyond the scope of the present paper and we leave it for future study. 
In particular, we consider regarding limits elsewhere~\cite{ZhevlakovEtAl},  
just collecting some general formulas in  Appendix~\ref{App_A}.

 \begin{figure}[!t]
\centering
\includegraphics[width=0.48\textwidth, trim={3.8cm 20.2cm 1cm 2cm},clip]{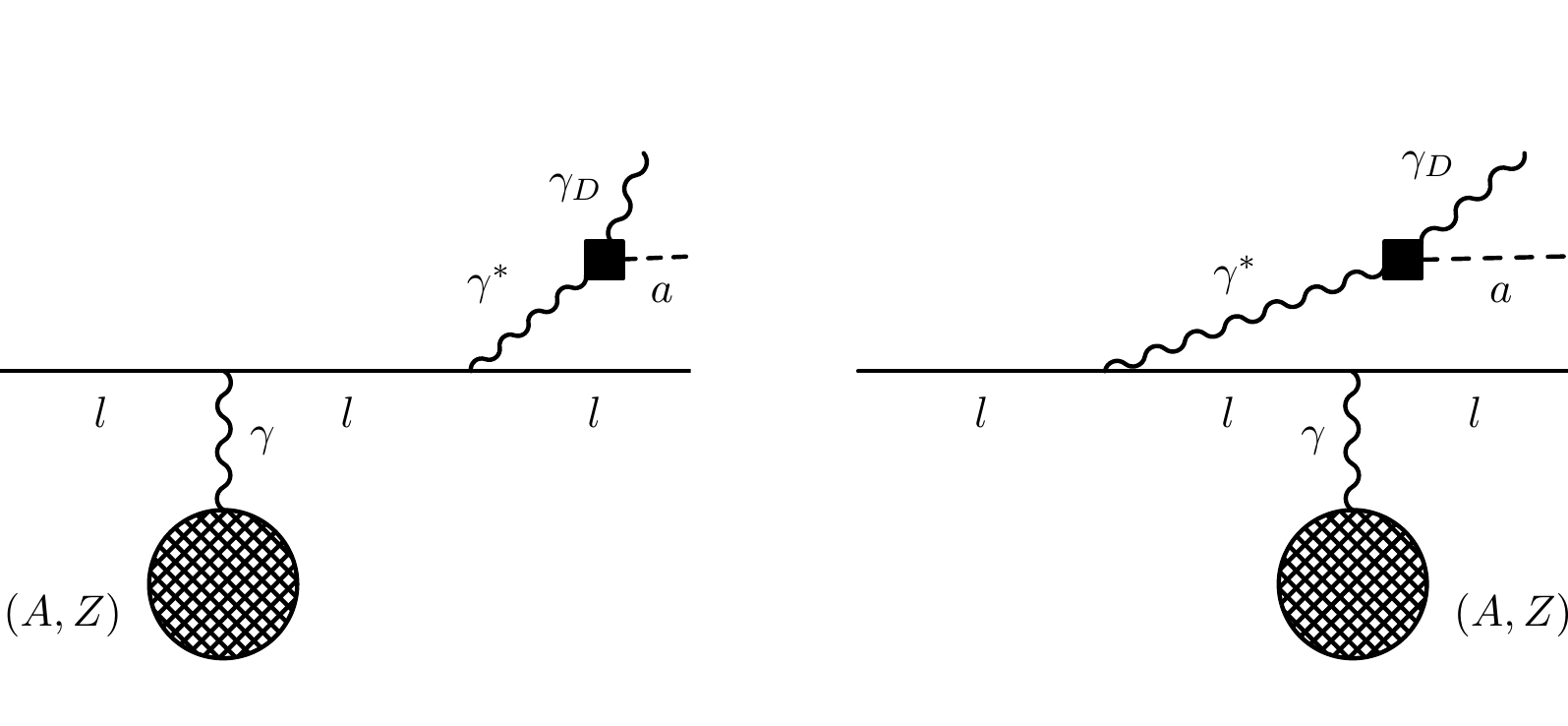}
\caption{ Feynman diagrams for the radiate ALP and dark photon production by an impinging lepton on a nucleus 
$Z$. The regarding process  represents the  missing energy signature for the minimal dark
ALP portal  scenario $\mathcal{L} \supset \frac{1}{2} g_{a\gamma \gamma_D} a F_{\mu \nu} 
\widetilde{F}_{\mu \mu}'$. Note that we neglect in the calculation
the diagrams with off-shell photon from the nucleus leg, since its contribution to the signal process 
is suppressed by a factor of $(Z m_l/M_Z)^2$ for both  electron and muon impinging on the nucleus.
\label{MinimalALPportalBremsFeynman}}
\end{figure}

%\begin{figure}[!tbh]	
\begin{figure}[!t]	
		\includegraphics[width=0.49\textwidth, trim={1.8cm 21.2cm 1cm 2cm},clip]{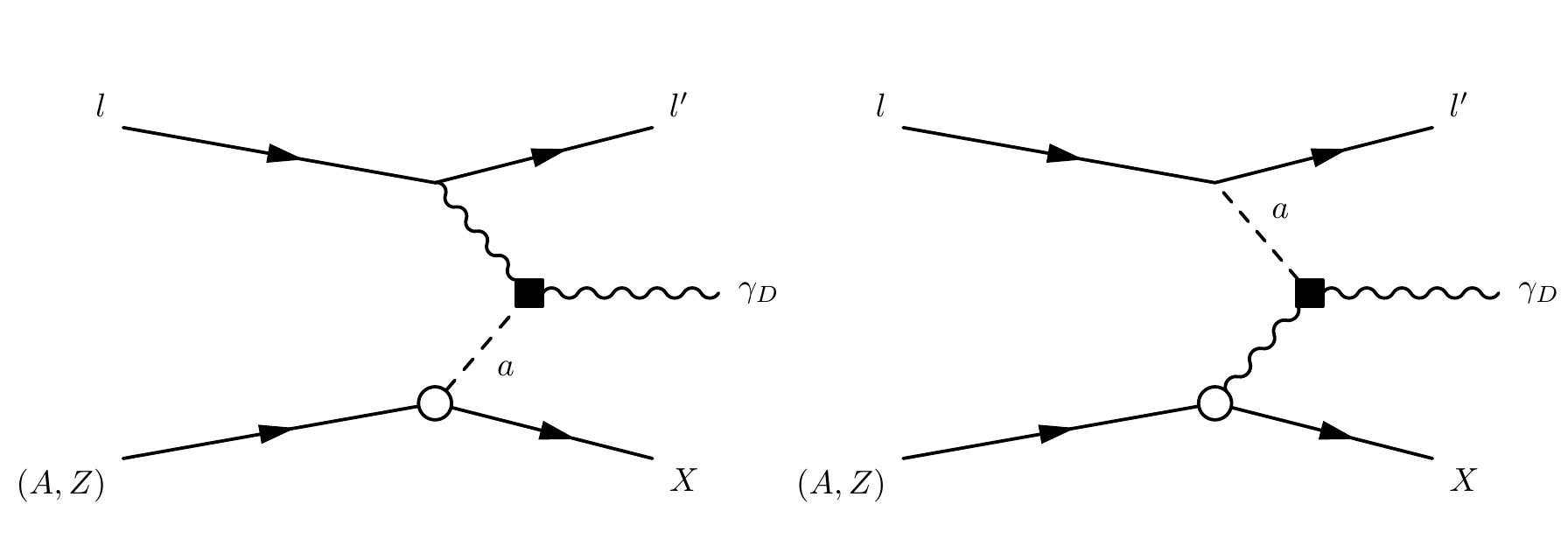}
	\caption{Feynman diagrams for processes of the lepton scattering off atomic target $(A,Z)$ 
	in which the SM and dark photon interact through the axion portal coupling for both ALP--hadrophilic 
	(left) and ALP--leptophilic (right) scenarios.} 
	\label{axionPhotonand_dark}
\end{figure}

   \begin{figure*}[!tbh]
\centering
\includegraphics[width=0.99\textwidth, trim={1.8cm 15.7cm 1cm 2.5cm},clip]{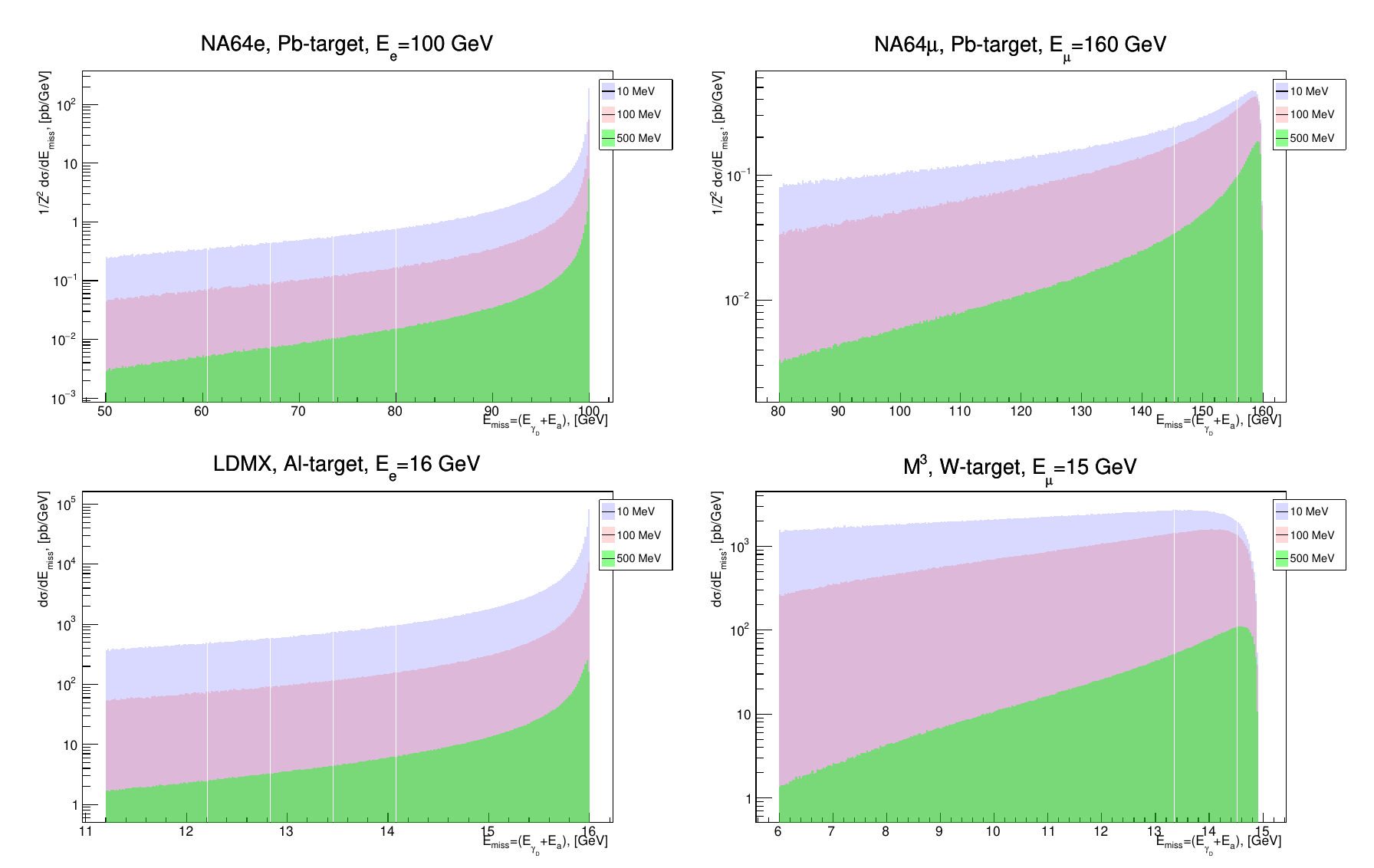}
\caption{  The differential spectra of the process $l Z \to l Z a \gamma_D(\to \chi \bar{\chi})$
as function of the  missing energy $E_{miss}=E_{\gamma_D}+E_a$  
 for various experimental fixed target facilities and for the set of dark photon masses, 
$m_{\gamma_D}=10\, \mbox{MeV}$, $m_{\gamma_D}=100\, \mbox{MeV}$ and $m_{\gamma_D}=500\, \mbox{MeV}$ 
in the framework  of minimal dark ALP portal scenario.
We set $g_{a\gamma\gamma_D}=1\, \mbox{GeV}^{-1}$ and $m_{a}=10 \, \mbox{keV}$. 
\label{DiffCSVariousExperimentsMinimalPortal}}
\end{figure*}
  
   \begin{figure}[t]
\centering
\includegraphics[width=0.5\textwidth]{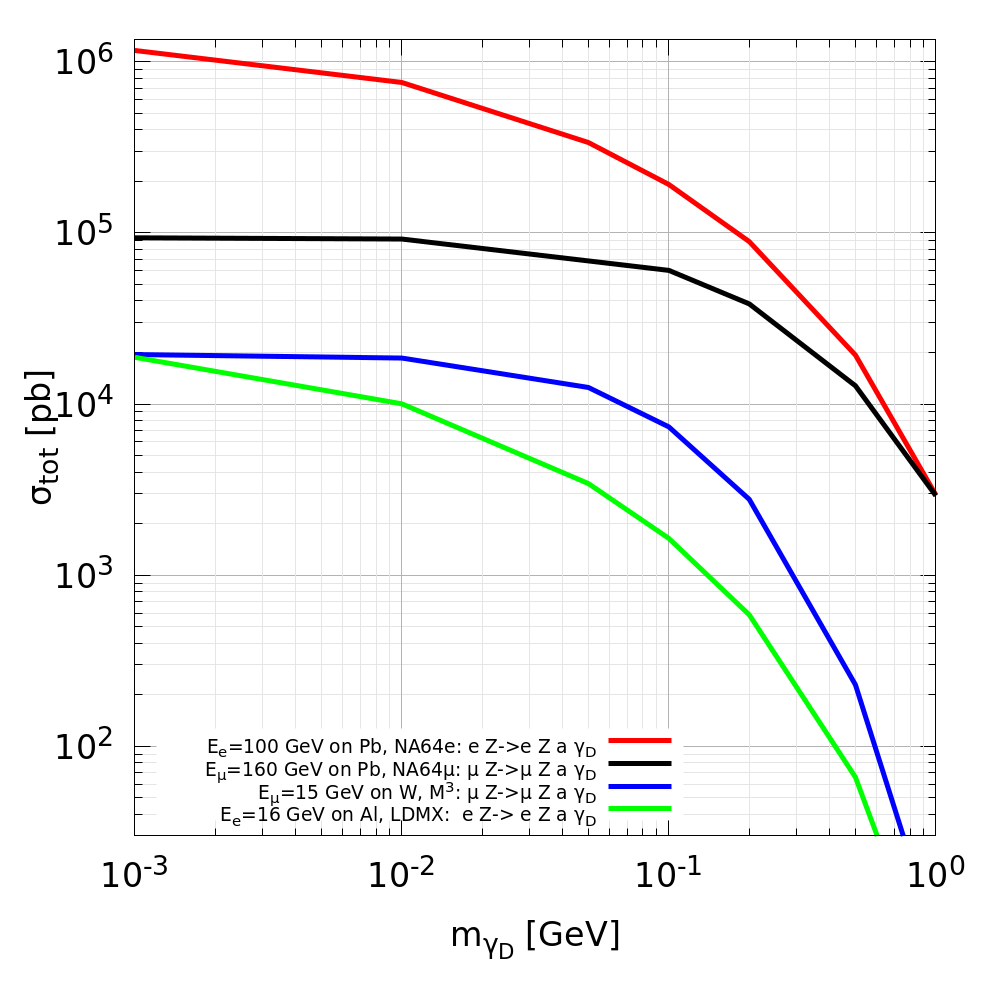}
\caption{ The total  cross-section as a function of the mass 
$m_{\gamma_D}$ for the minimal portal scenario, $m_a\simeq 10\,\mbox{keV}$. 
We set here  $g_{a\gamma\gamma_D}=1 \, \mbox{GeV}^{-1}$ and integrate the cross-section  
over the experimental cut range  $x_{min} \lesssim x \lesssim x_{max}$. Red line is the 
cross-section for the NA64$e$ experiment, black line corresponds to NA64$\mu$ experiment, 
blue line is the total cross-section for $\mbox{M}^3$ experiment and green line corresponds 
to the cross-section of the LDMX facility.  
\label{MinimalPortalFigCS}}
\end{figure}

\section{The missing energy signal
\label{ExperimentalBenchmark}}

In this section we discuss the  setups for the 
fixed target experiments such as 
NA64$e$~\cite{Gninenko:2017yus,Gninenko:2019qiv,Banerjee:2019pds,NA64:2021xzo,Andreev:2021fzd},
LDMX~\cite{Mans:2017vej,Berlin:2018bsc,LDMX:2018cma,Ankowski:2019mfd,Schuster:2021mlr,Akesson:2022vza}, 
NA64$\mu$~\cite{Gninenko:2014pea,Gninenko:2018tlp,Kirpichnikov:2021jev,Sieber:2021fue} and
$\mbox{M}^3$~\cite{Kahn:2018cqs,Capdevilla:2021kcf},  
which can potentially probe the invisible  signatures associated with a lepton missing energy process 
\begin{equation}
l Z \to Z l(E_{miss})\,,
    \label{generalMissinEnergyProcess}
\end{equation}
where $l=(e,\mu)$ is the label  either  for electron or muon beam and $Z$ designates the target nucleus.
For  instance, in the framework of the minimal dark ALP portal scenario  the  missing energy of  the lepton beam 
$E_{miss}$ can arise  from the production of ALP $a$ and dark photon  $\gamma_D$ by the off-shell photon 
$\gamma^*$ in the process  
$l Z \to l Z \gamma^* \to l Z a \gamma_D (\to \bar{\chi}\chi)$ 
(see, e.~g., Fig.~\ref{MinimalALPportalBremsFeynman} for detail). To calculate the regarding yield 
we use state--of--the--art {\tt CalcHEP} package~\cite{Belyaev:2012qa}.

In addition, for the hadron-specific and lepton-specific  couplings of the ALPs the missing
energy process $l Z \to l Z  \gamma_D(\to \bar{\chi} \chi)$ is possible.  This reaction implies that  dark photon 
production is associated with interaction of ALP with SM photon which is produced either by beam or target
(see, e.~g., Fig.~\ref{axionPhotonand_dark} for detail). 
To calculate cross--section of these processes we exploit the approximation of equivalent 
photon, also known as the Weizsacker--Williams (WW)~\cite{Budnev:1975poe,Engel:1995pu,Vysotsky:2018slo} approximation. 
This approach is common for DM production study in the beam dump and fixed target experiments. 
It implies replacing the fast moving charged particles by a specific photon distribution. 

The dark photon can decay through the different channels in the dark ALP portal scenario. 
In particular, as soon as $m_{\gamma_D} \gtrsim m_a$
the visible two--body decay through the ALP and photon
is kinetically accessible with a decay width~\cite{deNiverville:2019xsx} 
\begin{equation}
\Gamma_{\gamma_D \to a \gamma} = \frac{g_{a\gamma\gamma_D}^2}{96 \pi} 
m_{\gamma_D}^3 \left(1-\frac{ m_a^2}{m_{\gamma_D}^2}\right)^3 \,.
\end{equation}
Lagrangian~(\ref{MinimalSetupCoupling1}) implies that the invisible two-body decay of 
dark photon $\gamma_D$ into $\bar{\chi}\chi$ pair is also allowed with a decay width~\cite{Bondi:2021nfp}
\begin{equation}
\Gamma_{\gamma_D\to \bar{\chi}\chi } = \frac{g_D^2}{12 \pi} m_{\gamma_D} 
\left(1+\frac{2 m_\chi^2}{m_{\gamma_D}^2}\right) \left(1-\frac{4 m_\chi^2}{m_{\gamma_D}^2}\right)^{1/2} \,.  
\end{equation} 

In the present paper we will focus on the process of invisible channel of 
dark photon decay into pair of hidden dark fermions, 
$\gamma_D \to \bar{\chi} \chi$ with  $\mbox{Br}(\gamma_D \to \bar{\chi}\chi)\simeq 1$ for 
$m_\chi \ll m_{\gamma_D}$.  That implies the following  condition on 
decay widths 
$\Gamma_{\gamma_D \to  \bar{\chi}\chi} \gg \Gamma_{\gamma_D\to a \gamma}$ and as
a result this yields  $g_D \gg g_{a\gamma\gamma_D} m_{\gamma_{D}}$.  
Therefore it leads  to the rapid decay of dark photon into 
$\bar{\chi}\chi$ pair after its production.  In addition, in our analysis throughout the paper we keep the 
ALP mass $m_a$  well below  $m_{\gamma_D}$, such that $m_a \ll  m_{\gamma_D}$,  to get rid of the possible visible decay  signatures  $a\to \gamma \gamma_D$  in the detector of the fixed target facility. 

To begin with,  we estimate the number of missing energy events
for the lepton beam at fixed target as follows
 \begin{equation}
N_{sign} \simeq \mbox{LOT}\cdot \frac{\rho N_A}{A} L_T \!\!\!\int\limits^{x_{max}}_{x_{min}} \!\!
dx \frac{d \sigma (E_l)}{dx} \mbox{Br}(\gamma_D \to \chi \bar{\chi})\,,
\label{NumberOfMissingEv1}
 \end{equation}
where $E_l$ is the initial energy of the beam, $A$ is atomic weight number of target material, 
$N_A$ is Avogadro's number, $\mbox{LOT}$ is number of 
leptons accumulated on target, $\rho$ is target density, $L_T$ is the effective thickness of the 
target, $d \sigma/dx$ is differential cross--section for the specific missing energy channel $l Z \to  Z l(E_{miss})$, 
$x_{min}$ and $x_{max}$ are the minimal and maximal fraction of the lepton energy $x=E_{miss}/E_l$ that hidden 
particles carry away. The  cuts on $x$ are  determined by specific fixed--target facility.   

Let us discuss now the benchmark input parameters for the specific experimental setup. 

\subsection{The NA64$e$ experiment} The dark photon and/or ALP  
can be produced in the reaction of high-energy electrons of
$E_e=100\,\mbox{GeV}$ scattering off
nuclei  of an active lead ECAL target
\begin{equation}
e Z \to e Z \gamma_D (a) \,, 
\label{eZtoeZInvis1}
\end{equation}
followed by prompt $\gamma_D \to \bar{\chi} \chi$ decay into dark matter particles  ($\chi$). Thus,  
the fraction $E_{miss}=x E_e$ 
of the primary beam energy is carried away by the $\chi$'s ($a$) 
which penetrate the detector of NA64$e$ without energy deposition. 
The remaining part of the beam energy fraction $E_e^{rec}\equiv E_e (1 -x)$
is deposited in the ECAL target by the scattered electrons. Therefore, the 
occurrence of the hidden particles produced in the process~(\ref{generalMissinEnergyProcess}) 
leads to an excess of events with a single electromagnetic (EM) shower with energy $E^{rec}_e$ above the expected 
background (see, e.~g., Ref.~\cite{NA64:2016oww} for detail).
In the present analysis, we conservatively assume that the EM shower
is localized in the first radiation length of the lead detector 
$X_0=0.56\, \mbox{cm}$, such that the effective thickness of the target in~(\ref{NumberOfMissingEv1}) 
is  $L_T \simeq X_0$.  That implies that the dominant 
production of the hidden particles occurs within first radiation length of the active 
ECAL target~\cite{Chu:2018qrm}. The candidate events are requested to have the missing energy in 
the range $E_e^{rec} \lesssim 0.5 E_e$, implying that  $x_{min} =0.5$ in 
Eq.~(\ref{NumberOfMissingEv1}).

The ECAL target of NA64$e$ is a matrix of $6\times 6$ 
Shashlyk-type modules assembled from lead (Pb) 
$(\rho=11.34\,\mbox{g cm}^{-3}, A=207\,\mbox{g mole}^{-1}, Z=82)$ 
and scintillator (Sc) plates.  Note that production of hidden particles in the scintillator 
is subleading due to its larger radiation length, $X_0(Sc)\gg X_0(Pb)$, thus we ignore it in the  calculation.  

The NA64$e$ employs  the optimized
electron beam from the $H4$ beam line at the SPS. 
The  maximum intensity of the beam is $\simeq 10^7$ electrons per spill of 
$4.8\,\mbox{s}$, the number of good spills per day is estimated to be $4000$. 
Therefore, approximately $120$ days are required to accumulate 
$5 \times 10^{12}$ electrons on target (EOT) at the H4 electron beam line.

\subsection{The LDMX experiments} 
 
The LDMX is the proposed fixed target experiment at Fermilab that exploits the electron beam as
well as the  NA64$e$ facility.  The LDMX is designed to measure missing momentum of the electron,
thus probing of the  process~(\ref{eZtoeZInvis1}) at LDMX is complementary to NA64$e$. Moreover,
the missing momentum cuts and the active veto systems of both experiments make them essentially 
background free Ref.~\cite{NA64:2016oww,Berlin:2018bsc}.  
The proposed LDMX facility consists of a  target, silicon tracker system, electromagnetic and hadron 
calorimeter (for details, see, e.~g., Refs.~\cite{Berlin:2018bsc,Akesson:2022vza}).  
The most of the electron beam energy is lost due to the emission of 
the dark particles occurring in the thin 
upstream target. The missing momentum of the electron is
tagged by the silicon tracker, downstream electromagnetic and hadron calorimeter.  The final state 
electron missing energy cut is  chosen to be 
$E_e^{rec} \lesssim 0.3 E_e$, that corresponds to $x_{min} =0.7$ in  Eq.~(\ref{NumberOfMissingEv1}). 
In our analysis we carry out the calculation  for the aluminium-target (Al)
$(X_0=8.9~\mbox{cm}, \rho=2.7\,\mbox{g cm}^{-3}, A=27\,\mbox{g mole}^{-1}, Z=13)$ with a thickness of
$L_T \simeq 0.4 X_0$.  Note that  the LDMX plans to accumulate $\mbox{EOT}\simeq 10^{16}$ 
with  the beam energy up to $E_e= 16\,\mbox{GeV}$ for the final phase of running 
after 2027~\cite{Akesson:2022vza}.  

\subsection{ The NA64$\mu$ experiment}
The NA64$\mu$ facility~\cite{Kirpichnikov:2021jev,Sieber:2021fue}
is a complementary experiment to the NA64$e$ that searches for the dark sector particles
in the muon beam mode 
\begin{equation}
    \mu Z \to \mu Z \gamma_D (a).
    \label{muZtomuZGammaDa1}
\end{equation}
In our calculations we set the muon beam energy of NA64$\mu$ to be $E_\mu\simeq 160\,\mbox{GeV}$, the  
muon flux is chosen to be about $\mbox{MOT} \simeq 5\times 10^{13}$ for the projected statistics.
We consider the lead  shashlyk--type electromagnetic calorimeter that serves a target with a typical  
thickness of $L_{T}\simeq 40 X_0\simeq 22.5\, \mbox{cm}$.  We also neglect the muon stopping loss in the 
lead target~\cite{Chen:2017awl,Gninenko:2018ter}, since its  typical energy attenuation  is rather 
small $\langle dE_\mu/dx \rangle \simeq 12.7 \times 10^{-3} \mbox{GeV}/\mbox{cm}$ for the ultra--relativistic 
approach $E_\mu \simeq 160\, \mbox{GeV}$.

The NA64$\mu$ facility exploits to magnet spectrometers 
allowing for  precise measurements of momenta for incident and outgoing muons~\cite{Sieber:2021fue}. 
We set the following cut on the energy of recoiling muon
$E_\mu^{rec} \lesssim 0.5E_\mu \simeq 80\,\mbox{GeV}$, so that $x_{min} =0.5$. 

We also note that the intensity of $160\, \mbox{GeV}$ muons at the M2 beam line can be 
a higher by factor of $10$ than that for the 
electrons. Therefore, about the same running time of $120$  
days is required to accumulate  a much higher statistics of $5\times 10^{13}$ muons
on target (MOT) relative  to $\mbox{EOT}=5\times 10^{12}$ for NA64$e$.  

\subsection{The $\mbox{M}^3$ experiments} The $\mbox{M}^3$ (muon missing momentum) experiments at 
Fermilab~\cite{Kahn:2018cqs} is the projected modification of 
LDMX facility that is suitable for  probing the muon-specific missing energy signatures,  
$\mu Z \to \mu Z \gamma_{D} (a)$. 
In particular, it considers new physics  discovery  potential for the muon beam of 
$E_\mu  \simeq 15\, \mbox{GeV}$,  thick tungsten target (W) ($X_0\simeq 0.35\,\mbox{cm}, 
\rho=19.3\mbox{g cm}^{-3}, A\simeq 184\,\mbox{g mole}^{-1}, Z=74$) of thickness of 
$L_T=50X_0\simeq 17.5\,\mbox{cm}$ and  downstream detector to veto  SM backgrounds.

The  $\mbox{M}^3$ plans to accumulate $10^3$ MOT within  $\simeq 3$  months of data taking.
The signal missing momentum signature of the recoil muon is 
$E_\mu^{rec} \lesssim 9\,\mbox{GeV}$, meaning that $x_{min} =0.4$ 
in Eq.~(\ref{NumberOfMissingEv1}). The 
reported~\cite{Kahn:2018cqs} muon stopping loss  is $530\, \mbox{MeV}$ through a tungsten target 
of $50X_0$ for muons of $E_\mu \simeq 15\, \mbox{GeV}$. 
So that we neglect $\langle dE/dx \rangle$  in the signal estimate for the sake of  simplicity. Given that 
approach one can exploit  the  Eq.~(\ref{NumberOfMissingEv1})  for the signal yield estimate. 

\section{Minimal dark ALP portal scenario
\label{MinimalSetupSection}}

In this section for the minimal dark ALP portal 
setup~(\ref{MinimalSetupCoupling1}) we discuss the experimental 
signature of dark photon and  ALP production in the processes   
\begin{itemize}
    \item    $e(\mu)Z \to e(\mu) Z \gamma^* \to e(\mu) Z a \gamma_D$ that is shown in Fig.~\ref{MinimalALPportalBremsFeynman} for the fixed target experiments,
    \item  $J/\psi$ vector meson photoproduction,  $e Z \to e Z J/\psi $ followed by invisible decay $J/\psi \to a \gamma_D$, at NA64$e$ experiment (see e.~g.~Fig.~\ref{JpsiFeynam} for details), 
    \item $e^+e^- \to  \gamma  \gamma^* \to \gamma (\gamma^* \to a \gamma_D) \to \gamma  a \gamma_D$ that is shown in Fig.~\ref{BaBarFeynman} for BaBar experiment.
\end{itemize}
In particular, we estimate the rate of these processes and calculate the  
sensitivity curves from the null result of the experimental facilities.

\subsection{Radiative cross-sections of the fixed target facilities}

In order to estimate the sensitivity of the lepton fixed target experiments to probe minimal dark ALP portal missing 
energy signatures we calculate cross-sections by exploiting state-of-art 
{\tt CalcHEP} package~\cite{Belyaev:2012qa}.   In the SM model of {\tt CalcHEP}  we added both the
massive ALP $a$ and dark photon $\gamma_D$ as the  new particles by including the corresponding 
interaction  Lagrangian with photon  
$\mathcal{L} \supset (1/2) g_{a\gamma \gamma_D}a F_{\mu \nu} \widetilde{F}_{\mu\nu}'$.
We also added the target nucleus with atomic number $A$, charge $Z$ with spin $1/2$ particle and with mass $M_Z$  that  couples 
the SM photon via $U(1)$ vertex $ i e Z F(t) \gamma_\mu $,
where $t=-q^2>0$ is a nucleus transfer momentum squared, $F(t)$ is an elastic from-factor, 
that can be written as follows 
\begin{equation}
F(t)=\frac{a^2 t}{(1+a^2 t)}\frac{1}{(1+t/d)},
\label{FFdefinitio1}
\end{equation}
here $a=111 Z^{-1/3}/m_e$ and 
$d=0.164 A^{-2/3}\, \mbox{GeV}^2$ are the form-factors parameters of screening and nucleus size 
respectively \cite{Bjorken:2009mm,Tsai:1986tx}. Form factor specified in Eq.~(\ref{FFdefinitio1}) was implemented
in the {\tt C++} files for the expression of the matrix element squared, $|\mathcal{M}(lZ\to lZa\gamma_D)|^2$, 
that is generated  in the analytical session of the {\tt CalcHEP}.

Given the input parameters of the experiments discussed in 
Sec.~\ref{ExperimentalBenchmark} we carry out the integration of the 
exact tree-level amplitude squared $e(\mu)Z\to e(\mu)Za\gamma_D$ over the 
phase  space of the outgoing particles by exploiting the {\tt CalcHEP} 
package. In particular,  for the specific 
 chemical element of the target and typical energies of the lepton beam $E_{l}$, we calculate 
 $\sigma_{tot}$ as a function of the mass $m_{\gamma_D}$ in the range $1\, \mbox{MeV} \lesssim
 m_{\gamma_D}\lesssim 1\, \mbox{GeV}$ for $m_a\simeq 10\,\mbox{keV}$. 
 The numerical integration was performed by the {\tt VEGAS} importance sampling 
 algorithm with $N_{session} =10$~runs and $N_{calls}=10^6$ sampling points during each run.
 The grid adapting of the  {\tt VEGAS} algorithm was performed with a fairly good accuracy 
 of~$ \mathcal{O}(0.1)\%- \mathcal{O}(0.01)\%$ in the numerical session of {\tt CalcHEP}.

In Fig.~\ref{DiffCSVariousExperimentsMinimalPortal} we show the differential 
cross-sections as a function of the missing energy, $E_{miss}= E_{\gamma_D}+E_a$, in the 
signal box range $E_l^{th} \lesssim E_{miss} \lesssim E_l$ for the fixed target experiments  and 
various masses $m_{\gamma_D}$, here we denote $E_l^{th}= x_{min} E_l$.
Both the NA64$e$ and LDMX cross-sections have a peak at 
$E_{miss}\simeq E_e$. Which implies that the signal is strongly forward peak for 
$E_{miss} \gg m_{\gamma_D}, m_a$ and the dominant part of the beam energy 
transfers to the $a\gamma_D$ pair. However for both muon beam cross--sections of NA64$\mu$ and 
$\mbox{M}^3$ the peak at $E_{miss} \lesssim E_\mu$ is mitigated since the production rates of $a\gamma_D$ pair 
are in the soft bremsstrahlung--like regime as long as $m_{\gamma_D}, m_a \lesssim m_\mu$.

   \begin{figure}[t!]
\centering
\includegraphics[width=0.5\textwidth]{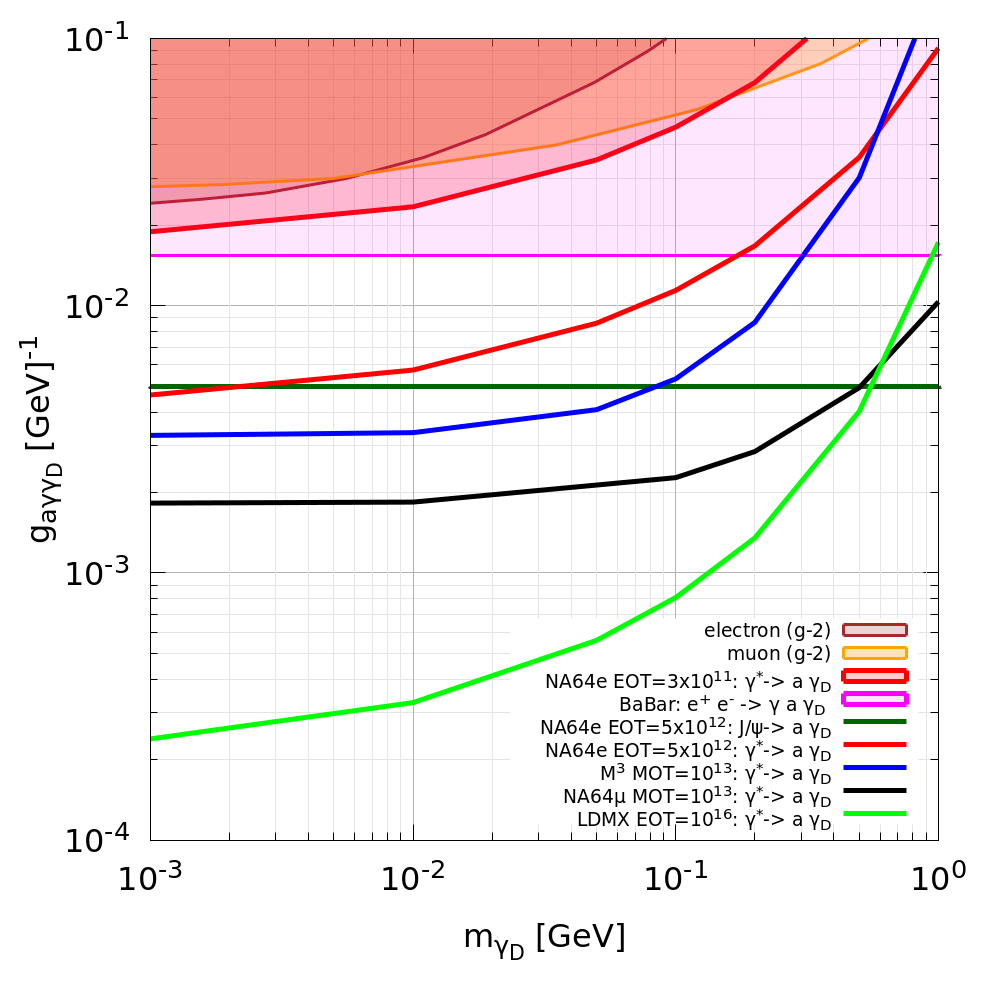}
\caption{The limits on $g_{a\gamma\gamma_D}$ coupling from the fixed-target experiments and BaBar for the minimal dark
ALP portal setup as a function of  the dark photon mass $m_{\gamma_D}$. For all sensitivity curves we 
imply that $\mbox{Br}(\gamma_D\to \chi \bar{\chi})\simeq 1$ and $m_a =10\, \mbox{keV}$.
Red line is the  projected sensitivity for NA64$e$ experiment, black line corresponds to NA64$\mu$
facility, blue line is the expected reach of $\mbox{M}^3$, and green line corresponds to the projected 
sensitivity of LDMX  facility. The shaded red region shows the parameter space  excluded   by the NA64$e$  
experiment for $\mbox{EOT}\simeq 3\times 10^{11}$. At $90\% CL$ the excluded region of NA64$e$ experiment
rules out the possible explanation of  $(g-2)$ muon (shaded orange region) and 
electron (shaded brown region) anomalies 
~\cite{Muong-2:2021ojo,Aoyama:2020ynm,Parker:2018vye} by the minimal ALP scenario  
at 2-loop level~\cite{deNiverville:2018hrc}. The BaBar pink shaded region refers to
search for the three body monophoton  process $e^+e^- \to a \gamma\gamma_D$ followed 
by  the rapid decay into DM pair $\gamma_D \to \chi \bar{\chi}$. 
Dark green line represents the expected reach of NA64$e$ experiment associated with $J/\psi$ meson 
photoproduction followed by invisible decay $J/\psi \to a \gamma_D$ for $\mbox{EOT} \simeq 5\times 10^{12}$.
\label{MinimalPortalFigExpectedReach} }
\end{figure} 

 In Fig.~\ref{MinimalPortalFigCS} the resulted total cross-sections are shown for NA64$e$, 
 NA64$\mu$, LDMX and $\mbox{M}^3$ experimental facilities.  
It is worth mentioning  that the NA64$e$ cross-section $\sigma^{tot}_{e}$ in 
Fig.~\ref{MinimalPortalFigCS} for the lead (Pb)  target ($Z\simeq 82$)  with  impinging electron beam  of $E_e=100\, \mbox{GeV}$  is  
generally larger by a factor of $\simeq 10$ than the lead cross-section $\sigma^{tot}_{\mu}$ for the muon beam of 
 $E_\mu = 160\, \mbox{GeV}$. Therefore, there is a cross-section  advantage of using the electron beam
 instead of muon beam in the  low mass region $m_{\gamma_D}\gtrsim 1\, \mbox{MeV}$ as long as 
 both electron and muon have
the typical energies of the order of $E_\mu\simeq E_e \simeq \mathcal{O}(100)\, \mbox{GeV}$. 
 On the other hand, $\sigma^{tot}_{e}$ decreases more 
 rapidly than $\sigma^{tot}_{\mu}$ as $m_{\gamma_D}$ increases towards $1\, \mbox{GeV}$
 in  the mass range of the interest 
 $ m_{\gamma_D} \lesssim 1\, \mbox{GeV}$. The latter one scales as 
 $\sigma_{\mu }^{tot}\propto g_{a\gamma \gamma_D}^2$ and depends weakly on $m_{\gamma_D}$ for the 
 bremsstrahlung-like  regime as long as $m_a, m_{\gamma_D}\ll m_\mu$. 
 
 In addition, one can see from 
 Fig.~\ref{MinimalPortalFigCS} that LDMX cross--section for the aluminium (Al) target ($Z\simeq 13$) is
 smaller than cross-section for the lead target of
 NA64$e$ facility, since the rate of $a\gamma_D$ production scales with $\propto Z^2$ and the 
 aluminium nucleus charge  is smaller by factor of~$82/13 \simeq 6.3$.  
 
However, if we compare the $\mbox{M}^3$ cross-section for the tungsten (W) target 
($Z\simeq 74$) with impinging muon beam of $E_\mu\simeq 15\, \mbox{GeV}$ and  
LDMX cross-section with impinging electrons of $E_e\simeq 16\,\mbox{GeV}$, one can conclude  
that the advantage of the electron beam exploiting in the low mass region $m_{\gamma_D}\gtrsim 1\, \mbox{MeV} $ 
is compensated  by the nucleus charge suppression. As the result, both cross-sections of the $\mbox{M}^3$ 
and LDMX are of the same order of the magnitude at $m_{\gamma_D}\simeq 1\, \mbox{MeV}$. 

\subsection{Limits from the fixed target experiments}

Using the formula (\ref{NumberOfMissingEv1}) for the number of produced $a\gamma_D$ pairs 
and the results on the production cross-sections, we find the expected bounds on the coupling $g_{a\gamma\gamma_D}$ 
for the minimal dark ALP portal scenario.  
We require  $N_{sign}\gtrsim 2.3$ that corresponds to the $90 \% CL$ exclusion limit on coupling $g_{a\gamma\gamma_D}$ 
for the background free case and null-result of the fixed target experiments.  
In Fig.~\ref{MinimalPortalFigExpectedReach} we show the expected reach of NA64$e$, LDMX, NA64$\mu$ 
and $\mbox{M}^3$.  Note that  projected  limits on $g_{a\gamma\gamma_D}$ from LDMX are fairly strong, even though 
the cross-section of $a\gamma_D$ pair production at LDMX is relatively small  
(see, e.~g., green line in Fig.~\ref{MinimalPortalFigCS}).  
The regarding LDMX sensitivity enchantment  can be explained by 
the large number of the projected accumulated statistics, $\mbox{EOT}\simeq 10^{16}$, by the final phase 
of  experimental  running. In addition, we note also from the $\mbox{M}^3$ and 
NA64$\mu$ cross-section shown in  Fig.~\ref{MinimalPortalFigCS}  that the signal of $a\gamma_D$ pair 
production  by muon beam  drops as the energy of muons decreases. In particular, compared to $\mbox{M}^3$ option 
with  $16\, \mbox{GeV}$ beam muons, a higher energy, e.~g.~$160\,\mbox{GeV}$, muons of NA64$\mu$ allows 
for probing  wider region in the parameter space of the minimal dark ALP portal scenario for $\mbox{MOT}=10^{13}$.  
In addition, we note that the projected limits of  NA64$e$ for $\mbox{EOT}\simeq 5\times 10^{12}$ can be ruled out by
other fixed target experiments.  

In Fig.~\ref{MinimalPortalFigExpectedReach} we show by the shaded red region the 
excluded limits of NA64$e$ for the current accumulated  statistics~\cite{Andreev:2021fzd} 
of $\mbox{EOT}\simeq 3 \times 10^{11}$.  This region rules out at $90\% CL$
the typical  parameter space of minimal dark ALP portal scenario~\cite{deNiverville:2018hrc}
that can explain  the $(g-2)_\mu$ and $(g-2)_e$ 
anomalies~\cite{ParticleDataGroup:2020ssz,Muong-2:2021ojo,Aoyama:2020ynm,Parker:2018vye} at two loop  
level. That contribution of ALP and dark photon  is analogous to neutral pion 
term~\cite{Blokland:2001pb} that contributes the 2--loop hadronic part of charged fermion $(g-2)_f$.

Concluding this subsection we note that there is detector advantage of exploiting the muons at NA64$\mu$ instead of 
electrons at NA64$e$, even though both experiments have a target of  total equivalent thickness 
of $40 X_0\simeq 22.5\, \mbox{cm}$.  The key point of that idea  is the  following \cite{Gninenko:2018ter}. 
The electron beam of $E_e\simeq 100\,\mbox{GeV}$ degrades  significantly within first 
radiation length $X_0$ of the  lead target. Contrary, the muons pass the target without significant loss of the 
energy, since their radiation length scales as $X_{0}^\mu \simeq (m_\mu/m_e)^2 X_0 \gg X_0$. This implies that
missing momentum signal of NA64$\mu$ is scaled as $N_{sign} \propto L_T$.  So that, one can 
improve the sensitivity of the muon beam mode by increasing the typical interaction length of the dump. 

 \begin{figure}[!h]
\centering
\includegraphics[width=0.35\textwidth, trim={2cm 24.2cm 12cm 2cm},clip]{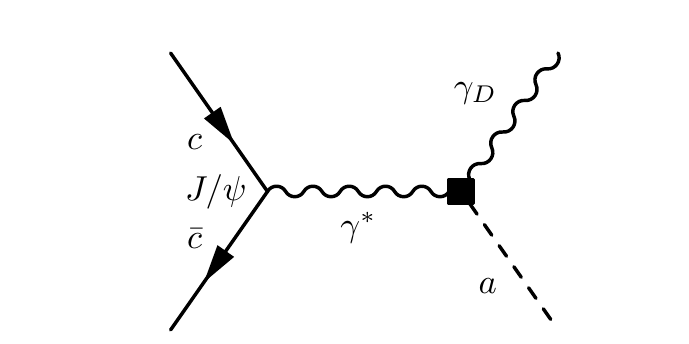}
\caption{Feynman diagrams for the radiative $J/\psi \to \gamma^* \to a \gamma_D$ decay.
\label{JpsiFeynam}}
\end{figure}

\subsection{Expected constraints from $J/\psi(1S) \to a \gamma_D$ decay at NA64$e$}
 In this subsection we follow Ref.~\cite{Schuster:2021mlr,Mangoni:2020zgq}
 to constrain the parameter space of the minimal ALP-portal 
 scenario from invisible decay of $J/\psi \to a \gamma_D$ at NA64$e$. 
 The matrix element of the charm quark and anti-quark transition process 
 $c(p_1) \bar{c}(p_2) \to \gamma^*(q) \to a(k_2) \gamma_D(k_1)$ 
 can be written as follows (see e.~g.~Fig.~\ref{JpsiFeynam} for detail)
 \begin{equation}
 \mathcal{M}_{c\bar{c}} = i e Q_c \bar{v}(p_2) \gamma^\mu u(p_1)
 \frac{g_{\mu \nu }}{q^2} g_{a\gamma\gamma_D} \epsilon^{\nu \lambda \rho \sigma} k^1_\lambda q^2_\rho \epsilon_{\sigma}(k^1),
 \label{AmplCCtoaGammaD1}
 \end{equation}
where $k_1, k_2$ and $p_1, p_2$ are, respectively, the four momenta of the ALP, dark photon and of the 
charm quarks.  In the center of mass frame one has, $p_1=p_2=p=(m_c,0,0,0)$ with $m_c$ being the mass of 
the charm quark, while $q$ in Eq.~(\ref{AmplCCtoaGammaD1}) is the photon four-momentum, such that 
$q^2=4m_c^2$, $Q_c=2/3$ is the  charge of the charm quark. By  integrating  
the averaged amplitude  squared $\overline{|\mathcal{M}_{cc}|^2}$ (see e.~g.~Eq.~(\ref{AmplCCtoaGammaD1}))
over the phase space of outgoing particles one can obtain  the invisible decay width  
\begin{equation}
\Gamma_{J/\psi\to a \gamma_D} =\frac{ |\psi_{J/\psi}(0)|^2}{8 \pi } 
g_{a\gamma\gamma_D}^2 e^2 Q_c^2\left( 1- \frac{m_{\gamma_D}^2}{M_{J/\psi}^2}\right)^3,
\label{DecayWidthJPSI1}
\end{equation}
where $M_{J/\psi}\simeq 3.1\, \mbox{GeV}$ is the mass of $J/\psi$ and 
$|\psi_{J/\psi}(0)|^2 \simeq 4.47\times 10^{-2}\, \mbox{GeV}^3$ is the squared radial 
wave-function~\cite{ParticleDataGroup:2020ssz} of $J/\psi$ at the origin $r=0$.   In order to calculate
$\overline{|\mathcal{M}_{cc}|^2}$ we exploit the state-of-the-art {\tt FeynCalc} package~\cite{Shtabovenko:2016sxi}
of {\tt Wolfram Mathematica}~\cite{Mathematica}. 
In  Eq.~(\ref{DecayWidthJPSI1}) we  neglect the ALP mass, $m_{\gamma_D} \gg m_a$. The authors provide 
in Tab.~II of  Ref.~\cite{Schuster:2021mlr}  the number  of expected $J/\psi$ vector mesons $N_{J/\psi} \simeq 1.1\times 10^5$  produced for projected statistics $\mbox{EOT}\simeq 5\times 10^{12}$ at NA64$e$. Therefore one can easily estimate the expected reach on 
$\mbox{Br}(J/\psi \to a \gamma_D ) \lesssim 2.3/N_{J/\psi}$ at $90\% CL$ that implies no signal events of  
$J/\psi$ invisible decays into $a\gamma_D$ pair at NA64$e$ for the background free case. 
Here we use the value for the total decay width
$\Gamma_{J/\psi}^{tot} \simeq 92.9\, \mbox{keV}$ from Ref.~\cite{ParticleDataGroup:2020ssz}. 
That yields the  expected  reach  on the coupling 
$g_{a\gamma \gamma_D}\lesssim 5\times 10^{-3}\, \mbox{GeV}^{-1}$ at $90\% CL$  for NA64$e$ in the 
mass range of interest $1\, \mbox{MeV} \lesssim m_{\gamma_D} \lesssim 1\, \mbox{GeV}$.
In Fig.~\ref{MinimalPortalFigExpectedReach}  we show the regarding expected limit that can rule out
the limit of NA64$e$ for $a\gamma_D$-pair  production in the process 
$e Z \to e Z  \gamma^*( \to a \gamma_D)$  for the same statistics 
$\mbox{EOT}\simeq 5\times 10^{12}$.  Note that in the present 
analysis  we  conservatively assume that dominant channel of missing energy events at 
NA64$e$ is  associated with photoproduction of $J/\psi$ meson $\gamma Z \to Z J/\psi$ followed by  
its rapid decay $J/\psi \to a \gamma_D$ in the detector of NA64$e$.  The production of 
$\rho, \omega$ and $\phi$ vector mesons at  NA64$e$  is expected to be 
subdominant~\cite{Arefyeva:2022eba,Schuster:2021mlr}.

\begin{figure}[!h]
\centering
	\includegraphics[width=0.3\textwidth, trim={2cm 23cm 12cm 2cm},clip]{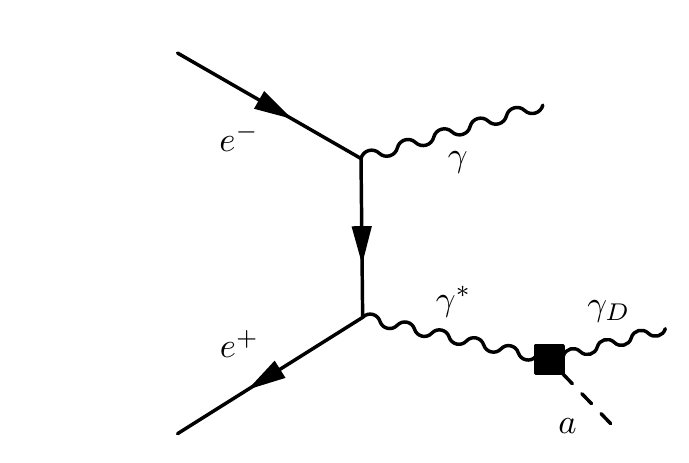} 
\caption{Feynman diagrams for the  monophoton three body 
process $e^+e^- \to a \gamma \gamma_D$ followed by invisible 
dark photon decay  into DM fermion pair, $\gamma_D \to \chi \bar{\chi}$.
That reaction is relevant for the BaBar constraints on $g_{a\gamma\gamma_D}$ coupling in the framework 
of the minimal dark ALP portal scenario.      
\label{BaBarFeynman}}
\end{figure}

\subsection{Bounds from monophoton BaBar data}

An authors of Ref.~\cite{deNiverville:2018hrc} provide an explicit 
analysis of the the mono-photon signal for the process 
$e^+ e^- \to a \gamma_{D}$ followed by decay  $\gamma_D \to a \gamma$ from BaBar data. 
They also provide the regarding exclusion limits (see e.~g.~Fig.~4 from 
Ref.~\cite{deNiverville:2018hrc}) in $(m_{\gamma_D}, g_{a\gamma\gamma_D})$ plane from
BaBar experiment.  These limits are not relevant for our minimal 
benchmark scenario, since we  suppose that $\gamma_D$ decays rapidly into  
$\chi \bar{\chi}$ pair with $\mbox{Br}(\gamma_D \to \chi \bar{\chi})\simeq 1$ and  thus visible monophoton
decays $\gamma_D\to a \gamma$ are suppressed, $\mbox{Br}(\gamma_D\to a \gamma) \ll 1$.
However, three body final state process $e^+ e^- \to a \gamma \gamma_D$ is 
kinematically  allowed for our analysis.  That reaction is  sub-leading relative to 
two-body final  state process  $e^+ e^- \to a \gamma_D$. The regarding suppression can be 
found in Fig.~3 of Ref.~\cite{deNiverville:2018hrc} which shows the total cross-sections of the relevant
processes as the function of mass $m_{\gamma_D}$. In particular, for the mass range 
$1\, \mbox{MeV} \lesssim m_{\gamma_D} \lesssim 1\, \mbox{GeV}$ and 
$g_{a\gamma\gamma_D} = 1\, \mbox{GeV}^{-1}$ one has $\sigma_{tot}(e^+e^-\to a \gamma_D)\simeq 1.2\times 10^{5} \mbox{pb}$ and  
$\sigma_{tot}(e^+e^-\to a\gamma \gamma_D) \simeq 6\times 10^{3}\, \mbox{pb}$ for two body
and three  body final state respectively. Therefore  it is straightforward 
to obtain the relevant for our analysis limit on $g_{a\gamma\gamma_D}^{(3-body)} $ from 
mono-photon bound   
$g_{a\gamma\gamma_D}^{(2-body)} \simeq 2\times 10^{-3} \, \mbox{GeV}^{-1}$ 
shown in Fig.~4 of Ref.~\cite{deNiverville:2018hrc}. Namely for
 the mass range of   interest $1\, \mbox{MeV} \lesssim m_{\gamma_D} \lesssim 1\, \mbox{GeV}$ this yields:
\begin{align}
& g_{a\gamma\gamma_D}^{(3-body)} \simeq  g_{a\gamma\gamma_D}^{(2-body)} \times \label{BaBarLim}
\\
& \times\left(\frac{\sigma_{tot}(e^+e^-\to a \gamma_D)}{\sigma_{tot}(e^+e^-\to a \gamma \gamma_D)} \right)^{1/2} \simeq 1.5\times 10^{-2}\, \mbox{GeV}^{-1}. \nn
\end{align}
As expected  \cite{deNiverville:2018hrc,deNiverville:2019xsx}, the experimental reach of the BaBar weakens by factor of 
approximately $\sqrt{20}\simeq 4.5$ for the $\mbox{Br}(\gamma_D\to \chi \bar{\chi})\simeq 1$.    
 In Fig.~\ref{MinimalPortalFigExpectedReach} we show the current 
 constraint from BaBar monophoton 
 signal $e^+e^- \to a \gamma \gamma_D$ by the shaded pink region. It rules out the current 
 experimental constraints  of the NA64$e$ experiment for $\mbox{EOT} \simeq 3\times 10^{11}$. 

\section{Non-minimal ALP-hadrophilic scenario
\label{NonMinimalHadrophilicSect}}

   In this section we calculate the cross-section
dark photon production in the process $l Z \to l Z \gamma_D$
that is shown in left side of the Fig.~\ref{axionPhotonand_dark} for 
the case of hadrophilic ALPs. That implies the benchmark coupling of the ALP and dark photons in the following form
\begin{align}
   \mathcal{L} \supset & \mathcal{L}_{\mbox{\scriptsize dark axion portal}} 
   + \bar{\chi}( \gamma^\mu  i \partial_\mu -  g_D \gamma^\mu  A_\mu'  + m_{\chi} )\chi 
 \nn   \\ 
 & + \sum_{N=n,p} a \bar{N} (g_\psi^s+ig_\psi^p \gamma_5) N \,.
    \label{HadroPhilicLagr1} 
\end{align}

For the sake of simplicity we consider benchmark universal scalar and pseudo-scalar coupling of ALP to nucleons, 
such that $g_p^s=g_n^s\equiv g^s$ and $g_p^p=g_n^p \equiv g^p$.

In order to calculate the cross-section of dark photon lepton-production at nucleus $(A,Z)$ 
where $A$ is nucleus mass number, $Z$ is charge of nucleus, we use the equivalent photon approximation, 
that implies replacing the fast moving leptons by the photons following a distribution~\cite{ALPtraum} 
\begin{equation}
    \gamma_l(x_\gamma,q_{\bot}^2) \simeq \frac{\alpha}{2 \pi}  \frac{1+(1-x_\gamma)^2}{x_\gamma}  \frac{q_{\bot}^2}{(q_{\bot}^2+x_\gamma^2 m_l^2)^2},
    \label{FluxPhbeamEl}
\end{equation}
where $E_\gamma$ is the energy of the photon emitted by incoming lepton,  $x$ is fraction of the photon energy defined as $ x_\gamma \equiv E_{\gamma} / E_l$, here $E_l$ is energy of
oncoming  lepton, $m_l$ is the lepton mass, $q_{\bot}$  is the photon  transfer momentum. 
Note that $q_{\bot}^2$ is typically very small, $q_{\bot} \ll E_l, E_\gamma$. 
The total cross section of dark photon lepton-production at nucleus can be written as 
\begin{equation}
\sigma_{l\,Z \to l Z \gamma_D} =\!\! \int \!\!dx \, dq_{\bot}^2   \gamma_l(x_\gamma,q_{\bot}^2) \int dt 
\frac{d\sigma_{\gamma Z \to Z \gamma_D}}{dt} ,
\end{equation}
%%%%%%%%%%%%%%%%%%%%%%%%%%%%%%%%%%
where the cross-section of dark photon production can be written as sum of 
partial  cross-sections for each nucleon
$N$ in nuclear of target
$$
\frac{d\sigma_{\gamma Z \to Z \gamma_D}}{dt} = A \frac{d\sigma_{\gamma N \to N \gamma_D}}{dt} ,
$$ 
%where $A$ is the mass number of the nucleus. 
The Lorentz invariant form of differential cross section of dark photon production due to photon scattering on nucleon is	
\begin{align}
\frac{d\sigma_{\gamma N \to N \gamma_D}}{dt} = \frac{\frac{1}{4}\sum\limits_{\rm pol} |M_{\gamma N}|^2}{16\pi \lambda(s,m_N^2,0)},
\end{align}
where  $\lambda(s,m_N^2,0) = (s-m_N^2)^2 = 4m_N^2 E_{\gamma}^2$ is the K\"allen function in the rest frame of 
initial nucleon. Then the square of the magnitude of matrix element is given by 
 \begin{align}
     |M_{\gamma N}|^2 &\simeq \frac{ g_{a\gamma\gamma_D}^2}{2}\frac{ (t-m_{\gamma_D}^2)^2 }{(t-m_a^2)^2} \, m_N^2 \, 
    \nn  \\ 
    &\times \Bigg[g_s^2-g_p^2  + (g_s^2+g_p^2) \frac{E_{2N}}{m_N}\Bigg] \,, \label{Amp2GammaNtoNGammaD1} 
    % \\
     \end{align}
 where $E_{2N} \simeq  m_N-t/(2m_N) $ is the energy of the outgoing nucleon in the laboratory frame, 
 $t$ is the nucleon transfer momentum squared, such that $t \equiv - 2 m_N(E_\gamma -E_{\gamma_D}),$
 for the nearly collinear emission of dark photon this yields
 $$t\simeq - E_\gamma^2 \theta_{\gamma_D}^2 -m_{\gamma_D}^4/(4 E_\gamma^2).$$
 It is worth mentioning  that for the small angles of dark photon production
the amplitude squared Eq.~(\ref{Amp2GammaNtoNGammaD1}) 
is suppressed if we  put $g_N^s=0$ and $g_N^p \neq 0$. 
The regarding suppression factor  is associated with the term $$|M_{\gamma N}|^2 \propto (g_N^p)^2(E_{2N}-m_N)/m_N \ll 1$$  
as long as $\theta_{\gamma_D} \ll 1$ and $E_{2N}\simeq m_N$. Otherwise, if we set $g_N^p=0$ 
and $g_N^s \neq 0$ one gets $$|M_{\gamma N}|^2 \propto (g_N^s)^2 E_{2N}/m_N.$$ 
As a result, the cross--section for the scalar--specific couplings of 
nucleons is enhanced by factor of approximately
$\simeq E_{2N}/(E_{2N}-m_N)$ relative to the pseudo--scalar--specific cross--section.  
 If one sets $g_p=g_s=g_N$ then for   
$t\ll m_a^2, m_{\gamma_D}^2$ we get the following  expression for the 
amplitude squared  
\eq 
\label{AmplitudeSquaredSmallTransfer1}
|M_{\gamma N}|^2 \simeq  g_{a\gamma\gamma_D}^2 \, g_N^2 \, m_N^2 \!     
\biggl(\frac{m_{\gamma_D}}{m_a}\biggr)^4 \! \biggl[ 1 + {\cal O}\Big(\frac{m_{\gamma_D}^2}{E_\gamma^2}\Big) \biggr]\, .\quad
\en 
 The Eq.~(\ref{AmplitudeSquaredSmallTransfer1})
implies that for the highly collinear emission of the  dark photon 
(negligible momentum transfer) its production rate grows as    
$ \propto (m_{\gamma_D}/m_a)^4$ for increasing  $m_{\gamma_D}$. In fact,
the nucleon momentum transfer can not be completely neglected 
and the realistic cross-section grows a bit more slowly than 
$(m_{\gamma_D}/m_a)^4$ as  
$m_{\gamma_D} \gtrsim \mathcal{O}(100)\, \mbox{MeV}$.

\begin{figure}[t!]	
		\includegraphics[width=0.48\textwidth, trim={0cm 0cm 0cm 0cm},clip]{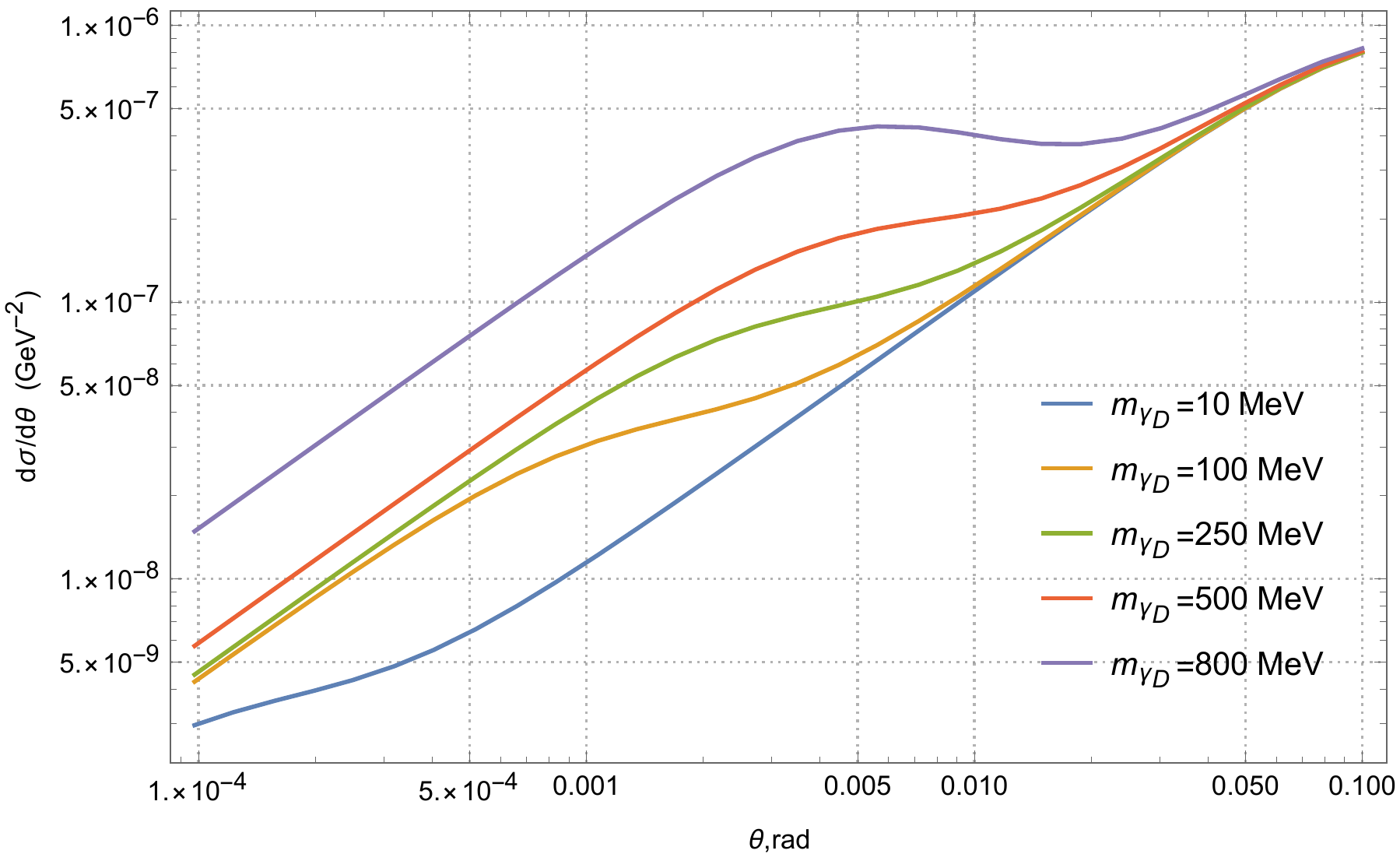}
	\caption{The differential cross--section by angle of dark photon production at fixed target of NA64$e$ 
	in hadrophilic-ALP portal for various dark photon masses $m_{\gamma_D}$ and for small values of angle $\theta$ 
	between dark photon and photons from beam. We set parameters as $m_a=10$ keV, $g^s=g^p=1$, 
	$g_{a\gamma\gamma_D}=1$ GeV$^{-1}$, $E_e=100$ GeV.} 
	\label{hadrofilic}
\end{figure}

 Finally, one can obtain  the double differential cross-section of 
 dark photon production
 \begin{equation}
              \frac{d \sigma_{l Z \to l Z \gamma_{D}}}{d E_{\gamma_D} d \theta_{\gamma_D}} \simeq  \frac{1}{E_{l}} 
  \int d q_{\bot}^2 \gamma_{l}(E_{\gamma_D}/E_l, q^2_{\bot}) \frac{d \sigma_{\gamma Z}}{d \theta_{\gamma_D}} \,.
  \label{DoubleDiffHadrophilic}
 \end{equation}
In Fig.~\ref{hadrofilic} we show the differential cross--section $(d\sigma/d\theta_{\gamma_D})_{l Z \to l Z \gamma_D}$
as the function of $\theta_{\gamma_D}$ in the range of small dark photon 
emission angle $\theta_{\gamma_D}\lesssim 0.1$.  As we mentioned
above, the larger masses  $m_{\gamma_D}$ imply the larger values of the  
differential cross--section.
That dependence can be described also in terms of specific Lorentz 
invariant characteristics.  Let us consider the  
characteristic function for the benchmark case~\cite{Byckling:1971vca}
\begin{align}
g_{\gamma_D}= &(s-m_N^2)/(s-m_{\gamma_D}^2+m_N^2)  \nn
\\
& \times \lambda^{\frac{1}{2}}(s,m_N^2,0)/\lambda^{\frac{1}{2}}(s,m^2_{\gamma_D},m_N^2),
\end{align}
that is associated with the typical angles of dark photon emission in the 
laboratory frame. It is straightforward to obtain that $g_{\gamma_D} > 1$ for 
$m_{\gamma_D} \lesssim 100\, \mbox{MeV}$, which means that $\theta_{\gamma_{D}} \ll 1$, i.e. the 
cross-section peaks forward. On the other hands, one can obtain that $g_{\gamma_D} < 1$ 
as soon as  $m_{\gamma_D} \gtrsim  1\, \mbox{GeV}$, therefore
the typical momentum of dark photon $\gamma_D$ production can be not collinear 
to the beam line in this case, i.~e.~the typical angles $\theta_{\gamma_D}$
can be as large as  $\theta_{\gamma_D}\gtrsim 1$.  In the present paper we study the elastic
production of dark photon, so that in the analysis  for simplicity 
 we conservatively set the typical maximum angle of dark photon emission to be 
$\theta_{\gamma_D}\lesssim \theta_{max}\equiv 0.1$.  
The total cross--section of the dark photon production is calculated for the regarding 
angle cut,  which however decreases significantly the signal rate of $\gamma_D$ emission. 

However, for the ALP--hadrophilic scenario it is worth to calculate   the inelastic photo-production 
cross--section of $\gamma_D$ that implies large nucleon transfer
momentum and  relatively wide emission angle $\theta_{\gamma_D} \gtrsim 1$.  
That analysis  would require the realistic simulation of  the experimental efficiency and the 
hadronic response in the detector.  This, however is beyond the scope of the present paper and 
we leave such  analysis for future study.

In Fig.~\ref{Boundshadrophilic} of Sec.~\ref{ResultsSection} the $90\, \% C.L.$ sensitivity curves of 
the fixed target experiments are shown 
for the combination of couplings $|g_{a\gamma\gamma_D} g_N|$. In order to plot theses curves we 
set $N_{sign}>2.3$,  implying Poisson statistics for the signal events, background free case and 
null result for the  DM detection.

\section{Non--minimal ALP--leptophilic scenario
\label{NonMinimalLeptophilicSect}}

 In this section we consider the process 
 \begin{equation}
     l(p)Z(\mathcal{P}_i) \to l(p') Z(\mathcal{P}_f)  \gamma_D(k),
     \label{leptophilicProcess1}
 \end{equation}
  shown in the right panel of 
Fig.~\ref{axionPhotonand_dark} for the leptophilic ALP, where 
$p=(E_l,{\bf p})$ and $p'=(E_l',{\bf p}')$ are the four momenta of
initial and outgoing leptons, respectively, $\mathcal{P}_i=(M_N, {\bf 0})$ 
and $\mathcal{P}_f=(\mathcal{P}_f^0,{\bf \mathcal{\bf P}}_f)$ are the four-momenta 
of initial and outgoing nuclei, respectively, here $M_N$ is the mass of the nucleus and 
$k=(E_{\gamma_D}, \bf{k})$ is the four-momentum of dark photon. 
The  benchmark Lagrangian of that 
simplified ALP portal scenario can be  written as follows
\begin{align}
   \mathcal{L} \supset & \mathcal{L}_{\mbox{\scriptsize dark axion portal}}
   + \bar{\chi}( \gamma^\mu  i \partial_\mu -  g_D \gamma^\mu  A_\mu'  + m_{\chi} )\chi 
 \nn   \\ 
 & + \sum_{l=e,\mu}  i  g^{p}_l a \bar{l}  \gamma_5 l \, ,
    \label{LeptoPhilicLagr1} 
\end{align}
where we consider only the pseudoscalar coupling of the ALP to the leptons $g^{p}_l$.
The numerical calculations reveal that the scalar coupling to leptons 
$\mathcal{L} \supset \sum_{l=e,\mu}  g_l^s a \bar{l} l $ yields the 
similar contribution to the  signal events if we set universal 
coupling as $g^{s}_l=g^{p}_l$. 

\begin{figure}[!tbh]
\centering
\includegraphics[width=0.49\textwidth]{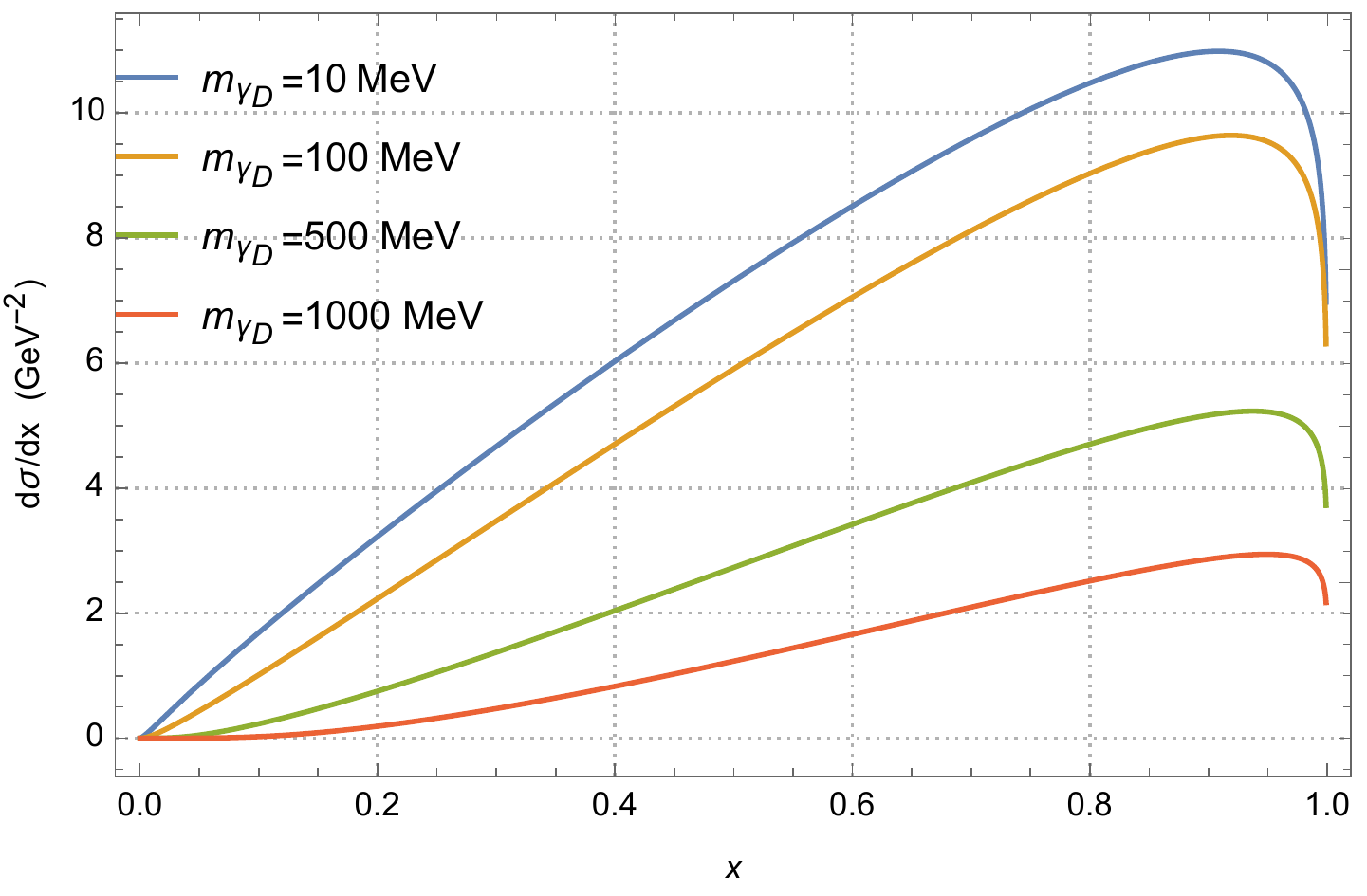}
\caption{The differential cross--section of dark photon production at NA64$e$
as a function of  fraction energy  $x=E_{\gamma_D}/E_l$ for the ALP-leptophilic
scenario and  for various masses $m_{\gamma_D}$.  We set $m_a = 10\,\mbox{keV}$,
$g^{p}_l=1$,  $g_{a\gamma \gamma_D}=1\,\mbox{GeV}^{-1}$ and $E_e=100\,\mbox{GeV}$.
\label{dsdxLeptophilic1}}
\end{figure} 

To calculate the  cross-section of the process (\ref{leptophilicProcess1}) we use the equivalent photon approximation~\cite{Kirpichnikov:2021jev,Bjorken:2009mm,Liu:2017htz}, 
that implies the factorization of $2\to 3$ rate into the product of photon flux and 
$2\to2$  cross--section of the Compton-like process $l(p) \gamma(q) \to l(p') \gamma_D(k)$
\begin{equation}
\frac{d \sigma_{l Z \to l Z \gamma_{D}}}{d(pk ) d (k \mathcal{P}_i) } \simeq      
\frac{\alpha \chi  }{\pi (p' \mathcal{P}_i)} \cdot \frac{d \sigma_{l  \gamma \to l \gamma_{D}}}{d(pk)} \Bigl|_{t=t_{min}},
\label{dsdxdthetaLeptophilic1}
\end{equation}
where $\chi$ is the effective photon flux from  
nucleus~\cite{Kirpichnikov:2021jev}, in~(\ref{dsdxdthetaLeptophilic1}) we assume
that the photon virtuality $t$ has its minimum $t_{min}$ when $\bf{q}$ is 
collinear with 
$\bf{k}-\bf{p}$. We define auxiliary Mandelstam variables and regarding identity
as follows 
\begin{equation}
\widetilde{u}=(p-k)^2-m_l^2, \quad \widetilde{s}=(p'+k)^2-m_l^2,
\label{uAndstildedef}
\end{equation}
\begin{equation}
t_2=(p-p')^2, \quad \widetilde{s}+\widetilde{u}+t_2 \simeq m^2_{\gamma_D},
\label{t2AnsSumdef}
\end{equation} 
then one can easily obtain the following expression for the differential 
cross-section in the Lorentz invariant notations~\cite{Kirpichnikov:2021jev,Bjorken:2009mm,Liu:2017htz}
\begin{equation}
\frac{d \sigma_{2\to 2}}{d (pk)} \simeq  
\frac{1}{8\pi \widetilde{s}^2} \cdot \overline{|\mathcal{M}_{2\to2}|^2}.
\end{equation}

Let us calculate now the amplitude of the relevant $2\to 2$ sub-process 
\begin{equation}
\mathcal{M}_{2\to2} = g^{p}_l \, g_{a\gamma\gamma_D} \, 
\frac{\bar{u}(p') \gamma_5 u(p)}{t_2-m_a^2} \, 
\epsilon_\mu(q) \epsilon^*_\nu(k) \epsilon^{\mu\nu q k} \,,        
\end{equation}
where $\epsilon^{\mu\nu q k} = q_\lambda k_\rho \epsilon^{\mu\nu \lambda \rho}$.  
As the result of averaging over polarizations one gets the following expression for the
amplitude  squared
\eq 
\overline{|\mathcal{M}_{2\to2}|^2} &=& \frac{1}{4} \sum_{\rm pol} |\mathcal{M}_{2\to2}|^2
\nonumber\\
&=& - \frac{1}{4} \frac{(g^{p}_l)^2 g_{a\gamma \gamma_D}^2}{(t_2-m_a^2)^2} 
\, t_2 \, \lambda(t_2, 0, m_{\gamma_D}^2) \,, 
\en 
where $\lambda(x,y,z) = x^2 + y^2 + z^2 - 2 x y - 2 x z - 2 y z$ is the kinematical 
triangle K\"allen function. These calculations are performed by exploiting 
the state--of--the--art {\tt FeynCalc} package~\cite{Shtabovenko:2016sxi}
of {\tt Wolfram Mathematica}~\cite{Mathematica}.

It is worth noting that longitudinal term $k_\mu k_\mu/m_{\gamma_D}^2$ in 
the dark photon  polarization tensor 
$\sum_i \epsilon^{*i}_\mu(k)  \epsilon^i_\nu(k)$ does not contribute to the
matrix element squared due to the current conservation.    
We label the energy fraction of $\gamma_D$ boson by $x=E_{\gamma_D}/E_l$ and the angle between  $\bf{k}$ and 
$\bf{p}$ by $\theta_{\gamma_D}$. Let us introduce the auxiliary function  $U$ as  follows
 \begin{equation}
U \equiv - \widetilde{u} \simeq E_l^2 \theta^2_{\gamma_D} x + m_{\gamma_D}^2 (1-x)/x + m_l^2 x,
\label{Udefinition}
 \end{equation}
  in~(\ref{Udefinition}) we keep only leading terms in $m_{\gamma_D}^2/E_{\gamma_D}^2$, 
 $m_{l}^2/E_{l}^2$, $m_{l}^2/E_{l}^{'2}$ and $\theta^2_{\gamma_D}$. In the  latter approach one has
 \begin{equation}
 t_{min}\simeq U^2/(4E_l^2 (1-x)^2),
 \end{equation}
 \begin{equation}
\widetilde{s} \simeq U/(1-x), \qquad t_2 \simeq - x U/(1-x)+m_{\gamma_D}^2.
 \end{equation}
Finally, we obtain the expression for the double differential 
cross-section 
\begin{equation}
\frac{d \sigma_{2\to 3}}{dx \, d\cos\theta_{\gamma_D}} 
\simeq \frac{\alpha \chi }{\pi(1-x)}\cdot  E_l^2 x \beta_{\gamma_D} \cdot \frac{d \sigma_{2\to 2}}{d (pk)},
\label{dsdxdcostheta2}
\end{equation}
 where $\beta_{\gamma_D}=(1-m_{\gamma_D}^2/(x E_l)^2)^{1/2}$ is the 
 velocity of dark photon in the laboratory frame.  The explicit analytical expression
 for the effective photon flux  $\chi$ is given in the Ref.~\cite{Kirpichnikov:2021jev} for
 the case of  elastic form-factor $G_{el}(t)$ that is proportional to 
 $\propto Z^2$. An inelastic form factor $G_{inel}(t)\propto Z$ and 
 for the heavy target nuclei $Z\propto \mathcal{O}(100)$, so that one can safely
 ignore it in the calculation below. 
 
The resulted cross-section can be rewritten as
 \eq
\frac{d \sigma_{2\to 3}}{dx d\cos\theta_{\gamma_D}} &\simeq&
\frac{\alpha \chi }{32 \pi^2} E_l^2\beta_{\gamma_D} 
(g^{p}_l)^2 g_{a\gamma \gamma_D}^2 \nonumber\\
&\times& \frac{x^3 \left[ x U-m_{\gamma_D}^2 (1-x) \right]}
{\left[ x U-(1-x)(m_{\gamma_D}^2-m_a^2)\right]^2} \,.
 \en 
As an example, in Fig.~(\ref{dsdxLeptophilic1}) we show the differential cross--sections of the 
 process $eZ\to  eZ\gamma_D$ as a function of $x$ for NA64$e$ experiment and  the 
 benchmark ALP--leptophilic setup~(\ref{LeptoPhilicLagr1}) with $m_a = 10\,\mbox{keV}$.
  The WW approximation  for the leptophilic cross--section 
 implies that the photon flux $\chi$ from nuclues is the function of $x$ and 
 $\theta_{\gamma_D}$, so that it is fairly accurate approach for the exact tree level 
 cross-section  (for detail, see e.~g.~Ref.~\cite{Kirpichnikov:2021jev} and references therein).  

In Fig.~\ref{BoundsLeptohilic} of Sec.~\ref{ResultsSection} the $90\, \%~C.L.$ sensitivity curves 
of the fixed target experiments are shown 
for the combination of couplings $|g_{a\gamma\gamma_D} g_l^p|$. 
To plot theses curves we put $N_{sign}>2.3$, implying null result for the  DM detection 
and  background free case.  

\section{Summary and discussion
\label{ResultsSection}}
%\label{Sec9}
In the present paper in addition to ALP contribution to lepton EDM we have calculated the novel limits 
on the combination of $CP$--even and $CP$--odd 
couplings of neutron with  ALPs from the current constraints on neutron EDM by taking into 
account its anomalous magnetic moment.  The regarding contribution to neutron EDM is associated 
with a light scalar/pseudoscalar boson exchange at 1-loop level. This contribution is proportional 
to $\sim\bar{\theta}$ of QCD parameter of $CP$--violation~\cite{Zhevlakov:2020bvr, Crewther:1979pi}.   
In addition we have considered the possible implication of dark axion portal couplings to the EDM of 
SM fermions. In particular, we have calculated at the 1-loop level the 
EDM, that can be induced by the three specific model-independent interactions: 
(i) $CP$--odd  Yukawa-like  couplings of ALP with fermions,  
(ii) $CP$--even  interaction of SM fermions and light vector boson, and 
(iii) $CP$--even dark axion/ALP portal coupling. Such Barr-Zee type diagrams have an ultraviolet (UV) 
logarithmic divergences, which can be calculated by using $\overline{MS}$ regularization scheme  
and removed by exploiting the local counter--terms of EDM type. We briefly discuss the regarding 
renormalization procedure in Appendix~\ref{App_A}. 

%\begin{widetext}
\onecolumngrid
\begin{center}
	%\begin{figure}[!tbh]
	\begin{figure}[t!]
		\centering
		\includegraphics[width=0.72\textwidth]{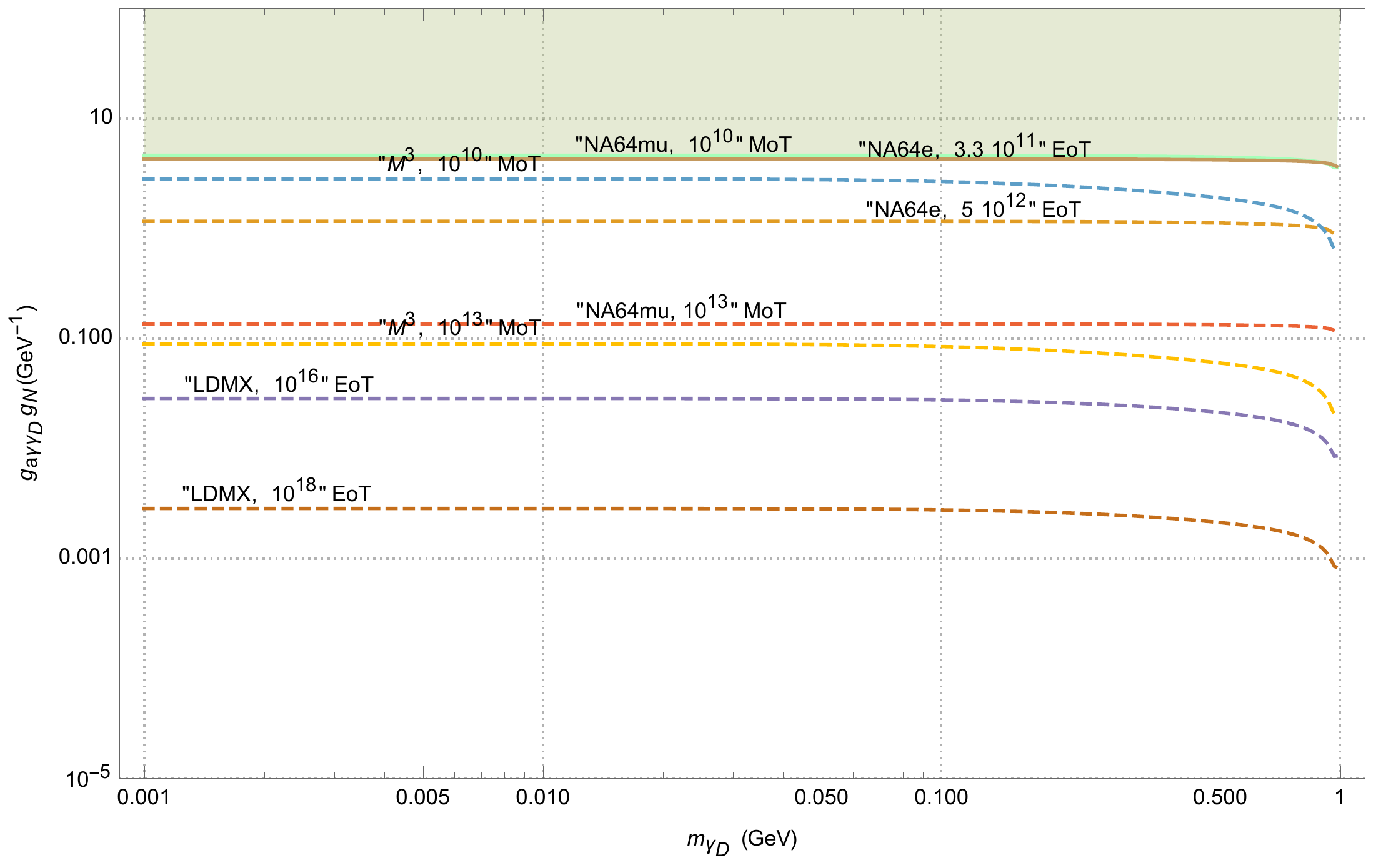}
		\caption{Bounds on combination of couplings ALPs with gauge fields $g_{a\gamma\gamma_D}$ and ALP couplings with 
			nucleons $g^s=g^p=g_N$ for hadrophilic channel from current and proposal statistics of the experiments NA64$e$, 
			NA64$\mu$, LDMX and M$^3$. We set here $\mbox{Br}(\gamma_D \to \chi \bar{\chi})\simeq 1$ and $m_a$=10 keV.
			\label{Boundshadrophilic}}
	\end{figure}
	%\end{center}
	%\end{widetext}
	%\begin{widetext}
	%\begin{center}
	%\begin{figure}[!tbh]
	\begin{figure}[t!]
		\centering
		\includegraphics[width=0.72\textwidth]{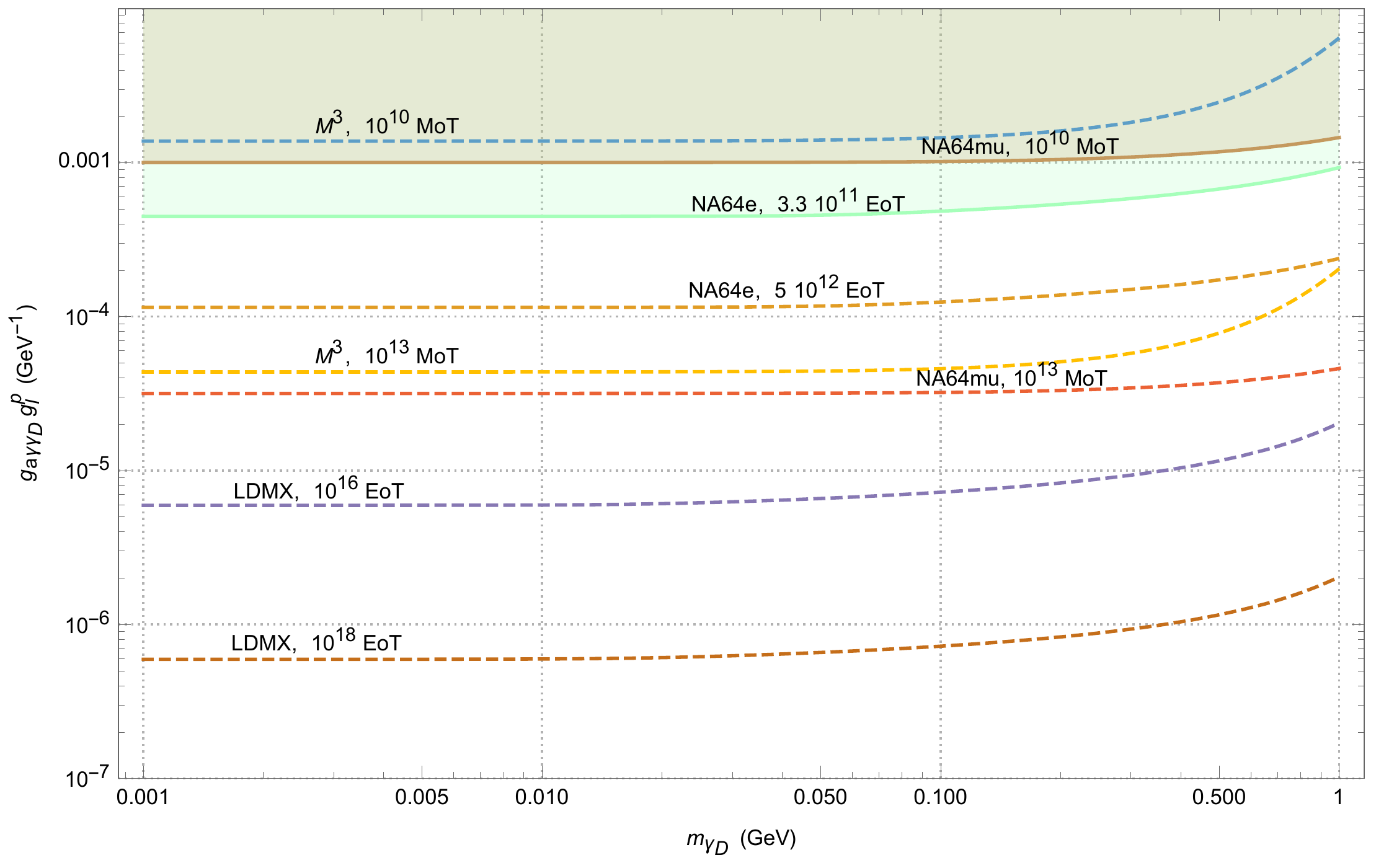}
		\caption{ Bounds for the combination of ALP couplings  with leptons $g^{p}_l$ and 
			with gauge fields $g_{a\gamma\gamma_D}$  for the leptophilic scenario from 
			current and proposal statistics of the NA64$e$, NA64$\mu$, LDMX and M$^3$ experiments. 
			We set here $m_a$=10 keV and $\mbox{Br}(\gamma_D\to\chi\bar{\chi})\simeq 1$. 
			\label{BoundsLeptohilic}}
	\end{figure}
\end{center}
%\end{widetext} 
\twocolumngrid

We have also discussed in detail the  probing the dark ALP portal scenario through the dark photon decaying 
predominantly to the DM particles, $\mbox{Br}(\gamma_D \to \chi \bar{\chi})\simeq 1$.
In the present paper we refer the relevant benchmark model as minimal dark ALP portal scenario (see bounds for 
$g_{a\gamma\gamma_D}$ coupling in Fig.~\ref{MinimalPortalFigExpectedReach}). 
This scenario implies that dark photon is $U_D(1)$ gauge field and serves the mediator between DM  and SM particles 
through the dark axion portal. In this scenario we imply that $m_{a}\ll m_{\gamma_D}$, therefore  the visible 
decay $a \to \gamma \gamma_D$ is kinematically forbidden.

We have studied in detail the missing energy signatures for the projected and existed 
lepton  fixed target facilities, such as NA64$e$, LDMX, NA64$\mu$ and $M^3$. In particular,
by using state-of-the-art {\tt CalcHEP} package we calculated the $a \gamma_D$ pair production  
cross-sections in the processes  $l Z \to l Z a \gamma_D$ followed by the invisible dark photon
decay into DM  particles  $\gamma_D\to \chi \bar{\chi}$ for the specific fixed target facility. 
We have calculated the sensitivity curves for these experiments using the null result for DM detection
for the existed and planned statistics of leptons accumulated on target, such called invisible mode. 

We have discussed in detail  the expected reach of the NA64$e$ experiment to probe the  dark photon emission via the 
missing energy process of vector meson production $e Z \to e Z J/\psi$ followed by invisible decay  
$J/\psi\to a \gamma_D(\to \chi\bar{\chi})$. We have shown that latter signal process
 dominates over the bremsstrahlung $a\gamma_D$ pair emission in the reaction  $e Z \to e Z  \gamma^*( \to a \gamma_D)$ 
 at NA64$e$. Thus expected reach of NA64$e$  from $J/\psi \to a \gamma_D$ can rule out the projected 
 bounds from the  off-shell photon emission for $\mbox{EOT}\simeq 5 \times 10^{12}$. 

We have recasted the BaBar monophoton data $e^+e^- \to a \gamma \gamma_D$ to derive constraints on the dark ALP portal
coupling, implying that the dark photon decays mainly to the DM fermion pair. 
We have shown that the existed BaBar bounds 
rule out the current NA64$e$ constraints for $\mbox{EOT}\simeq 3 \times 10^{11}$. 
 
 We have considered the modifications of the minimal dark ALP portal scenario by including  two 
 benchmark Lagrangians in the model: hadron-- and lepton--specific couplings which imply ALP Yukawa--like 
 interaction mainly with hadrons and leptons respectively.  For both hadron-- and lepton--specific 
 scenarios we have calculated the cross-sections of dark photon production $l Z \to l Z \gamma_D$
 by using WW approximation.  Calculation in WW approximation is in a reasonable agreement with exact 
 tree--level calculation performed with {\tt CalcHEP}.  

The calculations reveal the  differences between both hadrophilc and leptophilic  cross--sections of dark photon 
production. First,  the ALP--hadrophilic cross-section is relatively small 
since it scales to the first power of nucleon/atomic number $\propto A$. Contrary, the ALP--leptophilic rate of 
$\gamma_D$ production  is enhanced due to the factor $\propto Z^2$.  Second, the hadrophilc cross-section 
is sensitive of the cut on the angle of dark photon emission $\theta_{\gamma_D}$. Large $\theta_{\gamma_D}$ leads  
to the inelastic scattering that is associated with significant transfer momentum to nucleon. So that, in 
present analysis we  conservatively set the upper benchmark value 
$\theta_{\gamma_D} \lesssim \theta_{\gamma_D}^{max}\simeq 0.1$ rad in order to get rid of the inelastic 
interaction of ALP with nucleus matter. On the other hand, the leptophilic 
cross-section  $d\sigma_{2\to 3}/ dx$ depends weakly on $\theta^{max}_{\gamma_D}$,  since 
$d\sigma_{2\to3}/d \theta_{\gamma_D}$ has a  very  narrow  peak at the typical angles of dark photon  emission  
$\theta_{\gamma_D}\simeq m_l/E_l \ll 1$. 

Moreover, we want to note that total cross-section for the  hadrophilic scenario is suppressed relative to  
leptophilic one. The  suppression  factor is estimated to be of the order of  $\sim  10^{-7}\div 10^{-5}$.  
Thus the resulted   hadrophilc constraints are weaker than leptophilic bounds  for the same 
number of  leptons  accumulated on  target. That difference can be found in both 
Fig.~\ref{Boundshadrophilic} and  Fig.~\ref{BoundsLeptohilic} where the bounds on the 
combination of couplings for the hadrophilic and leptophilic 
scenarios are depicted respectively.  In the framework of considered benchmark scenarios, 
the most stringent  constraints on the couplings are expected from the projected experiment LDMX. 

In the future we plan to  consider also the possible 
benchmark dark ALP portal signatures along with both ALP portal and vector portal  scenarios.

\begin{acknowledgments} 
We would like to thank N.~Arefyeva, R.~Capdevilla, A.~Celentano, X.~Chu, P.~Crivelli, S.~Demidov, 
D.~Forbes, S.~Gninenko, D.~Gorbunov, Y.~Kahn, M.~Kirsanov, N.~Krasnikov, 
G.~Krnjaic,  L.~Molina Bueno, A.~Pimikov,  J.~Pradler,  A.~Pukhov, P.~Schuster, H.~Sieber  
and  F.~Tkachov\footnote{Deceased} for very helpful discussions and  correspondences.

This work was funded by BMBF (Germany) ``Verbundark photon projekt 05P2021 (ErUM-FSP T01) -
Run 3 von ALICE am LHC: Perturbative Berechnungen von Wirkungsquerschnitten
f\"ur ALICE" (F\"orderkennzeichen: 05P21VTCAA), by ANID PIA/APOYO AFB180002 (Chile),
by FONDECYT (Chile) under Grant No. 1191103, 
and by ANID$-$Millennium Program$-$ICN2019\_044 (Chile).  
This research was supported by TSU and TPU development programs. 
The work of A.S.Z. is supported by Grant of AYSS JINR (22-302-02).	
The work of D.V.K on description of the  dark matter missing energy signatures of NA64$e$ 
and regarding exclusion limits for the minimal dark axion portal  scenario is  supported
by the  Russian Science Foundation  RSF grant 21-12-00379.

\end{acknowledgments}	

\appendix

\section{Analytic form of loop integral}
\label{AppLoop}

Here we present analytic form for loop integrals (\ref{gy1}) and (\ref{gy2}) which are used for fermion EDM calculation:  
\eq
&&g_1(y)=1-y^2\log{y} -\frac{y(y^2-2)}{2\sqrt{y^2-4}}\times\\
&&\times\left(\arctan \left(\frac{y^2-2}{y\sqrt{y^2-4}}\right)-\arctan \left(\frac{y}{\sqrt{y^2-4}}\right)\right)\, ,\nonumber
\en
\eq
&&g_2(y)= k_n\Bigg\{\frac{3}{2}-y^2+(y^2-3)y^2\log{y}+ \frac{\sqrt{y^2-4}}{2}\times\nonumber\\
&& (y^2-2)y\left(\arctan{\frac{y}{\sqrt{y^2-4}}} +\arctan{\frac{2-y^2}{y\sqrt{y^2-4}}}\right)\\
&& +\frac{y^3}{2}\sqrt{y^2-4}\left(
\arctan{\frac{y}{\sqrt{y^2-4}}}-\arctan{\frac{2-y^2}{y\sqrt{y^2-4}}}
\right)\!\!\Bigg\}\, .\nonumber
\en

\section{Lepton EDM induced by the Barr-Zee diagrams with lepton-dark photon-axion loop}
\label{App_A}
Here we consider details of calculation of the correction to the lepton EDM induced by 
Barr-Zee diagrams with lepton-dark photon-axion loop. In order to remove the logarithmic divergence 
containing in the Barr-Zee diagrams we add the local counter-term, which is induced by the following 
Lagrangian
\eq
\mathcal{L}_{\rm ct} = g_{\rm ct}(\mu) \, F_{\mu\nu} \,
\bar{l} \, i \sigma^{\mu\nu} \, \gamma^5 \, l  \,,
\en
where $g_{\rm ct}(\mu)$ is the coupling constant depending on 
renormalization scale $\mu$. 
In particular, we choose $g_{\rm ct}(\mu)$ as
\eq
g_{\rm ct}(\mu) = g \, \biggl[ \frac{1}{\bar{\epsilon}} - \gamma \, \log\frac{m_l^2}{\mu^2}
\biggr] \,,
\en
where $m_l$ is the lepton mass, $g$ and $\gamma$ are the parameters, which
will be fixed to drop divergence in the Barr-Zee diagrams and
guarantee scale independence of the result for lepton EDM. Here,
$\bar\epsilon$ is the pole in the $\overline{\rm MS}$ renormalization scheme:
\eq
\frac{1}{\bar\epsilon} = \frac{1}{\epsilon} + \Gamma'(1) - \log(4\pi)
\en
Therefore, the contribution of the counter-term to the lepton EDM is
\eq
d_{l}^{\rm ct} = - 2 g \, \biggl[ \frac{1}{\bar\epsilon} + \gamma \, \log\frac{m_l^2}{\mu^2}
\biggr]
\en
The contribution of the Barr-Zee diagrams to the lepton EDM reads
\eq
d_{l}^{BZ} = \frac{G}{8 \pi^2} \, \biggl[ \frac{1}{\bar\epsilon} -
\int\limits_0^1 d^3\alpha \, \log\frac{\Delta}{\mu^2} \biggr] \,,
\en
where $G =  g_{a\gamma \gamma_D} \,  g_s \, e \epsilon  $ is the effective coupling,
which is the product of	three respective couplings between photon, dark	photon,	axion,	
and leptons (see e.~g.~Eq.~(\ref{LagrangianVectorAxionEDM}) for detail),
\eq
\Delta = m_{A'}^2 \alpha_1 + m_{a}^2 \alpha_2 + m_l^2 \alpha_3^2
\,,
\en
where 
\eq
\int\limits_0^1 d^3\alpha \equiv 
\int\limits_0^1 d\alpha_1 \int\limits_0^1 d\alpha_2 \int\limits_0^1 d\alpha_3
\, \delta\Big(1-\sum\limits_{i=1}^3 \alpha_i\Big) \,.
\en 
Result for the lepton EDM can be written as 
\eq\label{d_tot}
d_l &=& d_{l}^{\rm BZ} + d_{l}^{\rm ct} =
\frac{2}{\bar\epsilon} \, \biggl[
\frac{G}{16 \pi^2} - g \biggr] \\
&-& 2 \log\frac{m_l^2}{\mu^2} \,
\biggl[
\frac{G}{16 \pi^2} - g \gamma \biggr] - \frac{G}{4 \pi^2} \,
\int\limits_0^1 d^3\alpha \, \log\frac{\Delta}{\mu^2} \,. \nonumber
\en
From Eq.~(\ref{d_tot}) follows that the couplings $g$ and $\gamma$ must fixed
as
\eq
g = \frac{G}{16 \pi^2}\,, \qquad \gamma = 1 \,.
\en
Finally, one gets
\eq
d_l = - \frac{G}{4 \pi^2} \,
\int\limits_0^1 d^3\alpha \, \log\frac{\Delta}{\mu^2}  \,.
\en
After integration over two Feynman parameters we have 
\eq
d_l = - \frac{G}{4 \pi^2} \, \biggl[ - \frac{1}{2} + 
\frac{I(m_{A'}^2) - I(m_{a}^2)}{m_{A'}^2 - m_{a}^2} \biggr] 
\,, 
\en
where 
\eq
I(m^2) &=& \int\limits_0^1 dx \, 
\Big[m^2 (1-x) + m_l^2 x^2\Big]  \nonumber\\ 
&\times& \log\frac{m^2 (1-x) + m_l^2 x^2}{\mu^2} 
\en 
Next, it is interesting to consider a few limiting cases: 

(1) $m_l = 0$ 
\eq
d_l = - \frac{G}{4 \pi^2} \, \biggl[ - 1 + 
\frac{I_0(m_{A'}^2) - I_0(m_{a}^2)}{m_{A'}^2 - m_{a}^2} 
\biggr] 
\,, 
\en
where 
\eq
I_0(m^2) = m^2 (1-x) \, \log\frac{m^2}{\mu^2} \,. 
\en 

(2) $m_l = 0$, $m_{A'} = m_a = m$ 
\eq
d_l = - \frac{G}{8 \pi^2} \, \biggl[ - 1 
+ \log\frac{m^2}{\mu^2}\biggr] 
\en

\bibliography{bibl}

\begin{thebibliography}{106}%
\makeatletter
\providecommand \@ifxundefined [1]{%
 \@ifx{#1\undefined}
}%
\providecommand \@ifnum [1]{%
 \ifnum #1\expandafter \@firstoftwo
 \else \expandafter \@secondoftwo
 \fi
}%
\providecommand \@ifx [1]{%
 \ifx #1\expandafter \@firstoftwo
 \else \expandafter \@secondoftwo
 \fi
}%
\providecommand \natexlab [1]{#1}%
\providecommand \enquote  [1]{``#1''}%
\providecommand \bibnamefont  [1]{#1}%
\providecommand \bibfnamefont [1]{#1}%
\providecommand \citenamefont [1]{#1}%
\providecommand \href@noop [0]{\@secondoftwo}%
\providecommand \href [0]{\begingroup \@sanitize@url \@href}%
\providecommand \@href[1]{\@@startlink{#1}\@@href}%
\providecommand \@@href[1]{\endgroup#1\@@endlink}%
\providecommand \@sanitize@url [0]{\catcode `\\12\catcode `\$12\catcode
  `\&12\catcode `\#12\catcode `\^12\catcode `\_12\catcode `\%12\relax}%
\providecommand \@@startlink[1]{}%
\providecommand \@@endlink[0]{}%
\providecommand \url  [0]{\begingroup\@sanitize@url \@url }%
\providecommand \@url [1]{\endgroup\@href {#1}{\urlprefix }}%
\providecommand \urlprefix  [0]{URL }%
\providecommand \Eprint [0]{\href }%
\providecommand \doibase [0]{http://dx.doi.org/}%
\providecommand \selectlanguage [0]{\@gobble}%
\providecommand \bibinfo  [0]{\@secondoftwo}%
\providecommand \bibfield  [0]{\@secondoftwo}%
\providecommand \translation [1]{[#1]}%
\providecommand \BibitemOpen [0]{}%
\providecommand \bibitemStop [0]{}%
\providecommand \bibitemNoStop [0]{.\EOS\space}%
\providecommand \EOS [0]{\spacefactor3000\relax}%
\providecommand \BibitemShut  [1]{\csname bibitem#1\endcsname}%
\let\auto@bib@innerbib\@empty
%</preamble>
\bibitem [{\citenamefont {Peccei}\ and\ \citenamefont
  {Quinn}(1977)}]{Peccei:1977ur}%
  \BibitemOpen
  \bibfield  {author} {\bibinfo {author} {\bibfnamefont {R.~D.}\ \bibnamefont
  {Peccei}}\ and\ \bibinfo {author} {\bibfnamefont {H.~R.}\ \bibnamefont
  {Quinn}},\ }\href {\doibase 10.1103/PhysRevD.16.1791} {\bibfield  {journal}
  {\bibinfo  {journal} {Phys. Rev. D}\ }\textbf {\bibinfo {volume} {16}},\
  \bibinfo {pages} {1791} (\bibinfo {year} {1977})}\BibitemShut {NoStop}%
\bibitem [{\citenamefont {Di~Luzio}\ \emph {et~al.}(2020)\citenamefont
  {Di~Luzio}, \citenamefont {Giannotti}, \citenamefont {Nardi},\ and\
  \citenamefont {Visinelli}}]{DiLuzio:2020wdo}%
  \BibitemOpen
  \bibfield  {author} {\bibinfo {author} {\bibfnamefont {L.}~\bibnamefont
  {Di~Luzio}}, \bibinfo {author} {\bibfnamefont {M.}~\bibnamefont {Giannotti}},
  \bibinfo {author} {\bibfnamefont {E.}~\bibnamefont {Nardi}}, \ and\ \bibinfo
  {author} {\bibfnamefont {L.}~\bibnamefont {Visinelli}},\ }\href {\doibase
  10.1016/j.physrep.2020.06.002} {\bibfield  {journal} {\bibinfo  {journal}
  {Phys. Rept.}\ }\textbf {\bibinfo {volume} {870}},\ \bibinfo {pages} {1}
  (\bibinfo {year} {2020})},\ \Eprint {http://arxiv.org/abs/2003.01100}
  {arXiv:2003.01100 [hep-ph]} \BibitemShut {NoStop}%
\bibitem [{\citenamefont {Aoyama}\ \emph {et~al.}(2020)\citenamefont {Aoyama}
  \emph {et~al.}}]{Aoyama:2020ynm}%
  \BibitemOpen
  \bibfield  {author} {\bibinfo {author} {\bibfnamefont {T.}~\bibnamefont
  {Aoyama}} \emph {et~al.},\ }\href {\doibase 10.1016/j.physrep.2020.07.006}
  {\bibfield  {journal} {\bibinfo  {journal} {Phys. Rept.}\ }\textbf {\bibinfo
  {volume} {887}},\ \bibinfo {pages} {1} (\bibinfo {year} {2020})},\ \Eprint
  {http://arxiv.org/abs/2006.04822} {arXiv:2006.04822 [hep-ph]} \BibitemShut
  {NoStop}%
\bibitem [{\citenamefont {Dorokhov}\ \emph {et~al.}(2014)\citenamefont
  {Dorokhov}, \citenamefont {Radzhabov},\ and\ \citenamefont
  {Zhevlakov}}]{Dorokhov:2014iva}%
  \BibitemOpen
  \bibfield  {author} {\bibinfo {author} {\bibfnamefont {A.~E.}\ \bibnamefont
  {Dorokhov}}, \bibinfo {author} {\bibfnamefont {A.~E.}\ \bibnamefont
  {Radzhabov}}, \ and\ \bibinfo {author} {\bibfnamefont {A.~S.}\ \bibnamefont
  {Zhevlakov}},\ }\href {\doibase 10.1134/S0021364014140045} {\bibfield
  {journal} {\bibinfo  {journal} {JETP Lett.}\ }\textbf {\bibinfo {volume}
  {100}},\ \bibinfo {pages} {133} (\bibinfo {year} {2014})},\ \Eprint
  {http://arxiv.org/abs/1406.1019} {arXiv:1406.1019 [hep-ph]} \BibitemShut
  {NoStop}%
\bibitem [{\citenamefont {Boehm}\ and\ \citenamefont
  {Fayet}(2004)}]{Boehm:2003hm}%
  \BibitemOpen
  \bibfield  {author} {\bibinfo {author} {\bibfnamefont {C.}~\bibnamefont
  {Boehm}}\ and\ \bibinfo {author} {\bibfnamefont {P.}~\bibnamefont {Fayet}},\
  }\href {\doibase 10.1016/j.nuclphysb.2004.01.015} {\bibfield  {journal}
  {\bibinfo  {journal} {Nucl. Phys. B}\ }\textbf {\bibinfo {volume} {683}},\
  \bibinfo {pages} {219} (\bibinfo {year} {2004})},\ \Eprint
  {http://arxiv.org/abs/hep-ph/0305261} {arXiv:hep-ph/0305261} \BibitemShut
  {NoStop}%
\bibitem [{\citenamefont {Dolan}\ \emph {et~al.}(2015)\citenamefont {Dolan},
  \citenamefont {Kahlhoefer}, \citenamefont {McCabe},\ and\ \citenamefont
  {Schmidt-Hoberg}}]{Dolan:2014ska}%
  \BibitemOpen
  \bibfield  {author} {\bibinfo {author} {\bibfnamefont {M.~J.}\ \bibnamefont
  {Dolan}}, \bibinfo {author} {\bibfnamefont {F.}~\bibnamefont {Kahlhoefer}},
  \bibinfo {author} {\bibfnamefont {C.}~\bibnamefont {McCabe}}, \ and\ \bibinfo
  {author} {\bibfnamefont {K.}~\bibnamefont {Schmidt-Hoberg}},\ }\href
  {\doibase 10.1007/JHEP03(2015)171} {\bibfield  {journal} {\bibinfo  {journal}
  {JHEP}\ }\textbf {\bibinfo {volume} {03}},\ \bibinfo {pages} {171} (\bibinfo
  {year} {2015})},\ \bibinfo {note} {[Erratum: JHEP 07, 103 (2015)]},\ \Eprint
  {http://arxiv.org/abs/1412.5174} {arXiv:1412.5174 [hep-ph]} \BibitemShut
  {NoStop}%
\bibitem [{\citenamefont {Hochberg}\ \emph {et~al.}(2018)\citenamefont
  {Hochberg}, \citenamefont {Kuflik}, \citenamefont {Mcgehee}, \citenamefont
  {Murayama},\ and\ \citenamefont {Schutz}}]{Hochberg:2018rjs}%
  \BibitemOpen
  \bibfield  {author} {\bibinfo {author} {\bibfnamefont {Y.}~\bibnamefont
  {Hochberg}}, \bibinfo {author} {\bibfnamefont {E.}~\bibnamefont {Kuflik}},
  \bibinfo {author} {\bibfnamefont {R.}~\bibnamefont {Mcgehee}}, \bibinfo
  {author} {\bibfnamefont {H.}~\bibnamefont {Murayama}}, \ and\ \bibinfo
  {author} {\bibfnamefont {K.}~\bibnamefont {Schutz}},\ }\href {\doibase
  10.1103/PhysRevD.98.115031} {\bibfield  {journal} {\bibinfo  {journal} {Phys.
  Rev. D}\ }\textbf {\bibinfo {volume} {98}},\ \bibinfo {pages} {115031}
  (\bibinfo {year} {2018})},\ \Eprint {http://arxiv.org/abs/1806.10139}
  {arXiv:1806.10139 [hep-ph]} \BibitemShut {NoStop}%
\bibitem [{\citenamefont {Han}\ \emph {et~al.}(2021)\citenamefont {Han},
  \citenamefont {L\'opez-Ib\'a\~nez}, \citenamefont {Melis}, \citenamefont
  {Vives},\ and\ \citenamefont {Yang}}]{Han:2020dwo}%
  \BibitemOpen
  \bibfield  {author} {\bibinfo {author} {\bibfnamefont {C.}~\bibnamefont
  {Han}}, \bibinfo {author} {\bibfnamefont {M.~L.}\ \bibnamefont
  {L\'opez-Ib\'a\~nez}}, \bibinfo {author} {\bibfnamefont {A.}~\bibnamefont
  {Melis}}, \bibinfo {author} {\bibfnamefont {O.}~\bibnamefont {Vives}}, \ and\
  \bibinfo {author} {\bibfnamefont {J.~M.}\ \bibnamefont {Yang}},\ }\href
  {\doibase 10.1103/PhysRevD.103.035028} {\bibfield  {journal} {\bibinfo
  {journal} {Phys. Rev. D}\ }\textbf {\bibinfo {volume} {103}},\ \bibinfo
  {pages} {035028} (\bibinfo {year} {2021})},\ \Eprint
  {http://arxiv.org/abs/2007.08834} {arXiv:2007.08834 [hep-ph]} \BibitemShut
  {NoStop}%
\bibitem [{\citenamefont {Davoudiasl}\ \emph {et~al.}(2021)\citenamefont
  {Davoudiasl}, \citenamefont {Marcarelli},\ and\ \citenamefont
  {Neil}}]{Davoudiasl:2021mjy}%
  \BibitemOpen
  \bibfield  {author} {\bibinfo {author} {\bibfnamefont {H.}~\bibnamefont
  {Davoudiasl}}, \bibinfo {author} {\bibfnamefont {R.}~\bibnamefont
  {Marcarelli}}, \ and\ \bibinfo {author} {\bibfnamefont {E.~T.}\ \bibnamefont
  {Neil}},\ }\href@noop {} {\  (\bibinfo {year} {2021})},\ \Eprint
  {http://arxiv.org/abs/2112.04513} {arXiv:2112.04513 [hep-ph]} \BibitemShut
  {NoStop}%
\bibitem [{\citenamefont {Gninenko}\ and\ \citenamefont
  {Krasnikov}(2022)}]{Gninenko:2022ttd}%
  \BibitemOpen
  \bibfield  {author} {\bibinfo {author} {\bibfnamefont {S.~N.}\ \bibnamefont
  {Gninenko}}\ and\ \bibinfo {author} {\bibfnamefont {N.~V.}\ \bibnamefont
  {Krasnikov}},\ }\href@noop {} {\  (\bibinfo {year} {2022})},\ \Eprint
  {http://arxiv.org/abs/2202.04410} {arXiv:2202.04410 [hep-ph]} \BibitemShut
  {NoStop}%
\bibitem [{\citenamefont {Bauer}\ \emph {et~al.}(2021)\citenamefont {Bauer},
  \citenamefont {Neubert}, \citenamefont {Renner}, \citenamefont {Schnubel},\
  and\ \citenamefont {Thamm}}]{Bauer:2020jbp}%
  \BibitemOpen
  \bibfield  {author} {\bibinfo {author} {\bibfnamefont {M.}~\bibnamefont
  {Bauer}}, \bibinfo {author} {\bibfnamefont {M.}~\bibnamefont {Neubert}},
  \bibinfo {author} {\bibfnamefont {S.}~\bibnamefont {Renner}}, \bibinfo
  {author} {\bibfnamefont {M.}~\bibnamefont {Schnubel}}, \ and\ \bibinfo
  {author} {\bibfnamefont {A.}~\bibnamefont {Thamm}},\ }\href {\doibase
  10.1007/JHEP04(2021)063} {\bibfield  {journal} {\bibinfo  {journal} {JHEP}\
  }\textbf {\bibinfo {volume} {04}},\ \bibinfo {pages} {063} (\bibinfo {year}
  {2021})},\ \Eprint {http://arxiv.org/abs/2012.12272} {arXiv:2012.12272
  [hep-ph]} \BibitemShut {NoStop}%
\bibitem [{\citenamefont {Choi}\ \emph {et~al.}(2021)\citenamefont {Choi},
  \citenamefont {Im},\ and\ \citenamefont {Sub~Shin}}]{Choi:2020rgn}%
  \BibitemOpen
  \bibfield  {author} {\bibinfo {author} {\bibfnamefont {K.}~\bibnamefont
  {Choi}}, \bibinfo {author} {\bibfnamefont {S.~H.}\ \bibnamefont {Im}}, \ and\
  \bibinfo {author} {\bibfnamefont {C.}~\bibnamefont {Sub~Shin}},\ }\href
  {\doibase 10.1146/annurev-nucl-120720-031147} {\bibfield  {journal} {\bibinfo
   {journal} {Ann. Rev. Nucl. Part. Sci.}\ }\textbf {\bibinfo {volume} {71}},\
  \bibinfo {pages} {225} (\bibinfo {year} {2021})},\ \Eprint
  {http://arxiv.org/abs/2012.05029} {arXiv:2012.05029 [hep-ph]} \BibitemShut
  {NoStop}%
\bibitem [{\citenamefont {Dusaev}\ \emph {et~al.}(2020)\citenamefont {Dusaev},
  \citenamefont {Kirpichnikov},\ and\ \citenamefont
  {Kirsanov}}]{Dusaev:2020gxi}%
  \BibitemOpen
  \bibfield  {author} {\bibinfo {author} {\bibfnamefont {R.~R.}\ \bibnamefont
  {Dusaev}}, \bibinfo {author} {\bibfnamefont {D.~V.}\ \bibnamefont
  {Kirpichnikov}}, \ and\ \bibinfo {author} {\bibfnamefont {M.~M.}\
  \bibnamefont {Kirsanov}},\ }\href {\doibase 10.1103/PhysRevD.102.055018}
  {\bibfield  {journal} {\bibinfo  {journal} {Phys. Rev. D}\ }\textbf {\bibinfo
  {volume} {102}},\ \bibinfo {pages} {055018} (\bibinfo {year} {2020})},\
  \Eprint {http://arxiv.org/abs/2004.04469} {arXiv:2004.04469 [hep-ph]}
  \BibitemShut {NoStop}%
\bibitem [{\citenamefont {Banerjee}\ \emph {et~al.}(2020)\citenamefont
  {Banerjee} \emph {et~al.}}]{NA64:2020qwq}%
  \BibitemOpen
  \bibfield  {author} {\bibinfo {author} {\bibfnamefont {D.}~\bibnamefont
  {Banerjee}} \emph {et~al.} (\bibinfo {collaboration} {NA64}),\ }\href
  {\doibase 10.1103/PhysRevLett.125.081801} {\bibfield  {journal} {\bibinfo
  {journal} {Phys. Rev. Lett.}\ }\textbf {\bibinfo {volume} {125}},\ \bibinfo
  {pages} {081801} (\bibinfo {year} {2020})},\ \Eprint
  {http://arxiv.org/abs/2005.02710} {arXiv:2005.02710 [hep-ex]} \BibitemShut
  {NoStop}%
\bibitem [{\citenamefont {Ishida}\ \emph {et~al.}(2021)\citenamefont {Ishida},
  \citenamefont {Matsuzaki},\ and\ \citenamefont {Shigekami}}]{Ishida:2020oxl}%
  \BibitemOpen
  \bibfield  {author} {\bibinfo {author} {\bibfnamefont {H.}~\bibnamefont
  {Ishida}}, \bibinfo {author} {\bibfnamefont {S.}~\bibnamefont {Matsuzaki}}, \
  and\ \bibinfo {author} {\bibfnamefont {Y.}~\bibnamefont {Shigekami}},\ }\href
  {\doibase 10.1103/PhysRevD.103.095022} {\bibfield  {journal} {\bibinfo
  {journal} {Phys. Rev. D}\ }\textbf {\bibinfo {volume} {103}},\ \bibinfo
  {pages} {095022} (\bibinfo {year} {2021})},\ \Eprint
  {http://arxiv.org/abs/2006.02725} {arXiv:2006.02725 [hep-ph]} \BibitemShut
  {NoStop}%
\bibitem [{\citenamefont {Sakaki}\ and\ \citenamefont
  {Ueda}(2021)}]{Sakaki:2020mqb}%
  \BibitemOpen
  \bibfield  {author} {\bibinfo {author} {\bibfnamefont {Y.}~\bibnamefont
  {Sakaki}}\ and\ \bibinfo {author} {\bibfnamefont {D.}~\bibnamefont {Ueda}},\
  }\href {\doibase 10.1103/PhysRevD.103.035024} {\bibfield  {journal} {\bibinfo
   {journal} {Phys. Rev. D}\ }\textbf {\bibinfo {volume} {103}},\ \bibinfo
  {pages} {035024} (\bibinfo {year} {2021})},\ \Eprint
  {http://arxiv.org/abs/2009.13790} {arXiv:2009.13790 [hep-ph]} \BibitemShut
  {NoStop}%
\bibitem [{\citenamefont {Brdar}\ \emph {et~al.}(2021)\citenamefont {Brdar},
  \citenamefont {Dutta}, \citenamefont {Jang}, \citenamefont {Kim},
  \citenamefont {Shoemaker}, \citenamefont {Tabrizi}, \citenamefont
  {Thompson},\ and\ \citenamefont {Yu}}]{Brdar:2020dpr}%
  \BibitemOpen
  \bibfield  {author} {\bibinfo {author} {\bibfnamefont {V.}~\bibnamefont
  {Brdar}}, \bibinfo {author} {\bibfnamefont {B.}~\bibnamefont {Dutta}},
  \bibinfo {author} {\bibfnamefont {W.}~\bibnamefont {Jang}}, \bibinfo {author}
  {\bibfnamefont {D.}~\bibnamefont {Kim}}, \bibinfo {author} {\bibfnamefont
  {I.~M.}\ \bibnamefont {Shoemaker}}, \bibinfo {author} {\bibfnamefont
  {Z.}~\bibnamefont {Tabrizi}}, \bibinfo {author} {\bibfnamefont
  {A.}~\bibnamefont {Thompson}}, \ and\ \bibinfo {author} {\bibfnamefont
  {J.}~\bibnamefont {Yu}},\ }\href {\doibase 10.1103/PhysRevLett.126.201801}
  {\bibfield  {journal} {\bibinfo  {journal} {Phys. Rev. Lett.}\ }\textbf
  {\bibinfo {volume} {126}},\ \bibinfo {pages} {201801} (\bibinfo {year}
  {2021})},\ \Eprint {http://arxiv.org/abs/2011.07054} {arXiv:2011.07054
  [hep-ph]} \BibitemShut {NoStop}%
\bibitem [{\citenamefont {Salnikov}\ \emph {et~al.}(2021)\citenamefont
  {Salnikov}, \citenamefont {Satunin}, \citenamefont {Kirpichnikov},\ and\
  \citenamefont {Fitkevich}}]{Salnikov:2020urr}%
  \BibitemOpen
  \bibfield  {author} {\bibinfo {author} {\bibfnamefont {D.}~\bibnamefont
  {Salnikov}}, \bibinfo {author} {\bibfnamefont {P.}~\bibnamefont {Satunin}},
  \bibinfo {author} {\bibfnamefont {D.~V.}\ \bibnamefont {Kirpichnikov}}, \
  and\ \bibinfo {author} {\bibfnamefont {M.}~\bibnamefont {Fitkevich}},\ }\href
  {\doibase 10.1007/jhep03(2021)143} {\bibfield  {journal} {\bibinfo  {journal}
  {JHEP}\ }\textbf {\bibinfo {volume} {03}},\ \bibinfo {pages} {143} (\bibinfo
  {year} {2021})},\ \Eprint {http://arxiv.org/abs/2011.12871} {arXiv:2011.12871
  [hep-ph]} \BibitemShut {NoStop}%
\bibitem [{\citenamefont {Bogorad}\ \emph {et~al.}(2019)\citenamefont
  {Bogorad}, \citenamefont {Hook}, \citenamefont {Kahn},\ and\ \citenamefont
  {Soreq}}]{Bogorad:2019pbu}%
  \BibitemOpen
  \bibfield  {author} {\bibinfo {author} {\bibfnamefont {Z.}~\bibnamefont
  {Bogorad}}, \bibinfo {author} {\bibfnamefont {A.}~\bibnamefont {Hook}},
  \bibinfo {author} {\bibfnamefont {Y.}~\bibnamefont {Kahn}}, \ and\ \bibinfo
  {author} {\bibfnamefont {Y.}~\bibnamefont {Soreq}},\ }\href {\doibase
  10.1103/PhysRevLett.123.021801} {\bibfield  {journal} {\bibinfo  {journal}
  {Phys. Rev. Lett.}\ }\textbf {\bibinfo {volume} {123}},\ \bibinfo {pages}
  {021801} (\bibinfo {year} {2019})},\ \Eprint
  {http://arxiv.org/abs/1902.01418} {arXiv:1902.01418 [hep-ph]} \BibitemShut
  {NoStop}%
\bibitem [{\citenamefont {Kahn}\ \emph {et~al.}(2022)\citenamefont {Kahn},
  \citenamefont {Giaccone}, \citenamefont {Lunin}, \citenamefont {Netepenko},
  \citenamefont {Pilipenko},\ and\ \citenamefont {Wentzel}}]{Kahn:2022uko}%
  \BibitemOpen
  \bibfield  {author} {\bibinfo {author} {\bibfnamefont {Y.}~\bibnamefont
  {Kahn}}, \bibinfo {author} {\bibfnamefont {B.}~\bibnamefont {Giaccone}},
  \bibinfo {author} {\bibfnamefont {A.}~\bibnamefont {Lunin}}, \bibinfo
  {author} {\bibfnamefont {A.}~\bibnamefont {Netepenko}}, \bibinfo {author}
  {\bibfnamefont {R.}~\bibnamefont {Pilipenko}}, \ and\ \bibinfo {author}
  {\bibfnamefont {M.}~\bibnamefont {Wentzel}},\ }\href {\doibase
  10.1117/12.2616734} {\bibfield  {journal} {\bibinfo  {journal} {Proc. SPIE
  Int. Soc. Opt. Eng.}\ }\textbf {\bibinfo {volume} {12016}},\ \bibinfo {pages}
  {29} (\bibinfo {year} {2022})}\BibitemShut {NoStop}%
\bibitem [{\citenamefont {Darm\'e}\ \emph {et~al.}(2021)\citenamefont
  {Darm\'e}, \citenamefont {Giacchino}, \citenamefont {Nardi},\ and\
  \citenamefont {Raggi}}]{Darme:2020sjf}%
  \BibitemOpen
  \bibfield  {author} {\bibinfo {author} {\bibfnamefont {L.}~\bibnamefont
  {Darm\'e}}, \bibinfo {author} {\bibfnamefont {F.}~\bibnamefont {Giacchino}},
  \bibinfo {author} {\bibfnamefont {E.}~\bibnamefont {Nardi}}, \ and\ \bibinfo
  {author} {\bibfnamefont {M.}~\bibnamefont {Raggi}},\ }\href {\doibase
  10.1007/JHEP06(2021)009} {\bibfield  {journal} {\bibinfo  {journal} {JHEP}\
  }\textbf {\bibinfo {volume} {06}},\ \bibinfo {pages} {009} (\bibinfo {year}
  {2021})},\ \Eprint {http://arxiv.org/abs/2012.07894} {arXiv:2012.07894
  [hep-ph]} \BibitemShut {NoStop}%
\bibitem [{\citenamefont {Dev}\ \emph {et~al.}(2021)\citenamefont {Dev},
  \citenamefont {Kim}, \citenamefont {Sinha},\ and\ \citenamefont
  {Zhang}}]{Dev:2021ofc}%
  \BibitemOpen
  \bibfield  {author} {\bibinfo {author} {\bibfnamefont {P.~S.~B.}\
  \bibnamefont {Dev}}, \bibinfo {author} {\bibfnamefont {D.}~\bibnamefont
  {Kim}}, \bibinfo {author} {\bibfnamefont {K.}~\bibnamefont {Sinha}}, \ and\
  \bibinfo {author} {\bibfnamefont {Y.}~\bibnamefont {Zhang}},\ }\href
  {\doibase 10.1103/PhysRevD.104.035037} {\bibfield  {journal} {\bibinfo
  {journal} {Phys. Rev. D}\ }\textbf {\bibinfo {volume} {104}},\ \bibinfo
  {pages} {035037} (\bibinfo {year} {2021})},\ \Eprint
  {http://arxiv.org/abs/2101.08781} {arXiv:2101.08781 [hep-ph]} \BibitemShut
  {NoStop}%
\bibitem [{\citenamefont {Abramowicz}\ \emph {et~al.}(2021)\citenamefont
  {Abramowicz} \emph {et~al.}}]{Abramowicz:2021zja}%
  \BibitemOpen
  \bibfield  {author} {\bibinfo {author} {\bibfnamefont {H.}~\bibnamefont
  {Abramowicz}} \emph {et~al.},\ }\href {\doibase
  10.1140/epjs/s11734-021-00249-z} {\bibfield  {journal} {\bibinfo  {journal}
  {Eur. Phys. J. ST}\ }\textbf {\bibinfo {volume} {230}},\ \bibinfo {pages}
  {2445} (\bibinfo {year} {2021})},\ \Eprint {http://arxiv.org/abs/2102.02032}
  {arXiv:2102.02032 [hep-ex]} \BibitemShut {NoStop}%
\bibitem [{\citenamefont {Fortin}\ \emph {et~al.}(2021)\citenamefont {Fortin},
  \citenamefont {Guo}, \citenamefont {Harris}, \citenamefont {Kim},
  \citenamefont {Sinha},\ and\ \citenamefont {Sun}}]{Fortin:2021cog}%
  \BibitemOpen
  \bibfield  {author} {\bibinfo {author} {\bibfnamefont {J.-F.}\ \bibnamefont
  {Fortin}}, \bibinfo {author} {\bibfnamefont {H.-K.}\ \bibnamefont {Guo}},
  \bibinfo {author} {\bibfnamefont {S.~P.}\ \bibnamefont {Harris}}, \bibinfo
  {author} {\bibfnamefont {D.}~\bibnamefont {Kim}}, \bibinfo {author}
  {\bibfnamefont {K.}~\bibnamefont {Sinha}}, \ and\ \bibinfo {author}
  {\bibfnamefont {C.}~\bibnamefont {Sun}},\ }\href {\doibase
  10.1142/S0218271821300020} {\bibfield  {journal} {\bibinfo  {journal} {Int.
  J. Mod. Phys. D}\ }\textbf {\bibinfo {volume} {30}},\ \bibinfo {pages}
  {2130002} (\bibinfo {year} {2021})},\ \Eprint
  {http://arxiv.org/abs/2102.12503} {arXiv:2102.12503 [hep-ph]} \BibitemShut
  {NoStop}%
\bibitem [{\citenamefont {Asai}\ \emph {et~al.}(2021)\citenamefont {Asai},
  \citenamefont {Iwamoto}, \citenamefont {Sakaki},\ and\ \citenamefont
  {Ueda}}]{Asai:2021ehn}%
  \BibitemOpen
  \bibfield  {author} {\bibinfo {author} {\bibfnamefont {K.}~\bibnamefont
  {Asai}}, \bibinfo {author} {\bibfnamefont {S.}~\bibnamefont {Iwamoto}},
  \bibinfo {author} {\bibfnamefont {Y.}~\bibnamefont {Sakaki}}, \ and\ \bibinfo
  {author} {\bibfnamefont {D.}~\bibnamefont {Ueda}},\ }\href {\doibase
  10.1007/JHEP09(2021)183} {\bibfield  {journal} {\bibinfo  {journal} {JHEP}\
  }\textbf {\bibinfo {volume} {09}},\ \bibinfo {pages} {183} (\bibinfo {year}
  {2021})},\ \Eprint {http://arxiv.org/abs/2105.13768} {arXiv:2105.13768
  [hep-ph]} \BibitemShut {NoStop}%
\bibitem [{\citenamefont {Balkin}\ \emph {et~al.}(2021)\citenamefont {Balkin},
  \citenamefont {Krasny}, \citenamefont {Ma}, \citenamefont {Safdi},\ and\
  \citenamefont {Soreq}}]{Balkin:2021jdr}%
  \BibitemOpen
  \bibfield  {author} {\bibinfo {author} {\bibfnamefont {R.}~\bibnamefont
  {Balkin}}, \bibinfo {author} {\bibfnamefont {M.~W.}\ \bibnamefont {Krasny}},
  \bibinfo {author} {\bibfnamefont {T.}~\bibnamefont {Ma}}, \bibinfo {author}
  {\bibfnamefont {B.~R.}\ \bibnamefont {Safdi}}, \ and\ \bibinfo {author}
  {\bibfnamefont {Y.}~\bibnamefont {Soreq}},\ }\href {\doibase
  10.1002/andp.202100222} {\  (\bibinfo {year} {2021}),\
  10.1002/andp.202100222},\ \Eprint {http://arxiv.org/abs/2105.15072}
  {arXiv:2105.15072 [hep-ph]} \BibitemShut {NoStop}%
\bibitem [{\citenamefont {Blinov}\ \emph {et~al.}(2022)\citenamefont {Blinov},
  \citenamefont {Kowalczyk},\ and\ \citenamefont {Wynne}}]{Blinov:2021say}%
  \BibitemOpen
  \bibfield  {author} {\bibinfo {author} {\bibfnamefont {N.}~\bibnamefont
  {Blinov}}, \bibinfo {author} {\bibfnamefont {E.}~\bibnamefont {Kowalczyk}}, \
  and\ \bibinfo {author} {\bibfnamefont {M.}~\bibnamefont {Wynne}},\ }\href
  {\doibase 10.1007/JHEP02(2022)036} {\bibfield  {journal} {\bibinfo  {journal}
  {JHEP}\ }\textbf {\bibinfo {volume} {02}},\ \bibinfo {pages} {036} (\bibinfo
  {year} {2022})},\ \Eprint {http://arxiv.org/abs/2112.09814} {arXiv:2112.09814
  [hep-ph]} \BibitemShut {NoStop}%
\bibitem [{\citenamefont {Larin}\ \emph {et~al.}(2011)\citenamefont {Larin}
  \emph {et~al.}}]{PrimEx:2010fvg}%
  \BibitemOpen
  \bibfield  {author} {\bibinfo {author} {\bibfnamefont {I.}~\bibnamefont
  {Larin}} \emph {et~al.} (\bibinfo {collaboration} {PrimEx}),\ }\href
  {\doibase 10.1103/PhysRevLett.106.162303} {\bibfield  {journal} {\bibinfo
  {journal} {Phys. Rev. Lett.}\ }\textbf {\bibinfo {volume} {106}},\ \bibinfo
  {pages} {162303} (\bibinfo {year} {2011})},\ \Eprint
  {http://arxiv.org/abs/1009.1681} {arXiv:1009.1681 [nucl-ex]} \BibitemShut
  {NoStop}%
\bibitem [{\citenamefont {Gninenko}\ \emph {et~al.}(2016)\citenamefont
  {Gninenko}, \citenamefont {Krasnikov}, \citenamefont {Kirsanov},\ and\
  \citenamefont {Kirpichnikov}}]{Gninenko:2016kpg}%
  \BibitemOpen
  \bibfield  {author} {\bibinfo {author} {\bibfnamefont {S.~N.}\ \bibnamefont
  {Gninenko}}, \bibinfo {author} {\bibfnamefont {N.~V.}\ \bibnamefont
  {Krasnikov}}, \bibinfo {author} {\bibfnamefont {M.~M.}\ \bibnamefont
  {Kirsanov}}, \ and\ \bibinfo {author} {\bibfnamefont {D.~V.}\ \bibnamefont
  {Kirpichnikov}},\ }\href {\doibase 10.1103/PhysRevD.94.095025} {\bibfield
  {journal} {\bibinfo  {journal} {Phys. Rev. D}\ }\textbf {\bibinfo {volume}
  {94}},\ \bibinfo {pages} {095025} (\bibinfo {year} {2016})},\ \Eprint
  {http://arxiv.org/abs/1604.08432} {arXiv:1604.08432 [hep-ph]} \BibitemShut
  {NoStop}%
\bibitem [{\citenamefont {Banerjee}\ \emph {et~al.}(2017)\citenamefont
  {Banerjee} \emph {et~al.}}]{NA64:2016oww}%
  \BibitemOpen
  \bibfield  {author} {\bibinfo {author} {\bibfnamefont {D.}~\bibnamefont
  {Banerjee}} \emph {et~al.} (\bibinfo {collaboration} {NA64}),\ }\href
  {\doibase 10.1103/PhysRevLett.118.011802} {\bibfield  {journal} {\bibinfo
  {journal} {Phys. Rev. Lett.}\ }\textbf {\bibinfo {volume} {118}},\ \bibinfo
  {pages} {011802} (\bibinfo {year} {2017})},\ \Eprint
  {http://arxiv.org/abs/1610.02988} {arXiv:1610.02988 [hep-ex]} \BibitemShut
  {NoStop}%
\bibitem [{\citenamefont {Gninenko}\ \emph {et~al.}(2018)\citenamefont
  {Gninenko}, \citenamefont {Kirpichnikov}, \citenamefont {Kirsanov},\ and\
  \citenamefont {Krasnikov}}]{Gninenko:2017yus}%
  \BibitemOpen
  \bibfield  {author} {\bibinfo {author} {\bibfnamefont {S.~N.}\ \bibnamefont
  {Gninenko}}, \bibinfo {author} {\bibfnamefont {D.~V.}\ \bibnamefont
  {Kirpichnikov}}, \bibinfo {author} {\bibfnamefont {M.~M.}\ \bibnamefont
  {Kirsanov}}, \ and\ \bibinfo {author} {\bibfnamefont {N.~V.}\ \bibnamefont
  {Krasnikov}},\ }\href {\doibase 10.1016/j.physletb.2018.05.010} {\bibfield
  {journal} {\bibinfo  {journal} {Phys. Lett. B}\ }\textbf {\bibinfo {volume}
  {782}},\ \bibinfo {pages} {406} (\bibinfo {year} {2018})},\ \Eprint
  {http://arxiv.org/abs/1712.05706} {arXiv:1712.05706 [hep-ph]} \BibitemShut
  {NoStop}%
\bibitem [{\citenamefont {Gninenko}\ \emph
  {et~al.}(2019{\natexlab{a}})\citenamefont {Gninenko}, \citenamefont
  {Kirpichnikov}, \citenamefont {Kirsanov},\ and\ \citenamefont
  {Krasnikov}}]{Gninenko:2019qiv}%
  \BibitemOpen
  \bibfield  {author} {\bibinfo {author} {\bibfnamefont {S.~N.}\ \bibnamefont
  {Gninenko}}, \bibinfo {author} {\bibfnamefont {D.~V.}\ \bibnamefont
  {Kirpichnikov}}, \bibinfo {author} {\bibfnamefont {M.~M.}\ \bibnamefont
  {Kirsanov}}, \ and\ \bibinfo {author} {\bibfnamefont {N.~V.}\ \bibnamefont
  {Krasnikov}},\ }\href {\doibase 10.1016/j.physletb.2019.07.015} {\bibfield
  {journal} {\bibinfo  {journal} {Phys. Lett. B}\ }\textbf {\bibinfo {volume}
  {796}},\ \bibinfo {pages} {117} (\bibinfo {year} {2019}{\natexlab{a}})},\
  \Eprint {http://arxiv.org/abs/1903.07899} {arXiv:1903.07899 [hep-ph]}
  \BibitemShut {NoStop}%
\bibitem [{\citenamefont {Banerjee}\ \emph {et~al.}(2019)\citenamefont
  {Banerjee} \emph {et~al.}}]{Banerjee:2019pds}%
  \BibitemOpen
  \bibfield  {author} {\bibinfo {author} {\bibfnamefont {D.}~\bibnamefont
  {Banerjee}} \emph {et~al.},\ }\href {\doibase 10.1103/PhysRevLett.123.121801}
  {\bibfield  {journal} {\bibinfo  {journal} {Phys. Rev. Lett.}\ }\textbf
  {\bibinfo {volume} {123}},\ \bibinfo {pages} {121801} (\bibinfo {year}
  {2019})},\ \Eprint {http://arxiv.org/abs/1906.00176} {arXiv:1906.00176
  [hep-ex]} \BibitemShut {NoStop}%
\bibitem [{\citenamefont {Andreev}\ \emph
  {et~al.}(2021{\natexlab{a}})\citenamefont {Andreev} \emph
  {et~al.}}]{Andreev:2021fzd}%
  \BibitemOpen
  \bibfield  {author} {\bibinfo {author} {\bibfnamefont {Y.~M.}\ \bibnamefont
  {Andreev}} \emph {et~al.},\ }\href {\doibase 10.1103/PhysRevD.104.L091701}
  {\bibfield  {journal} {\bibinfo  {journal} {Phys. Rev. D}\ }\textbf {\bibinfo
  {volume} {104}},\ \bibinfo {pages} {L091701} (\bibinfo {year}
  {2021}{\natexlab{a}})},\ \Eprint {http://arxiv.org/abs/2108.04195}
  {arXiv:2108.04195 [hep-ex]} \BibitemShut {NoStop}%
\bibitem [{\citenamefont {Andreev}\ \emph
  {et~al.}(2021{\natexlab{b}})\citenamefont {Andreev} \emph
  {et~al.}}]{NA64:2021xzo}%
  \BibitemOpen
  \bibfield  {author} {\bibinfo {author} {\bibfnamefont {Y.~M.}\ \bibnamefont
  {Andreev}} \emph {et~al.} (\bibinfo {collaboration} {NA64}),\ }\href
  {\doibase 10.1103/PhysRevLett.126.211802} {\bibfield  {journal} {\bibinfo
  {journal} {Phys. Rev. Lett.}\ }\textbf {\bibinfo {volume} {126}},\ \bibinfo
  {pages} {211802} (\bibinfo {year} {2021}{\natexlab{b}})},\ \Eprint
  {http://arxiv.org/abs/2102.01885} {arXiv:2102.01885 [hep-ex]} \BibitemShut
  {NoStop}%
\bibitem [{\citenamefont {Blinov}\ \emph {et~al.}(2021)\citenamefont {Blinov},
  \citenamefont {Krnjaic},\ and\ \citenamefont {Tuckler}}]{Blinov:2020epi}%
  \BibitemOpen
  \bibfield  {author} {\bibinfo {author} {\bibfnamefont {N.}~\bibnamefont
  {Blinov}}, \bibinfo {author} {\bibfnamefont {G.}~\bibnamefont {Krnjaic}}, \
  and\ \bibinfo {author} {\bibfnamefont {D.}~\bibnamefont {Tuckler}},\ }\href
  {\doibase 10.1103/PhysRevD.103.035030} {\bibfield  {journal} {\bibinfo
  {journal} {Phys. Rev. D}\ }\textbf {\bibinfo {volume} {103}},\ \bibinfo
  {pages} {035030} (\bibinfo {year} {2021})},\ \Eprint
  {http://arxiv.org/abs/2010.03577} {arXiv:2010.03577 [hep-ph]} \BibitemShut
  {NoStop}%
\bibitem [{\citenamefont {Beattie}\ \emph {et~al.}(2018)\citenamefont {Beattie}
  \emph {et~al.}}]{Beattie:2018xsk}%
  \BibitemOpen
  \bibfield  {author} {\bibinfo {author} {\bibfnamefont {T.~D.}\ \bibnamefont
  {Beattie}} \emph {et~al.},\ }\href {\doibase 10.1016/j.nima.2018.04.006}
  {\bibfield  {journal} {\bibinfo  {journal} {Nucl. Instrum. Meth. A}\ }\textbf
  {\bibinfo {volume} {896}},\ \bibinfo {pages} {24} (\bibinfo {year} {2018})},\
  \Eprint {http://arxiv.org/abs/1801.03088} {arXiv:1801.03088
  [physics.ins-det]} \BibitemShut {NoStop}%
\bibitem [{\citenamefont {Essig}\ \emph {et~al.}(2013)\citenamefont {Essig}
  \emph {et~al.}}]{Essig:2013lka}%
  \BibitemOpen
  \bibfield  {author} {\bibinfo {author} {\bibfnamefont {R.}~\bibnamefont
  {Essig}} \emph {et~al.},\ }in\ \href@noop {} {\emph {\bibinfo {booktitle}
  {{Community Summer Study 2013}: {Snowmass on the Mississippi}}}}\ (\bibinfo
  {year} {2013})\ \Eprint {http://arxiv.org/abs/1311.0029} {arXiv:1311.0029
  [hep-ph]} \BibitemShut {NoStop}%
\bibitem [{\citenamefont {Arcadi}\ \emph {et~al.}(2020)\citenamefont {Arcadi},
  \citenamefont {Djouadi},\ and\ \citenamefont {Raidal}}]{Arcadi:2019lka}%
  \BibitemOpen
  \bibfield  {author} {\bibinfo {author} {\bibfnamefont {G.}~\bibnamefont
  {Arcadi}}, \bibinfo {author} {\bibfnamefont {A.}~\bibnamefont {Djouadi}}, \
  and\ \bibinfo {author} {\bibfnamefont {M.}~\bibnamefont {Raidal}},\ }\href
  {\doibase 10.1016/j.physrep.2019.11.003} {\bibfield  {journal} {\bibinfo
  {journal} {Phys. Rept.}\ }\textbf {\bibinfo {volume} {842}},\ \bibinfo
  {pages} {1} (\bibinfo {year} {2020})},\ \Eprint
  {http://arxiv.org/abs/1903.03616} {arXiv:1903.03616 [hep-ph]} \BibitemShut
  {NoStop}%
\bibitem [{\citenamefont {Fortuna}\ \emph {et~al.}(2021)\citenamefont
  {Fortuna}, \citenamefont {Roig},\ and\ \citenamefont
  {Wudka}}]{Fortuna:2020wwx}%
  \BibitemOpen
  \bibfield  {author} {\bibinfo {author} {\bibfnamefont {F.}~\bibnamefont
  {Fortuna}}, \bibinfo {author} {\bibfnamefont {P.}~\bibnamefont {Roig}}, \
  and\ \bibinfo {author} {\bibfnamefont {J.}~\bibnamefont {Wudka}},\ }\href
  {\doibase 10.1007/JHEP02(2021)223} {\bibfield  {journal} {\bibinfo  {journal}
  {JHEP}\ }\textbf {\bibinfo {volume} {02}},\ \bibinfo {pages} {223} (\bibinfo
  {year} {2021})},\ \Eprint {http://arxiv.org/abs/2008.10609} {arXiv:2008.10609
  [hep-ph]} \BibitemShut {NoStop}%
\bibitem [{\citenamefont {Buras}\ \emph {et~al.}(2021)\citenamefont {Buras},
  \citenamefont {Crivellin}, \citenamefont {Kirk}, \citenamefont {Manzari},\
  and\ \citenamefont {Montull}}]{Buras:2021btx}%
  \BibitemOpen
  \bibfield  {author} {\bibinfo {author} {\bibfnamefont {A.~J.}\ \bibnamefont
  {Buras}}, \bibinfo {author} {\bibfnamefont {A.}~\bibnamefont {Crivellin}},
  \bibinfo {author} {\bibfnamefont {F.}~\bibnamefont {Kirk}}, \bibinfo {author}
  {\bibfnamefont {C.~A.}\ \bibnamefont {Manzari}}, \ and\ \bibinfo {author}
  {\bibfnamefont {M.}~\bibnamefont {Montull}},\ }\href {\doibase
  10.1007/JHEP06(2021)068} {\bibfield  {journal} {\bibinfo  {journal} {JHEP}\
  }\textbf {\bibinfo {volume} {06}},\ \bibinfo {pages} {068} (\bibinfo {year}
  {2021})},\ \Eprint {http://arxiv.org/abs/2104.07680} {arXiv:2104.07680
  [hep-ph]} \BibitemShut {NoStop}%
\bibitem [{\citenamefont {Kachanovich}\ \emph {et~al.}(2022)\citenamefont
  {Kachanovich}, \citenamefont {Kovalenko}, \citenamefont {Kuleshov},
  \citenamefont {Lyubovitskij},\ and\ \citenamefont
  {Zhevlakov}}]{Kachanovich:2021eqa}%
  \BibitemOpen
  \bibfield  {author} {\bibinfo {author} {\bibfnamefont {A.}~\bibnamefont
  {Kachanovich}}, \bibinfo {author} {\bibfnamefont {S.}~\bibnamefont
  {Kovalenko}}, \bibinfo {author} {\bibfnamefont {S.}~\bibnamefont {Kuleshov}},
  \bibinfo {author} {\bibfnamefont {V.~E.}\ \bibnamefont {Lyubovitskij}}, \
  and\ \bibinfo {author} {\bibfnamefont {A.~S.}\ \bibnamefont {Zhevlakov}},\
  }\href {\doibase 10.1103/PhysRevD.105.075004} {\bibfield  {journal} {\bibinfo
   {journal} {Phys. Rev. D}\ }\textbf {\bibinfo {volume} {105}},\ \bibinfo
  {pages} {075004} (\bibinfo {year} {2022})},\ \Eprint
  {http://arxiv.org/abs/2111.12522} {arXiv:2111.12522 [hep-ph]} \BibitemShut
  {NoStop}%
\bibitem [{\citenamefont {Escudero}\ \emph {et~al.}(2017)\citenamefont
  {Escudero}, \citenamefont {Rius},\ and\ \citenamefont
  {Sanz}}]{Escudero:2016tzx}%
  \BibitemOpen
  \bibfield  {author} {\bibinfo {author} {\bibfnamefont {M.}~\bibnamefont
  {Escudero}}, \bibinfo {author} {\bibfnamefont {N.}~\bibnamefont {Rius}}, \
  and\ \bibinfo {author} {\bibfnamefont {V.}~\bibnamefont {Sanz}},\ }\href
  {\doibase 10.1007/JHEP02(2017)045} {\bibfield  {journal} {\bibinfo  {journal}
  {JHEP}\ }\textbf {\bibinfo {volume} {02}},\ \bibinfo {pages} {045} (\bibinfo
  {year} {2017})},\ \Eprint {http://arxiv.org/abs/1606.01258} {arXiv:1606.01258
  [hep-ph]} \BibitemShut {NoStop}%
\bibitem [{\citenamefont {Nomura}\ and\ \citenamefont
  {Thaler}(2009)}]{Nomura:2008ru}%
  \BibitemOpen
  \bibfield  {author} {\bibinfo {author} {\bibfnamefont {Y.}~\bibnamefont
  {Nomura}}\ and\ \bibinfo {author} {\bibfnamefont {J.}~\bibnamefont
  {Thaler}},\ }\href {\doibase 10.1103/PhysRevD.79.075008} {\bibfield
  {journal} {\bibinfo  {journal} {Phys. Rev. D}\ }\textbf {\bibinfo {volume}
  {79}},\ \bibinfo {pages} {075008} (\bibinfo {year} {2009})},\ \Eprint
  {http://arxiv.org/abs/0810.5397} {arXiv:0810.5397 [hep-ph]} \BibitemShut
  {NoStop}%
\bibitem [{\citenamefont {Kaneta}\ \emph
  {et~al.}(2017{\natexlab{a}})\citenamefont {Kaneta}, \citenamefont {Lee},\
  and\ \citenamefont {Yun}}]{Kaneta:2016wvf}%
  \BibitemOpen
  \bibfield  {author} {\bibinfo {author} {\bibfnamefont {K.}~\bibnamefont
  {Kaneta}}, \bibinfo {author} {\bibfnamefont {H.-S.}\ \bibnamefont {Lee}}, \
  and\ \bibinfo {author} {\bibfnamefont {S.}~\bibnamefont {Yun}},\ }\href
  {\doibase 10.1103/PhysRevLett.118.101802} {\bibfield  {journal} {\bibinfo
  {journal} {Phys. Rev. Lett.}\ }\textbf {\bibinfo {volume} {118}},\ \bibinfo
  {pages} {101802} (\bibinfo {year} {2017}{\natexlab{a}})},\ \Eprint
  {http://arxiv.org/abs/1611.01466} {arXiv:1611.01466 [hep-ph]} \BibitemShut
  {NoStop}%
\bibitem [{\citenamefont {Kaneta}\ \emph
  {et~al.}(2017{\natexlab{b}})\citenamefont {Kaneta}, \citenamefont {Lee},\
  and\ \citenamefont {Yun}}]{Kaneta:2017wfh}%
  \BibitemOpen
  \bibfield  {author} {\bibinfo {author} {\bibfnamefont {K.}~\bibnamefont
  {Kaneta}}, \bibinfo {author} {\bibfnamefont {H.-S.}\ \bibnamefont {Lee}}, \
  and\ \bibinfo {author} {\bibfnamefont {S.}~\bibnamefont {Yun}},\ }\href
  {\doibase 10.1103/PhysRevD.95.115032} {\bibfield  {journal} {\bibinfo
  {journal} {Phys. Rev. D}\ }\textbf {\bibinfo {volume} {95}},\ \bibinfo
  {pages} {115032} (\bibinfo {year} {2017}{\natexlab{b}})},\ \Eprint
  {http://arxiv.org/abs/1704.07542} {arXiv:1704.07542 [hep-ph]} \BibitemShut
  {NoStop}%
\bibitem [{\citenamefont {Guti\'errez}\ \emph {et~al.}(2021)\citenamefont
  {Guti\'errez}, \citenamefont {Kavanagh}, \citenamefont {Castell\'o-Mor},
  \citenamefont {Casas}, \citenamefont {Diego}, \citenamefont
  {Mart\'\i{}nez-Gonz\'alez},\ and\ \citenamefont
  {Cortabitarte}}]{Gutierrez:2021gol}%
  \BibitemOpen
  \bibfield  {author} {\bibinfo {author} {\bibfnamefont {J.~C.}\ \bibnamefont
  {Guti\'errez}}, \bibinfo {author} {\bibfnamefont {B.~J.}\ \bibnamefont
  {Kavanagh}}, \bibinfo {author} {\bibfnamefont {N.}~\bibnamefont
  {Castell\'o-Mor}}, \bibinfo {author} {\bibfnamefont {F.~J.}\ \bibnamefont
  {Casas}}, \bibinfo {author} {\bibfnamefont {J.~M.}\ \bibnamefont {Diego}},
  \bibinfo {author} {\bibfnamefont {E.}~\bibnamefont
  {Mart\'\i{}nez-Gonz\'alez}}, \ and\ \bibinfo {author} {\bibfnamefont {R.~V.}\
  \bibnamefont {Cortabitarte}},\ }\href@noop {} {\  (\bibinfo {year} {2021})},\
  \Eprint {http://arxiv.org/abs/2112.11387} {arXiv:2112.11387 [hep-ph]}
  \BibitemShut {NoStop}%
\bibitem [{\citenamefont {deNiverville}\ \emph {et~al.}(2018)\citenamefont
  {deNiverville}, \citenamefont {Lee},\ and\ \citenamefont
  {Seo}}]{deNiverville:2018hrc}%
  \BibitemOpen
  \bibfield  {author} {\bibinfo {author} {\bibfnamefont {P.}~\bibnamefont
  {deNiverville}}, \bibinfo {author} {\bibfnamefont {H.-S.}\ \bibnamefont
  {Lee}}, \ and\ \bibinfo {author} {\bibfnamefont {M.-S.}\ \bibnamefont
  {Seo}},\ }\href {\doibase 10.1103/PhysRevD.98.115011} {\bibfield  {journal}
  {\bibinfo  {journal} {Phys. Rev. D}\ }\textbf {\bibinfo {volume} {98}},\
  \bibinfo {pages} {115011} (\bibinfo {year} {2018})},\ \Eprint
  {http://arxiv.org/abs/1806.00757} {arXiv:1806.00757 [hep-ph]} \BibitemShut
  {NoStop}%
\bibitem [{\citenamefont {deNiverville}\ and\ \citenamefont
  {Lee}(2019)}]{deNiverville:2019xsx}%
  \BibitemOpen
  \bibfield  {author} {\bibinfo {author} {\bibfnamefont {P.}~\bibnamefont
  {deNiverville}}\ and\ \bibinfo {author} {\bibfnamefont {H.-S.}\ \bibnamefont
  {Lee}},\ }\href {\doibase 10.1103/PhysRevD.100.055017} {\bibfield  {journal}
  {\bibinfo  {journal} {Phys. Rev. D}\ }\textbf {\bibinfo {volume} {100}},\
  \bibinfo {pages} {055017} (\bibinfo {year} {2019})},\ \Eprint
  {http://arxiv.org/abs/1904.13061} {arXiv:1904.13061 [hep-ph]} \BibitemShut
  {NoStop}%
\bibitem [{\citenamefont {Alekhin}\ \emph {et~al.}(2016)\citenamefont {Alekhin}
  \emph {et~al.}}]{Alekhin:2015byh}%
  \BibitemOpen
  \bibfield  {author} {\bibinfo {author} {\bibfnamefont {S.}~\bibnamefont
  {Alekhin}} \emph {et~al.},\ }\href {\doibase 10.1088/0034-4885/79/12/124201}
  {\bibfield  {journal} {\bibinfo  {journal} {Rept. Prog. Phys.}\ }\textbf
  {\bibinfo {volume} {79}},\ \bibinfo {pages} {124201} (\bibinfo {year}
  {2016})},\ \Eprint {http://arxiv.org/abs/1504.04855} {arXiv:1504.04855
  [hep-ph]} \BibitemShut {NoStop}%
\bibitem [{\citenamefont {Feng}\ \emph {et~al.}(2022)\citenamefont {Feng} \emph
  {et~al.}}]{Feng:2022inv}%
  \BibitemOpen
  \bibfield  {author} {\bibinfo {author} {\bibfnamefont {J.~L.}\ \bibnamefont
  {Feng}} \emph {et~al.},\ }\href@noop {} {\  (\bibinfo {year} {2022})},\
  \Eprint {http://arxiv.org/abs/2203.05090} {arXiv:2203.05090 [hep-ex]}
  \BibitemShut {NoStop}%
\bibitem [{\citenamefont {Deniverville}\ \emph {et~al.}(2021)\citenamefont
  {Deniverville}, \citenamefont {Lee},\ and\ \citenamefont
  {Lee}}]{Deniverville:2020rbv}%
  \BibitemOpen
  \bibfield  {author} {\bibinfo {author} {\bibfnamefont {P.}~\bibnamefont
  {Deniverville}}, \bibinfo {author} {\bibfnamefont {H.-S.}\ \bibnamefont
  {Lee}}, \ and\ \bibinfo {author} {\bibfnamefont {Y.-M.}\ \bibnamefont
  {Lee}},\ }\href {\doibase 10.1103/PhysRevD.103.075006} {\bibfield  {journal}
  {\bibinfo  {journal} {Phys. Rev. D}\ }\textbf {\bibinfo {volume} {103}},\
  \bibinfo {pages} {075006} (\bibinfo {year} {2021})},\ \Eprint
  {http://arxiv.org/abs/2011.03276} {arXiv:2011.03276 [hep-ph]} \BibitemShut
  {NoStop}%
\bibitem [{\citenamefont {Domcke}\ \emph {et~al.}(2021)\citenamefont {Domcke},
  \citenamefont {Schmitz},\ and\ \citenamefont {You}}]{Domcke:2021yuz}%
  \BibitemOpen
  \bibfield  {author} {\bibinfo {author} {\bibfnamefont {V.}~\bibnamefont
  {Domcke}}, \bibinfo {author} {\bibfnamefont {K.}~\bibnamefont {Schmitz}}, \
  and\ \bibinfo {author} {\bibfnamefont {T.}~\bibnamefont {You}},\ }\href@noop
  {} {\  (\bibinfo {year} {2021})},\ \Eprint {http://arxiv.org/abs/2108.11295}
  {arXiv:2108.11295 [hep-ph]} \BibitemShut {NoStop}%
\bibitem [{\citenamefont {Ge}\ \emph {et~al.}(2021)\citenamefont {Ge},
  \citenamefont {Ma},\ and\ \citenamefont {Pasquini}}]{Ge:2021cjz}%
  \BibitemOpen
  \bibfield  {author} {\bibinfo {author} {\bibfnamefont {S.-F.}\ \bibnamefont
  {Ge}}, \bibinfo {author} {\bibfnamefont {X.-D.}\ \bibnamefont {Ma}}, \ and\
  \bibinfo {author} {\bibfnamefont {P.}~\bibnamefont {Pasquini}},\ }\href
  {\doibase 10.1140/epjc/s10052-021-09571-1} {\bibfield  {journal} {\bibinfo
  {journal} {Eur. Phys. J. C}\ }\textbf {\bibinfo {volume} {81}},\ \bibinfo
  {pages} {787} (\bibinfo {year} {2021})},\ \Eprint
  {http://arxiv.org/abs/2104.03276} {arXiv:2104.03276 [hep-ph]} \BibitemShut
  {NoStop}%
\bibitem [{\citenamefont {Kirpichnikov}\ \emph
  {et~al.}(2020{\natexlab{a}})\citenamefont {Kirpichnikov}, \citenamefont
  {Lyubovitskij},\ and\ \citenamefont {Zhevlakov}}]{Kirpichnikov:2020tcf}%
  \BibitemOpen
  \bibfield  {author} {\bibinfo {author} {\bibfnamefont {D.~V.}\ \bibnamefont
  {Kirpichnikov}}, \bibinfo {author} {\bibfnamefont {V.~E.}\ \bibnamefont
  {Lyubovitskij}}, \ and\ \bibinfo {author} {\bibfnamefont {A.~S.}\
  \bibnamefont {Zhevlakov}},\ }\href {\doibase 10.1103/PhysRevD.102.095024}
  {\bibfield  {journal} {\bibinfo  {journal} {Phys. Rev. D}\ }\textbf {\bibinfo
  {volume} {102}},\ \bibinfo {pages} {095024} (\bibinfo {year}
  {2020}{\natexlab{a}})},\ \Eprint {http://arxiv.org/abs/2002.07496}
  {arXiv:2002.07496 [hep-ph]} \BibitemShut {NoStop}%
\bibitem [{\citenamefont {Kirpichnikov}\ \emph
  {et~al.}(2020{\natexlab{b}})\citenamefont {Kirpichnikov}, \citenamefont
  {Lyubovitskij},\ and\ \citenamefont {Zhevlakov}}]{Kirpichnikov:2020lws}%
  \BibitemOpen
  \bibfield  {author} {\bibinfo {author} {\bibfnamefont {D.~V.}\ \bibnamefont
  {Kirpichnikov}}, \bibinfo {author} {\bibfnamefont {V.~E.}\ \bibnamefont
  {Lyubovitskij}}, \ and\ \bibinfo {author} {\bibfnamefont {A.~S.}\
  \bibnamefont {Zhevlakov}},\ }\href {\doibase 10.3390/particles3040047}
  {\bibfield  {journal} {\bibinfo  {journal} {Particles}\ }\textbf {\bibinfo
  {volume} {3}},\ \bibinfo {pages} {719} (\bibinfo {year}
  {2020}{\natexlab{b}})},\ \Eprint {http://arxiv.org/abs/2004.13656}
  {arXiv:2004.13656 [hep-ph]} \BibitemShut {NoStop}%
\bibitem [{\citenamefont {Gninenko}\ \emph {et~al.}(2015)\citenamefont
  {Gninenko}, \citenamefont {Krasnikov},\ and\ \citenamefont
  {Matveev}}]{Gninenko:2014pea}%
  \BibitemOpen
  \bibfield  {author} {\bibinfo {author} {\bibfnamefont {S.~N.}\ \bibnamefont
  {Gninenko}}, \bibinfo {author} {\bibfnamefont {N.~V.}\ \bibnamefont
  {Krasnikov}}, \ and\ \bibinfo {author} {\bibfnamefont {V.~A.}\ \bibnamefont
  {Matveev}},\ }\href {\doibase 10.1103/PhysRevD.91.095015} {\bibfield
  {journal} {\bibinfo  {journal} {Phys. Rev. D}\ }\textbf {\bibinfo {volume}
  {91}},\ \bibinfo {pages} {095015} (\bibinfo {year} {2015})},\ \Eprint
  {http://arxiv.org/abs/1412.1400} {arXiv:1412.1400 [hep-ph]} \BibitemShut
  {NoStop}%
\bibitem [{\citenamefont {Gninenko}\ and\ \citenamefont
  {Krasnikov}(2018)}]{Gninenko:2018tlp}%
  \BibitemOpen
  \bibfield  {author} {\bibinfo {author} {\bibfnamefont {S.~N.}\ \bibnamefont
  {Gninenko}}\ and\ \bibinfo {author} {\bibfnamefont {N.~V.}\ \bibnamefont
  {Krasnikov}},\ }\href {\doibase 10.1016/j.physletb.2018.06.043} {\bibfield
  {journal} {\bibinfo  {journal} {Phys. Lett. B}\ }\textbf {\bibinfo {volume}
  {783}},\ \bibinfo {pages} {24} (\bibinfo {year} {2018})},\ \Eprint
  {http://arxiv.org/abs/1801.10448} {arXiv:1801.10448 [hep-ph]} \BibitemShut
  {NoStop}%
\bibitem [{\citenamefont {Kirpichnikov}\ \emph {et~al.}(2021)\citenamefont
  {Kirpichnikov}, \citenamefont {Sieber}, \citenamefont {Bueno}, \citenamefont
  {Crivelli},\ and\ \citenamefont {Kirsanov}}]{Kirpichnikov:2021jev}%
  \BibitemOpen
  \bibfield  {author} {\bibinfo {author} {\bibfnamefont {D.~V.}\ \bibnamefont
  {Kirpichnikov}}, \bibinfo {author} {\bibfnamefont {H.}~\bibnamefont
  {Sieber}}, \bibinfo {author} {\bibfnamefont {L.~M.}\ \bibnamefont {Bueno}},
  \bibinfo {author} {\bibfnamefont {P.}~\bibnamefont {Crivelli}}, \ and\
  \bibinfo {author} {\bibfnamefont {M.~M.}\ \bibnamefont {Kirsanov}},\ }\href
  {\doibase 10.1103/PhysRevD.104.076012} {\bibfield  {journal} {\bibinfo
  {journal} {Phys. Rev. D}\ }\textbf {\bibinfo {volume} {104}},\ \bibinfo
  {pages} {076012} (\bibinfo {year} {2021})},\ \Eprint
  {http://arxiv.org/abs/2107.13297} {arXiv:2107.13297 [hep-ph]} \BibitemShut
  {NoStop}%
\bibitem [{\citenamefont {Sieber}\ \emph {et~al.}(2022)\citenamefont {Sieber},
  \citenamefont {Banerjee}, \citenamefont {Crivelli}, \citenamefont {Depero},
  \citenamefont {Gninenko}, \citenamefont {Kirpichnikov}, \citenamefont
  {Kirsanov}, \citenamefont {Poliakov},\ and\ \citenamefont
  {Molina~Bueno}}]{Sieber:2021fue}%
  \BibitemOpen
  \bibfield  {author} {\bibinfo {author} {\bibfnamefont {H.}~\bibnamefont
  {Sieber}}, \bibinfo {author} {\bibfnamefont {D.}~\bibnamefont {Banerjee}},
  \bibinfo {author} {\bibfnamefont {P.}~\bibnamefont {Crivelli}}, \bibinfo
  {author} {\bibfnamefont {E.}~\bibnamefont {Depero}}, \bibinfo {author}
  {\bibfnamefont {S.~N.}\ \bibnamefont {Gninenko}}, \bibinfo {author}
  {\bibfnamefont {D.~V.}\ \bibnamefont {Kirpichnikov}}, \bibinfo {author}
  {\bibfnamefont {M.~M.}\ \bibnamefont {Kirsanov}}, \bibinfo {author}
  {\bibfnamefont {V.}~\bibnamefont {Poliakov}}, \ and\ \bibinfo {author}
  {\bibfnamefont {L.}~\bibnamefont {Molina~Bueno}},\ }\href {\doibase
  10.1103/PhysRevD.105.052006} {\bibfield  {journal} {\bibinfo  {journal}
  {Phys. Rev. D}\ }\textbf {\bibinfo {volume} {105}},\ \bibinfo {pages}
  {052006} (\bibinfo {year} {2022})},\ \Eprint
  {http://arxiv.org/abs/2110.15111} {arXiv:2110.15111 [hep-ex]} \BibitemShut
  {NoStop}%
\bibitem [{\citenamefont {Mans}(2017)}]{Mans:2017vej}%
  \BibitemOpen
  \bibfield  {author} {\bibinfo {author} {\bibfnamefont {J.}~\bibnamefont
  {Mans}} (\bibinfo {collaboration} {LDMX}),\ }\href {\doibase
  10.1051/epjconf/201714201020} {\bibfield  {journal} {\bibinfo  {journal} {EPJ
  Web Conf.}\ }\textbf {\bibinfo {volume} {142}},\ \bibinfo {pages} {01020}
  (\bibinfo {year} {2017})}\BibitemShut {NoStop}%
\bibitem [{\citenamefont {Berlin}\ \emph {et~al.}(2019)\citenamefont {Berlin},
  \citenamefont {Blinov}, \citenamefont {Krnjaic}, \citenamefont {Schuster},\
  and\ \citenamefont {Toro}}]{Berlin:2018bsc}%
  \BibitemOpen
  \bibfield  {author} {\bibinfo {author} {\bibfnamefont {A.}~\bibnamefont
  {Berlin}}, \bibinfo {author} {\bibfnamefont {N.}~\bibnamefont {Blinov}},
  \bibinfo {author} {\bibfnamefont {G.}~\bibnamefont {Krnjaic}}, \bibinfo
  {author} {\bibfnamefont {P.}~\bibnamefont {Schuster}}, \ and\ \bibinfo
  {author} {\bibfnamefont {N.}~\bibnamefont {Toro}},\ }\href {\doibase
  10.1103/PhysRevD.99.075001} {\bibfield  {journal} {\bibinfo  {journal} {Phys.
  Rev. D}\ }\textbf {\bibinfo {volume} {99}},\ \bibinfo {pages} {075001}
  (\bibinfo {year} {2019})},\ \Eprint {http://arxiv.org/abs/1807.01730}
  {arXiv:1807.01730 [hep-ph]} \BibitemShut {NoStop}%
\bibitem [{\citenamefont {\r{A}kesson}\ \emph {et~al.}(2018)\citenamefont
  {\r{A}kesson} \emph {et~al.}}]{LDMX:2018cma}%
  \BibitemOpen
  \bibfield  {author} {\bibinfo {author} {\bibfnamefont {T.}~\bibnamefont
  {\r{A}kesson}} \emph {et~al.} (\bibinfo {collaboration} {LDMX}),\ }\href@noop
  {} {\  (\bibinfo {year} {2018})},\ \Eprint {http://arxiv.org/abs/1808.05219}
  {arXiv:1808.05219 [hep-ex]} \BibitemShut {NoStop}%
\bibitem [{\citenamefont {Ankowski}\ \emph {et~al.}(2020)\citenamefont
  {Ankowski}, \citenamefont {Friedland}, \citenamefont {Li}, \citenamefont
  {Moreno}, \citenamefont {Schuster}, \citenamefont {Toro},\ and\ \citenamefont
  {Tran}}]{Ankowski:2019mfd}%
  \BibitemOpen
  \bibfield  {author} {\bibinfo {author} {\bibfnamefont {A.~M.}\ \bibnamefont
  {Ankowski}}, \bibinfo {author} {\bibfnamefont {A.}~\bibnamefont {Friedland}},
  \bibinfo {author} {\bibfnamefont {S.~W.}\ \bibnamefont {Li}}, \bibinfo
  {author} {\bibfnamefont {O.}~\bibnamefont {Moreno}}, \bibinfo {author}
  {\bibfnamefont {P.}~\bibnamefont {Schuster}}, \bibinfo {author}
  {\bibfnamefont {N.}~\bibnamefont {Toro}}, \ and\ \bibinfo {author}
  {\bibfnamefont {N.}~\bibnamefont {Tran}},\ }\href {\doibase
  10.1103/PhysRevD.101.053004} {\bibfield  {journal} {\bibinfo  {journal}
  {Phys. Rev. D}\ }\textbf {\bibinfo {volume} {101}},\ \bibinfo {pages}
  {053004} (\bibinfo {year} {2020})},\ \Eprint
  {http://arxiv.org/abs/1912.06140} {arXiv:1912.06140 [hep-ph]} \BibitemShut
  {NoStop}%
\bibitem [{\citenamefont {Schuster}\ \emph {et~al.}(2022)\citenamefont
  {Schuster}, \citenamefont {Toro},\ and\ \citenamefont
  {Zhou}}]{Schuster:2021mlr}%
  \BibitemOpen
  \bibfield  {author} {\bibinfo {author} {\bibfnamefont {P.}~\bibnamefont
  {Schuster}}, \bibinfo {author} {\bibfnamefont {N.}~\bibnamefont {Toro}}, \
  and\ \bibinfo {author} {\bibfnamefont {K.}~\bibnamefont {Zhou}},\ }\href
  {\doibase 10.1103/PhysRevD.105.035036} {\bibfield  {journal} {\bibinfo
  {journal} {Phys. Rev. D}\ }\textbf {\bibinfo {volume} {105}},\ \bibinfo
  {pages} {035036} (\bibinfo {year} {2022})},\ \Eprint
  {http://arxiv.org/abs/2112.02104} {arXiv:2112.02104 [hep-ph]} \BibitemShut
  {NoStop}%
\bibitem [{\citenamefont {\r{A}kesson}\ \emph {et~al.}(2022)\citenamefont
  {\r{A}kesson} \emph {et~al.}}]{Akesson:2022vza}%
  \BibitemOpen
  \bibfield  {author} {\bibinfo {author} {\bibfnamefont {T.}~\bibnamefont
  {\r{A}kesson}} \emph {et~al.},\ }in\ \href@noop {} {\emph {\bibinfo
  {booktitle} {{2022 Snowmass Summer Study}}}}\ (\bibinfo {year} {2022})\
  \Eprint {http://arxiv.org/abs/2203.08192} {arXiv:2203.08192 [hep-ex]}
  \BibitemShut {NoStop}%
\bibitem [{\citenamefont {Kahn}\ \emph {et~al.}(2018)\citenamefont {Kahn},
  \citenamefont {Krnjaic}, \citenamefont {Tran},\ and\ \citenamefont
  {Whitbeck}}]{Kahn:2018cqs}%
  \BibitemOpen
  \bibfield  {author} {\bibinfo {author} {\bibfnamefont {Y.}~\bibnamefont
  {Kahn}}, \bibinfo {author} {\bibfnamefont {G.}~\bibnamefont {Krnjaic}},
  \bibinfo {author} {\bibfnamefont {N.}~\bibnamefont {Tran}}, \ and\ \bibinfo
  {author} {\bibfnamefont {A.}~\bibnamefont {Whitbeck}},\ }\href {\doibase
  10.1007/JHEP09(2018)153} {\bibfield  {journal} {\bibinfo  {journal} {JHEP}\
  }\textbf {\bibinfo {volume} {09}},\ \bibinfo {pages} {153} (\bibinfo {year}
  {2018})},\ \Eprint {http://arxiv.org/abs/1804.03144} {arXiv:1804.03144
  [hep-ph]} \BibitemShut {NoStop}%
\bibitem [{\citenamefont {Capdevilla}\ \emph {et~al.}(2021)\citenamefont
  {Capdevilla}, \citenamefont {Curtin}, \citenamefont {Kahn},\ and\
  \citenamefont {Krnjaic}}]{Capdevilla:2021kcf}%
  \BibitemOpen
  \bibfield  {author} {\bibinfo {author} {\bibfnamefont {R.}~\bibnamefont
  {Capdevilla}}, \bibinfo {author} {\bibfnamefont {D.}~\bibnamefont {Curtin}},
  \bibinfo {author} {\bibfnamefont {Y.}~\bibnamefont {Kahn}}, \ and\ \bibinfo
  {author} {\bibfnamefont {G.}~\bibnamefont {Krnjaic}},\ }\href@noop {} {\
  (\bibinfo {year} {2021})},\ \Eprint {http://arxiv.org/abs/2112.08377}
  {arXiv:2112.08377 [hep-ph]} \BibitemShut {NoStop}%
\bibitem [{\citenamefont {Holdom}(1986)}]{Holdom:1985ag}%
  \BibitemOpen
  \bibfield  {author} {\bibinfo {author} {\bibfnamefont {B.}~\bibnamefont
  {Holdom}},\ }\href {\doibase 10.1016/0370-2693(86)91377-8} {\bibfield
  {journal} {\bibinfo  {journal} {Phys. Lett. B}\ }\textbf {\bibinfo {volume}
  {166}},\ \bibinfo {pages} {196} (\bibinfo {year} {1986})}\BibitemShut
  {NoStop}%
\bibitem [{\citenamefont {Gunion}\ \emph {et~al.}(2000)\citenamefont {Gunion},
  \citenamefont {Haber}, \citenamefont {Kane},\ and\ \citenamefont
  {Dawson}}]{Gunion:1989we}%
  \BibitemOpen
  \bibfield  {author} {\bibinfo {author} {\bibfnamefont {J.~F.}\ \bibnamefont
  {Gunion}}, \bibinfo {author} {\bibfnamefont {H.~E.}\ \bibnamefont {Haber}},
  \bibinfo {author} {\bibfnamefont {G.~L.}\ \bibnamefont {Kane}}, \ and\
  \bibinfo {author} {\bibfnamefont {S.}~\bibnamefont {Dawson}},\ }\href@noop {}
  {\emph {\bibinfo {title} {{The Higgs Hunter's Guide}}}},\ Vol.~\bibinfo
  {volume} {80}\ (\bibinfo {year} {2000})\BibitemShut {NoStop}%
\bibitem [{\citenamefont {Djouadi}(2008)}]{Djouadi:2005gj}%
  \BibitemOpen
  \bibfield  {author} {\bibinfo {author} {\bibfnamefont {A.}~\bibnamefont
  {Djouadi}},\ }\href {\doibase 10.1016/j.physrep.2007.10.005} {\bibfield
  {journal} {\bibinfo  {journal} {Phys. Rept.}\ }\textbf {\bibinfo {volume}
  {459}},\ \bibinfo {pages} {1} (\bibinfo {year} {2008})},\ \Eprint
  {http://arxiv.org/abs/hep-ph/0503173} {arXiv:hep-ph/0503173} \BibitemShut
  {NoStop}%
\bibitem [{\citenamefont {Branco}\ \emph {et~al.}(2012)\citenamefont {Branco},
  \citenamefont {Ferreira}, \citenamefont {Lavoura}, \citenamefont {Rebelo},
  \citenamefont {Sher},\ and\ \citenamefont {Silva}}]{Branco:2011iw}%
  \BibitemOpen
  \bibfield  {author} {\bibinfo {author} {\bibfnamefont {G.~C.}\ \bibnamefont
  {Branco}}, \bibinfo {author} {\bibfnamefont {P.~M.}\ \bibnamefont
  {Ferreira}}, \bibinfo {author} {\bibfnamefont {L.}~\bibnamefont {Lavoura}},
  \bibinfo {author} {\bibfnamefont {M.~N.}\ \bibnamefont {Rebelo}}, \bibinfo
  {author} {\bibfnamefont {M.}~\bibnamefont {Sher}}, \ and\ \bibinfo {author}
  {\bibfnamefont {J.~P.}\ \bibnamefont {Silva}},\ }\href {\doibase
  10.1016/j.physrep.2012.02.002} {\bibfield  {journal} {\bibinfo  {journal}
  {Phys. Rept.}\ }\textbf {\bibinfo {volume} {516}},\ \bibinfo {pages} {1}
  (\bibinfo {year} {2012})},\ \Eprint {http://arxiv.org/abs/1106.0034}
  {arXiv:1106.0034 [hep-ph]} \BibitemShut {NoStop}%
\bibitem [{\citenamefont {Chun}\ and\ \citenamefont
  {Mondal}(2021)}]{Chun:2021rtk}%
  \BibitemOpen
  \bibfield  {author} {\bibinfo {author} {\bibfnamefont {E.~J.}\ \bibnamefont
  {Chun}}\ and\ \bibinfo {author} {\bibfnamefont {T.}~\bibnamefont {Mondal}},\
  }\href {\doibase 10.1007/JHEP07(2021)044} {\bibfield  {journal} {\bibinfo
  {journal} {JHEP}\ }\textbf {\bibinfo {volume} {07}},\ \bibinfo {pages} {044}
  (\bibinfo {year} {2021})},\ \Eprint {http://arxiv.org/abs/2104.03701}
  {arXiv:2104.03701 [hep-ph]} \BibitemShut {NoStop}%
\bibitem [{\citenamefont {Dzuba}\ \emph {et~al.}(2018)\citenamefont {Dzuba},
  \citenamefont {Flambaum}, \citenamefont {Samsonov},\ and\ \citenamefont
  {Stadnik}}]{Dzuba:2018anu}%
  \BibitemOpen
  \bibfield  {author} {\bibinfo {author} {\bibfnamefont {V.~A.}\ \bibnamefont
  {Dzuba}}, \bibinfo {author} {\bibfnamefont {V.~V.}\ \bibnamefont {Flambaum}},
  \bibinfo {author} {\bibfnamefont {I.~B.}\ \bibnamefont {Samsonov}}, \ and\
  \bibinfo {author} {\bibfnamefont {Y.~V.}\ \bibnamefont {Stadnik}},\ }\href
  {\doibase 10.1103/PhysRevD.98.035048} {\bibfield  {journal} {\bibinfo
  {journal} {Phys. Rev. D}\ }\textbf {\bibinfo {volume} {98}},\ \bibinfo
  {pages} {035048} (\bibinfo {year} {2018})},\ \Eprint
  {http://arxiv.org/abs/1805.01234} {arXiv:1805.01234 [physics.atom-ph]}
  \BibitemShut {NoStop}%
\bibitem [{\citenamefont {Flambaum}\ \emph {et~al.}(2009)\citenamefont
  {Flambaum}, \citenamefont {Lambert},\ and\ \citenamefont
  {Pospelov}}]{Flambaum:2009mz}%
  \BibitemOpen
  \bibfield  {author} {\bibinfo {author} {\bibfnamefont {V.}~\bibnamefont
  {Flambaum}}, \bibinfo {author} {\bibfnamefont {S.}~\bibnamefont {Lambert}}, \
  and\ \bibinfo {author} {\bibfnamefont {M.}~\bibnamefont {Pospelov}},\ }\href
  {\doibase 10.1103/PhysRevD.80.105021} {\bibfield  {journal} {\bibinfo
  {journal} {Phys. Rev. D}\ }\textbf {\bibinfo {volume} {80}},\ \bibinfo
  {pages} {105021} (\bibinfo {year} {2009})},\ \Eprint
  {http://arxiv.org/abs/0902.3217} {arXiv:0902.3217 [hep-ph]} \BibitemShut
  {NoStop}%
\bibitem [{\citenamefont {Yan}\ \emph {et~al.}(2019)\citenamefont {Yan},
  \citenamefont {Sun}, \citenamefont {Peng}, \citenamefont {Guo}, \citenamefont
  {Liu}, \citenamefont {Peng},\ and\ \citenamefont {Zheng}}]{Yan:2019dar}%
  \BibitemOpen
  \bibfield  {author} {\bibinfo {author} {\bibfnamefont {H.}~\bibnamefont
  {Yan}}, \bibinfo {author} {\bibfnamefont {G.~A.}\ \bibnamefont {Sun}},
  \bibinfo {author} {\bibfnamefont {S.~M.}\ \bibnamefont {Peng}}, \bibinfo
  {author} {\bibfnamefont {H.}~\bibnamefont {Guo}}, \bibinfo {author}
  {\bibfnamefont {B.~Q.}\ \bibnamefont {Liu}}, \bibinfo {author} {\bibfnamefont
  {M.}~\bibnamefont {Peng}}, \ and\ \bibinfo {author} {\bibfnamefont
  {H.}~\bibnamefont {Zheng}},\ }\href {\doibase 10.1140/epjc/s10052-019-7442-8}
  {\bibfield  {journal} {\bibinfo  {journal} {Eur. Phys. J. C}\ }\textbf
  {\bibinfo {volume} {79}},\ \bibinfo {pages} {971} (\bibinfo {year}
  {2019})}\BibitemShut {NoStop}%
\bibitem [{\citenamefont {O'Hare}\ and\ \citenamefont
  {Vitagliano}(2020)}]{OHare:2020wah}%
  \BibitemOpen
  \bibfield  {author} {\bibinfo {author} {\bibfnamefont {C.~A.~J.}\
  \bibnamefont {O'Hare}}\ and\ \bibinfo {author} {\bibfnamefont
  {E.}~\bibnamefont {Vitagliano}},\ }\href {\doibase
  10.1103/PhysRevD.102.115026} {\bibfield  {journal} {\bibinfo  {journal}
  {Phys. Rev. D}\ }\textbf {\bibinfo {volume} {102}},\ \bibinfo {pages}
  {115026} (\bibinfo {year} {2020})},\ \Eprint
  {http://arxiv.org/abs/2010.03889} {arXiv:2010.03889 [hep-ph]} \BibitemShut
  {NoStop}%
\bibitem [{\citenamefont {Zyla}\ \emph
  {et~al.}(2020{\natexlab{a}})\citenamefont {Zyla} \emph {et~al.}}]{PDG20}%
  \BibitemOpen
  \bibfield  {author} {\bibinfo {author} {\bibfnamefont {P.~A.}\ \bibnamefont
  {Zyla}} \emph {et~al.} (\bibinfo {collaboration} {Particle Data Group}),\
  }\href {\doibase 10.1093/ptep/ptaa104} {\bibfield  {journal} {\bibinfo
  {journal} {PTEP}\ }\textbf {\bibinfo {volume} {2020}},\ \bibinfo {pages}
  {083C01} (\bibinfo {year} {2020}{\natexlab{a}})}\BibitemShut {NoStop}%
\bibitem [{\citenamefont {Zhevlakov}\ and\ \citenamefont
  {Lyubovitskij}(2020)}]{Zhevlakov:2020bvr}%
  \BibitemOpen
  \bibfield  {author} {\bibinfo {author} {\bibfnamefont {A.~S.}\ \bibnamefont
  {Zhevlakov}}\ and\ \bibinfo {author} {\bibfnamefont {V.~E.}\ \bibnamefont
  {Lyubovitskij}},\ }\href {\doibase 10.1103/PhysRevD.101.115041} {\bibfield
  {journal} {\bibinfo  {journal} {Phys. Rev. D}\ }\textbf {\bibinfo {volume}
  {101}},\ \bibinfo {pages} {115041} (\bibinfo {year} {2020})},\ \Eprint
  {http://arxiv.org/abs/2003.12217} {arXiv:2003.12217 [hep-ph]} \BibitemShut
  {NoStop}%
\bibitem [{\citenamefont {Crewther}\ \emph {et~al.}(1979)\citenamefont
  {Crewther}, \citenamefont {Di~Vecchia}, \citenamefont {Veneziano},\ and\
  \citenamefont {Witten}}]{Crewther:1979pi}%
  \BibitemOpen
  \bibfield  {author} {\bibinfo {author} {\bibfnamefont {R.~J.}\ \bibnamefont
  {Crewther}}, \bibinfo {author} {\bibfnamefont {P.}~\bibnamefont
  {Di~Vecchia}}, \bibinfo {author} {\bibfnamefont {G.}~\bibnamefont
  {Veneziano}}, \ and\ \bibinfo {author} {\bibfnamefont {E.}~\bibnamefont
  {Witten}},\ }\href {\doibase 10.1016/0370-2693(79)90128-X} {\bibfield
  {journal} {\bibinfo  {journal} {Phys. Lett. B}\ }\textbf {\bibinfo {volume}
  {88}},\ \bibinfo {pages} {123} (\bibinfo {year} {1979})},\ \bibinfo {note}
  {[Erratum: Phys.Lett.B 91, 487 (1980)]}\BibitemShut {NoStop}%
\bibitem [{\citenamefont {Zhevlakov}\ \emph
  {et~al.}(2019{\natexlab{a}})\citenamefont {Zhevlakov}, \citenamefont
  {Gorchtein}, \citenamefont {Hiller~Blin}, \citenamefont {Gutsche},\ and\
  \citenamefont {Lyubovitskij}}]{Zhevlakov:2018rwo}%
  \BibitemOpen
  \bibfield  {author} {\bibinfo {author} {\bibfnamefont {A.~S.}\ \bibnamefont
  {Zhevlakov}}, \bibinfo {author} {\bibfnamefont {M.}~\bibnamefont
  {Gorchtein}}, \bibinfo {author} {\bibfnamefont {A.~N.}\ \bibnamefont
  {Hiller~Blin}}, \bibinfo {author} {\bibfnamefont {T.}~\bibnamefont
  {Gutsche}}, \ and\ \bibinfo {author} {\bibfnamefont {V.~E.}\ \bibnamefont
  {Lyubovitskij}},\ }\href {\doibase 10.1103/PhysRevD.99.031703} {\bibfield
  {journal} {\bibinfo  {journal} {Phys. Rev. D}\ }\textbf {\bibinfo {volume}
  {99}},\ \bibinfo {pages} {031703} (\bibinfo {year} {2019}{\natexlab{a}})},\
  \Eprint {http://arxiv.org/abs/1812.00171} {arXiv:1812.00171 [hep-ph]}
  \BibitemShut {NoStop}%
\bibitem [{\citenamefont {Zhevlakov}\ \emph
  {et~al.}(2019{\natexlab{b}})\citenamefont {Zhevlakov}, \citenamefont
  {Gutsche},\ and\ \citenamefont {Lyubovitskij}}]{Zhevlakov:2019ymi}%
  \BibitemOpen
  \bibfield  {author} {\bibinfo {author} {\bibfnamefont {A.~S.}\ \bibnamefont
  {Zhevlakov}}, \bibinfo {author} {\bibfnamefont {T.}~\bibnamefont {Gutsche}},
  \ and\ \bibinfo {author} {\bibfnamefont {V.~E.}\ \bibnamefont
  {Lyubovitskij}},\ }\href {\doibase 10.1103/PhysRevD.99.115004} {\bibfield
  {journal} {\bibinfo  {journal} {Phys. Rev. D}\ }\textbf {\bibinfo {volume}
  {99}},\ \bibinfo {pages} {115004} (\bibinfo {year} {2019}{\natexlab{b}})},\
  \Eprint {http://arxiv.org/abs/1904.08154} {arXiv:1904.08154 [hep-ph]}
  \BibitemShut {NoStop}%
\bibitem [{\citenamefont {Gutsche}\ \emph {et~al.}(2017)\citenamefont
  {Gutsche}, \citenamefont {Hiller~Blin}, \citenamefont {Kovalenko},
  \citenamefont {Kuleshov}, \citenamefont {Lyubovitskij}, \citenamefont
  {Vicente~Vacas},\ and\ \citenamefont {Zhevlakov}}]{Gutsche:2016jap}%
  \BibitemOpen
  \bibfield  {author} {\bibinfo {author} {\bibfnamefont {T.}~\bibnamefont
  {Gutsche}}, \bibinfo {author} {\bibfnamefont {A.~N.}\ \bibnamefont
  {Hiller~Blin}}, \bibinfo {author} {\bibfnamefont {S.}~\bibnamefont
  {Kovalenko}}, \bibinfo {author} {\bibfnamefont {S.}~\bibnamefont {Kuleshov}},
  \bibinfo {author} {\bibfnamefont {V.~E.}\ \bibnamefont {Lyubovitskij}},
  \bibinfo {author} {\bibfnamefont {M.~J.}\ \bibnamefont {Vicente~Vacas}}, \
  and\ \bibinfo {author} {\bibfnamefont {A.}~\bibnamefont {Zhevlakov}},\ }\href
  {\doibase 10.1103/PhysRevD.95.036022} {\bibfield  {journal} {\bibinfo
  {journal} {Phys. Rev. D}\ }\textbf {\bibinfo {volume} {95}},\ \bibinfo
  {pages} {036022} (\bibinfo {year} {2017})},\ \Eprint
  {http://arxiv.org/abs/1612.02276} {arXiv:1612.02276 [hep-ph]} \BibitemShut
  {NoStop}%
\bibitem [{\citenamefont {Maison}\ and\ \citenamefont
  {Skripnikov}(2022)}]{Maison:2022zaz}%
  \BibitemOpen
  \bibfield  {author} {\bibinfo {author} {\bibfnamefont {D.~E.}\ \bibnamefont
  {Maison}}\ and\ \bibinfo {author} {\bibfnamefont {L.~V.}\ \bibnamefont
  {Skripnikov}},\ }\href {\doibase 10.1103/PhysRevA.105.032813} {\bibfield
  {journal} {\bibinfo  {journal} {Phys. Rev. A}\ }\textbf {\bibinfo {volume}
  {105}},\ \bibinfo {pages} {032813} (\bibinfo {year} {2022})},\ \Eprint
  {http://arxiv.org/abs/2201.12574} {arXiv:2201.12574 [physics.atom-ph]}
  \BibitemShut {NoStop}%
\bibitem [{\citenamefont {Kirpichnikov}\ \emph {et~al.}()\citenamefont
  {Kirpichnikov}, \citenamefont {Lyubovitskij},\ and\ \citenamefont
  {Zhevlakov}}]{ZhevlakovEtAl}%
  \BibitemOpen
  \bibfield  {author} {\bibinfo {author} {\bibfnamefont {D.~V.}\ \bibnamefont
  {Kirpichnikov}}, \bibinfo {author} {\bibfnamefont {V.~E.}\ \bibnamefont
  {Lyubovitskij}}, \ and\ \bibinfo {author} {\bibfnamefont {A.~S.}\
  \bibnamefont {Zhevlakov}},\ }\href@noop {} {\bibinfo  {journal} {In
  preparation}\ }\BibitemShut {NoStop}%
\bibitem [{\citenamefont {Belyaev}\ \emph {et~al.}(2013)\citenamefont
  {Belyaev}, \citenamefont {Christensen},\ and\ \citenamefont
  {Pukhov}}]{Belyaev:2012qa}%
  \BibitemOpen
\bibfield  {journal} {  }\bibfield  {author} {\bibinfo {author} {\bibfnamefont
  {A.}~\bibnamefont {Belyaev}}, \bibinfo {author} {\bibfnamefont {N.~D.}\
  \bibnamefont {Christensen}}, \ and\ \bibinfo {author} {\bibfnamefont
  {A.}~\bibnamefont {Pukhov}},\ }\href {\doibase 10.1016/j.cpc.2013.01.014}
  {\bibfield  {journal} {\bibinfo  {journal} {Comput. Phys. Commun.}\ }\textbf
  {\bibinfo {volume} {184}},\ \bibinfo {pages} {1729} (\bibinfo {year}
  {2013})},\ \Eprint {http://arxiv.org/abs/1207.6082} {arXiv:1207.6082
  [hep-ph]} \BibitemShut {NoStop}%
\bibitem [{\citenamefont {Budnev}\ \emph {et~al.}(1975)\citenamefont {Budnev},
  \citenamefont {Ginzburg}, \citenamefont {Meledin},\ and\ \citenamefont
  {Serbo}}]{Budnev:1975poe}%
  \BibitemOpen
  \bibfield  {author} {\bibinfo {author} {\bibfnamefont {V.~M.}\ \bibnamefont
  {Budnev}}, \bibinfo {author} {\bibfnamefont {I.~F.}\ \bibnamefont
  {Ginzburg}}, \bibinfo {author} {\bibfnamefont {G.~V.}\ \bibnamefont
  {Meledin}}, \ and\ \bibinfo {author} {\bibfnamefont {V.~G.}\ \bibnamefont
  {Serbo}},\ }\href {\doibase 10.1016/0370-1573(75)90009-5} {\bibfield
  {journal} {\bibinfo  {journal} {Phys. Rept.}\ }\textbf {\bibinfo {volume}
  {15}},\ \bibinfo {pages} {181} (\bibinfo {year} {1975})}\BibitemShut
  {NoStop}%
\bibitem [{\citenamefont {Engel}\ \emph {et~al.}(1996)\citenamefont {Engel},
  \citenamefont {Schiller},\ and\ \citenamefont {Serbo}}]{Engel:1995pu}%
  \BibitemOpen
  \bibfield  {author} {\bibinfo {author} {\bibfnamefont {R.}~\bibnamefont
  {Engel}}, \bibinfo {author} {\bibfnamefont {A.}~\bibnamefont {Schiller}}, \
  and\ \bibinfo {author} {\bibfnamefont {V.~G.}\ \bibnamefont {Serbo}},\ }\href
  {\doibase 10.1007/s002880050214} {\bibfield  {journal} {\bibinfo  {journal}
  {Z. Phys. C}\ }\textbf {\bibinfo {volume} {71}},\ \bibinfo {pages} {651}
  (\bibinfo {year} {1996})},\ \Eprint {http://arxiv.org/abs/hep-ph/9511262}
  {arXiv:hep-ph/9511262} \BibitemShut {NoStop}%
\bibitem [{\citenamefont {Vysotsky}\ and\ \citenamefont
  {Zhemchugov}(2019)}]{Vysotsky:2018slo}%
  \BibitemOpen
  \bibfield  {author} {\bibinfo {author} {\bibfnamefont {M.~I.}\ \bibnamefont
  {Vysotsky}}\ and\ \bibinfo {author} {\bibfnamefont {E.}~\bibnamefont
  {Zhemchugov}},\ }\href {\doibase 10.3367/UFNe.2018.07.038389} {\bibfield
  {journal} {\bibinfo  {journal} {Phys. Usp.}\ }\textbf {\bibinfo {volume}
  {62}},\ \bibinfo {pages} {910} (\bibinfo {year} {2019})},\ \Eprint
  {http://arxiv.org/abs/1806.07238} {arXiv:1806.07238 [hep-ph]} \BibitemShut
  {NoStop}%
\bibitem [{\citenamefont {Bondi}\ \emph {et~al.}(2021)\citenamefont {Bondi},
  \citenamefont {Celentano}, \citenamefont {Dusaev}, \citenamefont
  {Kirpichnikov}, \citenamefont {Kirsanov}, \citenamefont {Krasnikov},
  \citenamefont {Marsicano},\ and\ \citenamefont {Shchukin}}]{Bondi:2021nfp}%
  \BibitemOpen
  \bibfield  {author} {\bibinfo {author} {\bibfnamefont {M.}~\bibnamefont
  {Bondi}}, \bibinfo {author} {\bibfnamefont {A.}~\bibnamefont {Celentano}},
  \bibinfo {author} {\bibfnamefont {R.~R.}\ \bibnamefont {Dusaev}}, \bibinfo
  {author} {\bibfnamefont {D.~V.}\ \bibnamefont {Kirpichnikov}}, \bibinfo
  {author} {\bibfnamefont {M.~M.}\ \bibnamefont {Kirsanov}}, \bibinfo {author}
  {\bibfnamefont {N.~V.}\ \bibnamefont {Krasnikov}}, \bibinfo {author}
  {\bibfnamefont {L.}~\bibnamefont {Marsicano}}, \ and\ \bibinfo {author}
  {\bibfnamefont {D.}~\bibnamefont {Shchukin}},\ }\href {\doibase
  10.1016/j.cpc.2021.108129} {\bibfield  {journal} {\bibinfo  {journal}
  {Comput. Phys. Commun.}\ }\textbf {\bibinfo {volume} {269}},\ \bibinfo
  {pages} {108129} (\bibinfo {year} {2021})},\ \Eprint
  {http://arxiv.org/abs/2101.12192} {arXiv:2101.12192 [hep-ph]} \BibitemShut
  {NoStop}%
\bibitem [{\citenamefont {Chu}\ \emph {et~al.}(2019)\citenamefont {Chu},
  \citenamefont {Pradler},\ and\ \citenamefont {Semmelrock}}]{Chu:2018qrm}%
  \BibitemOpen
  \bibfield  {author} {\bibinfo {author} {\bibfnamefont {X.}~\bibnamefont
  {Chu}}, \bibinfo {author} {\bibfnamefont {J.}~\bibnamefont {Pradler}}, \ and\
  \bibinfo {author} {\bibfnamefont {L.}~\bibnamefont {Semmelrock}},\ }\href
  {\doibase 10.1103/PhysRevD.99.015040} {\bibfield  {journal} {\bibinfo
  {journal} {Phys. Rev. D}\ }\textbf {\bibinfo {volume} {99}},\ \bibinfo
  {pages} {015040} (\bibinfo {year} {2019})},\ \Eprint
  {http://arxiv.org/abs/1811.04095} {arXiv:1811.04095 [hep-ph]} \BibitemShut
  {NoStop}%
\bibitem [{\citenamefont {Chen}\ \emph {et~al.}(2017)\citenamefont {Chen},
  \citenamefont {Pospelov},\ and\ \citenamefont {Zhong}}]{Chen:2017awl}%
  \BibitemOpen
  \bibfield  {author} {\bibinfo {author} {\bibfnamefont {C.-Y.}\ \bibnamefont
  {Chen}}, \bibinfo {author} {\bibfnamefont {M.}~\bibnamefont {Pospelov}}, \
  and\ \bibinfo {author} {\bibfnamefont {Y.-M.}\ \bibnamefont {Zhong}},\ }\href
  {\doibase 10.1103/PhysRevD.95.115005} {\bibfield  {journal} {\bibinfo
  {journal} {Phys. Rev. D}\ }\textbf {\bibinfo {volume} {95}},\ \bibinfo
  {pages} {115005} (\bibinfo {year} {2017})},\ \Eprint
  {http://arxiv.org/abs/1701.07437} {arXiv:1701.07437 [hep-ph]} \BibitemShut
  {NoStop}%
\bibitem [{\citenamefont {Gninenko}\ \emph
  {et~al.}(2019{\natexlab{b}})\citenamefont {Gninenko}, \citenamefont
  {Kirpichnikov},\ and\ \citenamefont {Krasnikov}}]{Gninenko:2018ter}%
  \BibitemOpen
  \bibfield  {author} {\bibinfo {author} {\bibfnamefont {S.~N.}\ \bibnamefont
  {Gninenko}}, \bibinfo {author} {\bibfnamefont {D.~V.}\ \bibnamefont
  {Kirpichnikov}}, \ and\ \bibinfo {author} {\bibfnamefont {N.~V.}\
  \bibnamefont {Krasnikov}},\ }\href {\doibase 10.1103/PhysRevD.100.035003}
  {\bibfield  {journal} {\bibinfo  {journal} {Phys. Rev. D}\ }\textbf {\bibinfo
  {volume} {100}},\ \bibinfo {pages} {035003} (\bibinfo {year}
  {2019}{\natexlab{b}})},\ \Eprint {http://arxiv.org/abs/1810.06856}
  {arXiv:1810.06856 [hep-ph]} \BibitemShut {NoStop}%
\bibitem [{\citenamefont {Bjorken}\ \emph {et~al.}(2009)\citenamefont
  {Bjorken}, \citenamefont {Essig}, \citenamefont {Schuster},\ and\
  \citenamefont {Toro}}]{Bjorken:2009mm}%
  \BibitemOpen
  \bibfield  {author} {\bibinfo {author} {\bibfnamefont {J.~D.}\ \bibnamefont
  {Bjorken}}, \bibinfo {author} {\bibfnamefont {R.}~\bibnamefont {Essig}},
  \bibinfo {author} {\bibfnamefont {P.}~\bibnamefont {Schuster}}, \ and\
  \bibinfo {author} {\bibfnamefont {N.}~\bibnamefont {Toro}},\ }\href {\doibase
  10.1103/PhysRevD.80.075018} {\bibfield  {journal} {\bibinfo  {journal} {Phys.
  Rev. D}\ }\textbf {\bibinfo {volume} {80}},\ \bibinfo {pages} {075018}
  (\bibinfo {year} {2009})},\ \Eprint {http://arxiv.org/abs/0906.0580}
  {arXiv:0906.0580 [hep-ph]} \BibitemShut {NoStop}%
\bibitem [{\citenamefont {Tsai}(1986)}]{Tsai:1986tx}%
  \BibitemOpen
  \bibfield  {author} {\bibinfo {author} {\bibfnamefont {Y.-S.}\ \bibnamefont
  {Tsai}},\ }\href {\doibase 10.1103/PhysRevD.34.1326} {\bibfield  {journal}
  {\bibinfo  {journal} {Phys. Rev. D}\ }\textbf {\bibinfo {volume} {34}},\
  \bibinfo {pages} {1326} (\bibinfo {year} {1986})}\BibitemShut {NoStop}%
\bibitem [{\citenamefont {Abi}\ \emph {et~al.}(2021)\citenamefont {Abi} \emph
  {et~al.}}]{Muong-2:2021ojo}%
  \BibitemOpen
  \bibfield  {author} {\bibinfo {author} {\bibfnamefont {B.}~\bibnamefont
  {Abi}} \emph {et~al.} (\bibinfo {collaboration} {Muon g-2}),\ }\href
  {\doibase 10.1103/PhysRevLett.126.141801} {\bibfield  {journal} {\bibinfo
  {journal} {Phys. Rev. Lett.}\ }\textbf {\bibinfo {volume} {126}},\ \bibinfo
  {pages} {141801} (\bibinfo {year} {2021})},\ \Eprint
  {http://arxiv.org/abs/2104.03281} {arXiv:2104.03281 [hep-ex]} \BibitemShut
  {NoStop}%
\bibitem [{\citenamefont {Parker}\ \emph {et~al.}(2018)\citenamefont {Parker},
  \citenamefont {Yu}, \citenamefont {Zhong}, \citenamefont {Estey},\ and\
  \citenamefont {M\"uller}}]{Parker:2018vye}%
  \BibitemOpen
  \bibfield  {author} {\bibinfo {author} {\bibfnamefont {R.~H.}\ \bibnamefont
  {Parker}}, \bibinfo {author} {\bibfnamefont {C.}~\bibnamefont {Yu}}, \bibinfo
  {author} {\bibfnamefont {W.}~\bibnamefont {Zhong}}, \bibinfo {author}
  {\bibfnamefont {B.}~\bibnamefont {Estey}}, \ and\ \bibinfo {author}
  {\bibfnamefont {H.}~\bibnamefont {M\"uller}},\ }\href {\doibase
  10.1126/science.aap7706} {\bibfield  {journal} {\bibinfo  {journal}
  {Science}\ }\textbf {\bibinfo {volume} {360}},\ \bibinfo {pages} {191}
  (\bibinfo {year} {2018})},\ \Eprint {http://arxiv.org/abs/1812.04130}
  {arXiv:1812.04130 [physics.atom-ph]} \BibitemShut {NoStop}%
\bibitem [{\citenamefont {Zyla}\ \emph
  {et~al.}(2020{\natexlab{b}})\citenamefont {Zyla} \emph
  {et~al.}}]{ParticleDataGroup:2020ssz}%
  \BibitemOpen
  \bibfield  {author} {\bibinfo {author} {\bibfnamefont {P.~A.}\ \bibnamefont
  {Zyla}} \emph {et~al.} (\bibinfo {collaboration} {Particle Data Group}),\
  }\href {\doibase 10.1093/ptep/ptaa104} {\bibfield  {journal} {\bibinfo
  {journal} {PTEP}\ }\textbf {\bibinfo {volume} {2020}},\ \bibinfo {pages}
  {083C01} (\bibinfo {year} {2020}{\natexlab{b}})}\BibitemShut {NoStop}%
\bibitem [{\citenamefont {Blokland}\ \emph {et~al.}(2002)\citenamefont
  {Blokland}, \citenamefont {Czarnecki},\ and\ \citenamefont
  {Melnikov}}]{Blokland:2001pb}%
  \BibitemOpen
  \bibfield  {author} {\bibinfo {author} {\bibfnamefont {I.~R.}\ \bibnamefont
  {Blokland}}, \bibinfo {author} {\bibfnamefont {A.}~\bibnamefont {Czarnecki}},
  \ and\ \bibinfo {author} {\bibfnamefont {K.}~\bibnamefont {Melnikov}},\
  }\href {\doibase 10.1103/PhysRevLett.88.071803} {\bibfield  {journal}
  {\bibinfo  {journal} {Phys. Rev. Lett.}\ }\textbf {\bibinfo {volume} {88}},\
  \bibinfo {pages} {071803} (\bibinfo {year} {2002})},\ \Eprint
  {http://arxiv.org/abs/hep-ph/0112117} {arXiv:hep-ph/0112117} \BibitemShut
  {NoStop}%
\bibitem [{\citenamefont {Mangoni}(2020)}]{Mangoni:2020zgq}%
  \BibitemOpen
  \bibfield  {author} {\bibinfo {author} {\bibfnamefont {A.}~\bibnamefont
  {Mangoni}},\ }\emph {\bibinfo {title} {{Hadronic decays of the J/psi
  meson}}},\ \href@noop {} {Ph.D. thesis},\ \bibinfo  {school} {Universita' Di
  Perugia} (\bibinfo {year} {2020}),\ \Eprint {http://arxiv.org/abs/2002.09675}
  {arXiv:2002.09675 [hep-ph]} \BibitemShut {NoStop}%
\bibitem [{\citenamefont {Shtabovenko}\ \emph {et~al.}(2016)\citenamefont
  {Shtabovenko}, \citenamefont {Mertig},\ and\ \citenamefont
  {Orellana}}]{Shtabovenko:2016sxi}%
  \BibitemOpen
  \bibfield  {author} {\bibinfo {author} {\bibfnamefont {V.}~\bibnamefont
  {Shtabovenko}}, \bibinfo {author} {\bibfnamefont {R.}~\bibnamefont {Mertig}},
  \ and\ \bibinfo {author} {\bibfnamefont {F.}~\bibnamefont {Orellana}},\
  }\href {\doibase 10.1016/j.cpc.2016.06.008} {\bibfield  {journal} {\bibinfo
  {journal} {Comput. Phys. Commun.}\ }\textbf {\bibinfo {volume} {207}},\
  \bibinfo {pages} {432} (\bibinfo {year} {2016})},\ \Eprint
  {http://arxiv.org/abs/1601.01167} {arXiv:1601.01167 [hep-ph]} \BibitemShut
  {NoStop}%
\bibitem [{\citenamefont {Inc.}()}]{Mathematica}%
  \BibitemOpen
  \bibfield  {author} {\bibinfo {author} {\bibfnamefont {W.~R.}\ \bibnamefont
  {Inc.}},\ }\href {https://www.wolfram.com/mathematica} {\enquote {\bibinfo
  {title} {Mathematica, {V}ersion 13.0.0},}\ }\bibinfo {note} {Champaign, IL,
  2021}\BibitemShut {NoStop}%
\bibitem [{\citenamefont {Arefyeva}\ \emph {et~al.}(2022)\citenamefont
  {Arefyeva}, \citenamefont {Gninenko}, \citenamefont {Gorbunov},\ and\
  \citenamefont {Kirpichnikov}}]{Arefyeva:2022eba}%
  \BibitemOpen
  \bibfield  {author} {\bibinfo {author} {\bibfnamefont {N.}~\bibnamefont
  {Arefyeva}}, \bibinfo {author} {\bibfnamefont {S.}~\bibnamefont {Gninenko}},
  \bibinfo {author} {\bibfnamefont {D.}~\bibnamefont {Gorbunov}}, \ and\
  \bibinfo {author} {\bibfnamefont {D.}~\bibnamefont {Kirpichnikov}},\
  }\href@noop {} {\  (\bibinfo {year} {2022})},\ \Eprint
  {http://arxiv.org/abs/2204.03984} {arXiv:2204.03984 [hep-ph]} \BibitemShut
  {NoStop}%
\bibitem [{\citenamefont {D\"obrich}\ \emph {et~al.}(2016)\citenamefont
  {D\"obrich}, \citenamefont {Jaeckel}, \citenamefont {Kahlhoefer},
  \citenamefont {Ringwald},\ and\ \citenamefont {Schmidt-Hoberg}}]{ALPtraum}%
  \BibitemOpen
  \bibfield  {author} {\bibinfo {author} {\bibfnamefont {B.}~\bibnamefont
  {D\"obrich}}, \bibinfo {author} {\bibfnamefont {J.}~\bibnamefont {Jaeckel}},
  \bibinfo {author} {\bibfnamefont {F.}~\bibnamefont {Kahlhoefer}}, \bibinfo
  {author} {\bibfnamefont {A.}~\bibnamefont {Ringwald}}, \ and\ \bibinfo
  {author} {\bibfnamefont {K.}~\bibnamefont {Schmidt-Hoberg}},\ }\href
  {\doibase 10.1007/JHEP02(2016)018} {\bibfield  {journal} {\bibinfo  {journal}
  {JHEP}\ }\textbf {\bibinfo {volume} {02}},\ \bibinfo {pages} {018} (\bibinfo
  {year} {2016})},\ \Eprint {http://arxiv.org/abs/1512.03069} {arXiv:1512.03069
  [hep-ph]} \BibitemShut {NoStop}%
\bibitem [{\citenamefont {Byckling}\ and\ \citenamefont
  {Kajantie}(1971)}]{Byckling:1971vca}%
  \BibitemOpen
  \bibfield  {author} {\bibinfo {author} {\bibfnamefont {E.}~\bibnamefont
  {Byckling}}\ and\ \bibinfo {author} {\bibfnamefont {K.}~\bibnamefont
  {Kajantie}},\ }\href@noop {} {\emph {\bibinfo {title} {{Particle
  Kinematics}}}}\ (\bibinfo  {publisher} {University of Jyvaskyla},\ \bibinfo
  {address} {Jyvaskyla, Finland},\ \bibinfo {year} {1971})\BibitemShut
  {NoStop}%
\bibitem [{\citenamefont {Liu}\ and\ \citenamefont
  {Miller}(2017)}]{Liu:2017htz}%
  \BibitemOpen
  \bibfield  {author} {\bibinfo {author} {\bibfnamefont {Y.-S.}\ \bibnamefont
  {Liu}}\ and\ \bibinfo {author} {\bibfnamefont {G.~A.}\ \bibnamefont
  {Miller}},\ }\href {\doibase 10.1103/PhysRevD.96.016004} {\bibfield
  {journal} {\bibinfo  {journal} {Phys. Rev. D}\ }\textbf {\bibinfo {volume}
  {96}},\ \bibinfo {pages} {016004} (\bibinfo {year} {2017})},\ \Eprint
  {http://arxiv.org/abs/1705.01633} {arXiv:1705.01633 [hep-ph]} \BibitemShut
  {NoStop}%
\end{thebibliography}%

\end{document}